\documentclass[pra,twocolumn,superscriptaddress,longbibliography]{revtex4-2}
\usepackage{array}[=2016-10-06]
\usepackage{latexsym}
\usepackage{cmap}
\usepackage[utf8]{inputenc}
\usepackage{nicefrac}
\usepackage[english]{babel}
\usepackage{amsmath,amssymb}
\usepackage{xcolor}
\usepackage{graphicx}
\usepackage{multirow}
\usepackage{dcolumn}
\usepackage{bm}
\usepackage{braket}
\usepackage{amsthm}
\usepackage[normalem]{ulem}
\usepackage[colorlinks=false]{hyperref}
\usepackage{booktabs}
\usepackage{float}
\usepackage{enumitem}
\usepackage{appendix}
\usepackage{chngcntr}
\usepackage{apptools}
\usepackage{multirow}
\usepackage{siunitx}
\usepackage{physics}
\usepackage[version=4]{mhchem}
\usepackage{mathtools}
\usepackage{leftindex}
\usepackage[titles]{tocloft}
\usepackage{pifont}
\usepackage[caption=false]{subfig}
\usepackage[export]{adjustbox}

\makeatletter
\newcommand{\setonecolumn}{%
  \par\ignorespaces
  \setbox\widetext@top\vbox{%
   \hb@xt@\hsize{%
    \leaders\hrule\hfil
    \vrule\@height6\p@
   }%
  }%
  \setbox\widetext@bot\hb@xt@\hsize{%
    \vrule\@depth6\p@
    \leaders\hrule\hfil
  }%
  \onecolumngrid
  \vskip10\p@
  \dimen@\ht\widetext@top\advance\dimen@\dp\widetext@top
  \cleaders\box\widetext@top\vskip\dimen@
  \vskip6\p@
  \prep@math@patch
}
\newcommand{\settwocolumn}{%
  \par
  \vskip6\p@
  \setbox\widetext@bot\vbox{%
   \hb@xt@\hsize{\hfil\box\widetext@bot}%
  }%
  \dimen@\ht\widetext@bot\advance\dimen@\dp\widetext@bot
  \cleaders\box\widetext@bot\vskip\dimen@
  \vskip8.5\p@
  \twocolumngrid\global\@ignoretrue
  \@endpetrue
}
\makeatother

\newcolumntype{P}{>{\raggedright\let\newline\\\arraybackslash\hspace{0pt}}p}
\AtBeginDocument{\RenewCommandCopy\qty\SI}
\DeclareSIUnit{\Ry}{Ry}

\newcommand{\todo}[1]{%
  \expandafter\ifx\expandafter\relax\detokenize{#1}\relax%
    \textcolor{red}{[TODO]}%
  \else%
    \textcolor{red}{[TODO: #1]}%
  \fi%
}

\newcommand{\citett}[1]{\citeauthor{#1}~\cite{#1}}

\newtheorem{proposition}{Proposition}

\newcommand{\ltw}[1]{\lceil\log_2 #1\rceil}

\graphicspath{{./figures_final/} {./}}

\begin{document}

\title{Benchmarking Quantum Simulation of Chemical Hamiltonians using the Sorted-List Encoding}
\author{Calvin Ku}
\affiliation{Hon Hai Research Institute, Taipei, Taiwan}
\affiliation{Department of Mechanical Engineering, City University of Hong Kong, Kowloon, Hong Kong SAR 999077, China}

\author{Yu-Cheng Chen}
\affiliation{Hon Hai Research Institute, Taipei, Taiwan}

\author{Alice Hu}
\email{alicehu@cityu.edu.hk}
\affiliation{Department of Mechanical Engineering, City University of Hong Kong, Kowloon, Hong Kong SAR 999077, China}
\affiliation{Department of Material Science and Engineering, City University of Hong Kong, Kowloon, Hong Kong SAR 999077, China}

\author{Min-Hsiu Hsieh}
\email{min-hsiu.hsieh@foxconn.com}
\affiliation{Hon Hai Research Institute, Taipei, Taiwan}

\begin{abstract}
    Quantum Phase Estimation (QPE) is a cornerstone algorithm for fault-tolerant quantum computation, especially for electronic structure calculations of chemical systems.
    Optimal simulation relies on a complex trade-offs across many parameters including Hamiltonian simulation techniques, basis sets, and the fermion-to-qubit encodings.
    Here, we characterize the trade-offs and quantify the quantum resource costs of the sorted-list encoding as a particle-conserving, low-qubit alternative to the Jordan-Wigner encoding.
    We identify specific regimes, across different simulation techniques and basis sets, where the sorted-list encoding would be favorable compared to existing methods.
    Our findings are further supported through numerical benchmarks of real-world chemical systems.
    We found the sorted-list encoding to be a viable alternative to the Jordan-Wigner encoding for the compact molecular orbital basis when the electron-filling ratio is low, which typically occurs when high-precision results are required.
    In the plane-wave basis, we found similar asymptotic gate and qubit scaling between the sorted-list and the first-quantized encoding, although the first-quantized encoding still retains lower constant factors.
\end{abstract}

\maketitle

\section{Introduction}
\label{sec:intro}
Significant advancements in quantum computing have opened up unprecedented opportunities for solving complex problems, particularly in the realm of quantum chemistry.
The ability to accurately model and simulate molecular systems on quantum computers holds the potential to revolutionize drug discovery, materials science, and fundamental chemical research.
Central to these simulations is the effective implementation of quantum chemical Hamiltonians
onto quantum circuits, a task primarily addressed by two prominent methodologies:
Trotterization~\cite{lowComplexityImplementingTrotter2023,guntherPhaseEstimationPartially2026} and the qubitization approaches~\cite{childsHamiltonianSimulationUsing2012,kothariEfficientAlgorithmsQuantum2014}.
As the major costs of most quantum simulation algorithms involves repeated applications of these methods~\cite{lowComplexityImplementingTrotter2023,guntherPhaseEstimationPartially2026,childsHamiltonianSimulationUsing2012,kothariEfficientAlgorithmsQuantum2014,lowHamiltonianSimulationQubitization2019,berryQubitizationArbitraryBasis2019,vonburgQuantumComputingEnhanced2021,leeEvenMoreEfficient2021,poulinTrotterStepSize2014,babbushChemicalBasisTrotterSuzuki2015}, it is crucial to have efficient implementations of both.

A key advantage of Trotterization lies in its conceptual simplicity and its minimal use of ancilla qubits.
However, its computational cost typically scales polynomially with the inverse of the desired error.
Furthermore, tight theoretical error bounds can be challenging to obtain, often being several times larger than experimental error calculations~\cite{babbushChemicalBasisTrotterSuzuki2015}.
Consequently, Trotterization is often considered more suitable for small to medium-sized system on near-term partially fault-tolerant architectures.
On the other hand, qubitization methods fare better in terms of their error scaling.
As an example, Hamiltonian evolution based on qubitization~\cite{lowHamiltonianSimulationQubitization2019} achieves computational costs with logarithmic dependence on the reciprocal of the error parameter.
This comes at a cost of a more complex circuit constructions, often with multiple Quantum Read-Only Memory (QROM) implementations, and requiring additional ancilla qubits.
This makes qubitization-based approaches particularly well-suited for larger systems in future fault-tolerant quantum computers.

\begin{figure*}[!t]
  \centering
  \includegraphics[width=\linewidth]{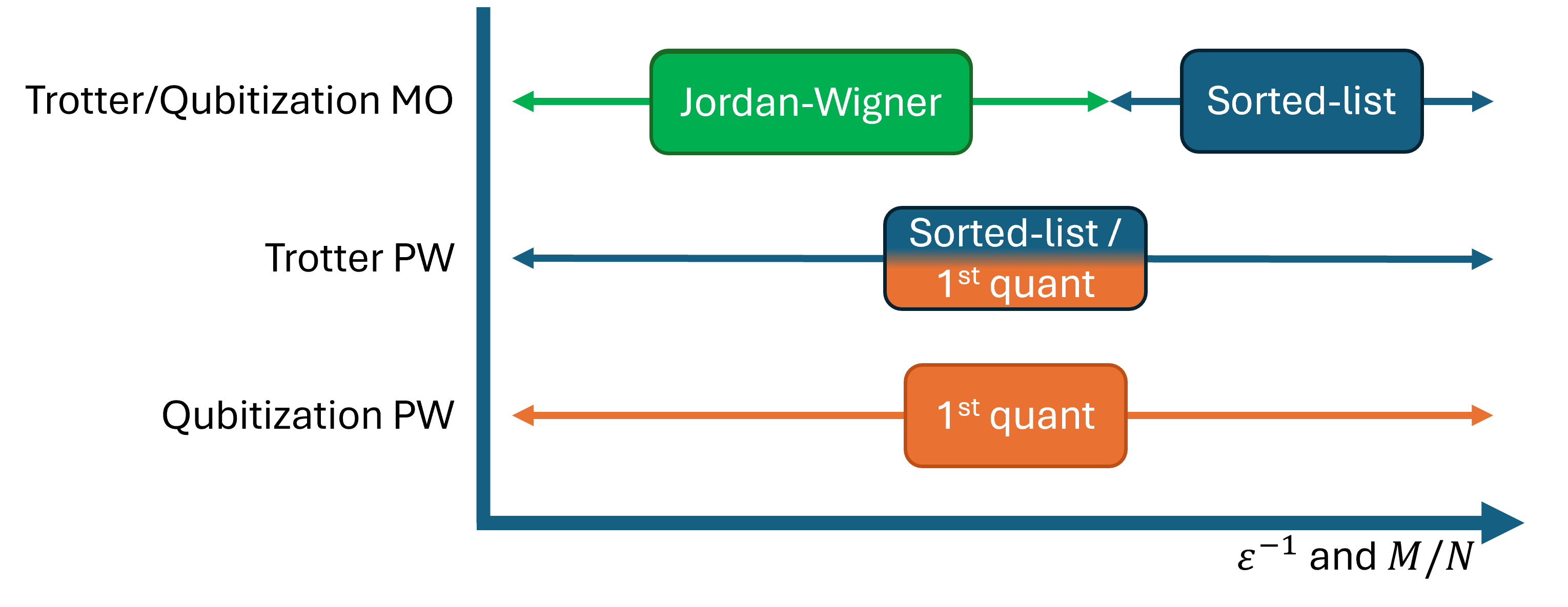}
  \caption{
    Phase diagram on the choice of encoding schemes for performing energy estimation algorithm using quantum computers.
    The $x$-axis represents increasing accuracy requirement which necessitates smaller precision parameter $\varepsilon$ and smaller electron filling ratios $N/M$.
  The optimal encoding schemes differs with the choice of basis-set (molecular orbitals or plane-waves) and between Trotterization and qubitization techniques.
  }
  \label{fig:intro_figure}
\end{figure*}

To accurately quantify the costs associated with both methods, it is essential to consider the various parameters affecting the computational costs and qubit requirements.
The major parameters include the fermion-to-qubit encoding and the basis set used for the calculations.
For fermion-to-qubit encodings, the Jordan-Wigner (JW)~\cite{jordanUeberPaulischeAequivalenzverbot1928} encoding has been the standard for previous cost analyses for Trotterization~\cite{guntherPhaseEstimationPartially2026,clintonNeartermQuantumSimulation2024,zhaoEntanglementAcceleratesQuantum2025} and qubitization~\cite{berryQubitizationArbitraryBasis2019,vonburgQuantumComputingEnhanced2021,leeEvenMoreEfficient2021}.
An obvious step to improve qubit costs is to use an encoding that leverages the particle conservation symmetry inherent in quantum chemical Hamiltonians.
To this end, the first-quantized encoding~\cite{kassalPolynomialtimeQuantumAlgorithm2008,babbushQuantumSimulationChemistry2019,suFaultTolerantQuantumSimulations2021} offers an improvement in qubit cost in the $N\ll M$ regime, scaling as $\mathcal{O}(N\log M)$ compared to $\mathcal{O}(M)$ for the Jordan-Wigner encoding, where $M$ is the number of orbitals and $N$ is the number of electrons.
For second-quantized encodings, the sorted-list encoding~\cite{carolanSuccinctFermionData2024} shows reasonable theoretical improvements over the Jordan-Wigner encoding, achieving $\mathcal{O}(N\log M)$ scaling in both qubit count and gate cost to implement a fermionic operator.
As a result, the sorted-list encoding could offer advantages over the Jordan-Wigner encoding when $N \ll M$ and in the limit of large $M$.
Furthermore, the sorted-list encoding can be efficiently converted to the first-quantized encoding and vice-versa, allowing algorithms designed from different quantization to be applied in the same simulation~\cite{kuOptimizingQuantumChemistry2026}.
In this work, we quantify the constant factors associated with the sorted-list encoding and demonstrated a qubit advantage over the Jordan-Wigner encoding for both Trotterization and qubitization circuits with minimal to moderate increases in circuit costs, as summarized in \autoref{fig:intro_figure}.

Another critical parameter affecting the costs associated with Trotterization and qubitization is the choice of the basis set.
The molecular orbital (MO) basis provides a compact representation of the chemical system, requiring a relatively smaller number of orbitals compared to the plane-wave basis.
Since MOs are constructed as eigenvectors of mean-field Hartree-Fock calculations, their associated orbital energies simplify the implementation of active space methods for ground-state calculations.
This technique removes certain orbitals and electrons from the quantum calculation, with a small trade-off in ground-state accuracy, by assuming that orbitals associated with the lowest and highest energies are always occupied or unoccupied, respectively, rather than being in a superposition.
However, the MO basis comes with its own trade-offs.
For qubitization, the arbitrary nature of the $h_{pq}$ and $h_{pqrs}$ coefficients necessitates extensive use of QROMs for their implementation.
While these costs can be somewhat reduced through rank-reduction techniques~\cite{berryQubitizationArbitraryBasis2019,vonburgQuantumComputingEnhanced2021,leeEvenMoreEfficient2021}, this involves expensive classical preprocessing.

On the other hand, plane-wave basis sets possess analytical coefficients that can be efficiently computed within quantum circuits.
This significantly benefits the qubitization circuits, as they rely on QROMs to a lesser extent compared to MO basis sets, achieving sublinear scaling in gate cost with respect to the number of plane-waves~\cite{babbushQuantumSimulationChemistry2019,suFaultTolerantQuantumSimulations2021}.
While existing literature~\cite{suFaultTolerantQuantumSimulations2021,suNearlyTightTrotterization2021} suggests such low scaling with respect to the number of plane waves is exclusive to the first-quantized representation, we show in \autoref{prop:qubitization_sorted_list} that the second-quantized qubitization algorithm can achieve equivalent asymptotic gate and qubit cost scaling with the first-quantized algorithm.
The structured nature of the plane-wave basis also greatly benefits Trotterization circuits, as all Hamiltonian terms can be grouped into two unitaries: one for kinetic terms and another for potential terms, with efficient quantum Fourier transform circuits enabling conversion between the plane-wave and its dual basis.
This is because the kinetic term of the Hamiltonian is diagonal in the plane-wave basis, and the potential terms are diagonal in the plane-wave dual basis.
This advantage comes at the trade-off of requiring a much larger number of plane waves, typically around 1000 times the number of MO orbitals, to achieve the same level of precision~\cite{suFaultTolerantQuantumSimulations2021}.
Moreover, active space techniques cannot be easily used to reduce the number of electrons in this context; instead, pseudopotentials are often employed to replace core electrons with approximate potentials.

In this paper, we present a comprehensive cost analysis of Trotterization and qubitization techniques for ground-state energy estimation, while showcasing the advantages of the sorted-list encoding for the quantum simulation of chemical systems.
  Specifically, we integrate Trotterization with robust phase estimation (RPE)~\cite{guntherPhaseEstimationPartially2026} and the qubitization with quantum phase estimation (QPE)~\cite{babbushEncodingElectronicSpectra2018} in order to estimate the resource requirements of such algorithms in several model chemical systems.
  Our analysis elucidates the impact of different encoding schemes and basis sets on both computational cost and qubit count, providing a quantitative characterization of their respective trade-offs and identifying regimes where each of encoding scheme provides a computational advantage.
  Our key findings are summarized in \autoref{fig:intro_figure}, where we found that the sorted-list encoding provides a viable alternative to the Jordan-Wigner encoding in the MO basis when highly accurate calculations are needed.
  On the other hand, the sorted-list encoding achieves the same asymptotic scaling as the first-quantized methods, though the latter still retains smaller constant factors in the qubitization circuits.

\subsection{Fermion-to-qubit encoding schemes}
We first provide a brief introduction on the fermion-to-qubit encoding schemes discussed in this paper, which includes the Jordan-Wigner, the first-quantized, and the sorted-list encoding.

\subsubsection{Jordan-Wigner encoding}
\label{sec:prev_encoding_jw}
The Jordan-Wigner~\cite{jordanUeberPaulischeAequivalenzverbot1928} encoding is a second-quantized approach that maps fermionic Fock states to qubit states.
Given a system of $M$ spin orbitals, a computational basis state of $N$ electrons is described by a Slater determinant, which is an anti-symmetrized product of $N$ single-particle wavefunctions:
\begin{equation*}
  \ket{\Psi(\vb{x}_1, \dots, \vb{x}_N)} = \frac{1}{\sqrt{N!}} \det
  \begin{bmatrix}
    \psi_{p_1}(\vb{x}_1) & \cdots & \psi_{p_N}(\vb{x}_1) \\
    \vdots & \ddots & \vdots \\
    \psi_{p_1}(\vb{x}_N) & \cdots & \psi_{p_N}(\vb{x}_N)
  \end{bmatrix}.
\end{equation*}
In the second-quantized representation, such a state is represented by the occupation of $M$ available modes, $\ket{f_1, f_2, \dots, f_M}$, where $f_p\in\{0,1\}$ indicates the occupancy of the $p$-th orbital.
The JW encoding simply maps these occupancy representation to qubits:
\begin{equation*}
  \mathcal{E}_{JW}(\ket{f_1, f_2, \dots, f_M}) = \ket{f_1}\otimes\ket{f_2}\otimes\cdots\otimes\ket{f_M}.
\end{equation*}

Hamiltonians in the second-quantized representation typically take the form:
\begin{equation*}
  H = \sum_{pq}^M h_{pq}a_p^\dagger a_q + \sum_{pqrs}^M h_{pqrs}a_p^\dagger a_q^\dagger a_r a_s,
\end{equation*}
where the creation and annihilation operators, $a_p^\dagger$ and $a_p$, controls the occupancy of the orbitals, while also accounting for the parity of the previous occupations:
\begin{align*}
  a_p^\dagger \ket{f_1, \dots, 0_p, \dots, f_M} &= (-1)^{\sum_{j < p}f_j}\\
                                                &\times\ket{f_1, \dots, 1_p, \dots, f_M}\\
  a_p^\dagger \ket{f_1, \dots, 1_p, \dots, f_M} &= 0\\
  a_p         \ket{f_1, \dots, 0_p, \dots, f_M} &= 0\\
  a_p         \ket{f_1, \dots, 1_p, \dots, f_M} &= (-1)^{\sum_{j < p}f_j}\\
                                                &\times\ket{f_1, \dots, 0_p, \dots, f_M}
\end{align*}

The JW encoding maps these operators to Pauli matrices as follows, using $X\pm iY$ to swap the orbital occupation and $Z$ strings to calculate parity.
\begin{align*}
  \mathcal{E}_{JW}(a_p^\dagger) &= \frac{1}{2}(X_p-iY_p)\otimes Z_{p-1}\otimes\cdots\otimes Z_1,\\
  \mathcal{E}_{JW}(a_p        ) &= \frac{1}{2}(X_p+iY_p)\otimes Z_{p-1}\otimes\cdots\otimes Z_1.
\end{align*}
Under this scheme, $M$ qubits are required to represent $M$ orbitals.
While the mapping is conceptually straightforward, the non-local nature of the parity string results in operators with weight $\mathcal{O}(M)$, which can lead to significant overhead in quantum simulations.

\subsubsection{First-Quantized Encoding}
\label{sec:prev_encoding_fq}
In contrast to the second-quantized representation, the first-quantized representation maps each computational basis state as a tensor product of single-particle wavefunctions, called Hartree products.
\begin{equation*}
  \ket{\Psi} = \ket{\psi_{p_1}(\vb{x}_1)}\ket{\psi_{p_2}(\vb{x}_2)}\cdots\ket{\psi_{p_N}(\vb{x}_N)},\\
\end{equation*}
where the tensor product $\otimes$ is omitted for brevity.
Unlike the Slater determinant used in the JW encoding, a single Hartree product does not inherently satisfy the Pauli exclusion principle.
As physical fermionic states must be anti-symmetric under the exchange of any two particles, the wavefunction must be expressed as a linear combination of $N!$ Hartree products.
Such antisymmetrization must be handled explicitly during the quantum simulation process.
Fortunately, Hamiltonian evolution preserve this antisymmetric property as the Hamiltonian commutes with the antisymmetrization operator.
Therefore, explicit antisymmetrization is only required during state initialization~\cite{babbushQuantumSimulationChemistry2019}.

Hamiltonians in the first-quantized representation generally adopt the following form
\begin{equation*}
  H = \sum_{i}^N\sum_{pq}^M h_{pq}\dyad{p}{q}_i + \sum_{i\ne j}^N\sum_{pqrs}^M h_{pqrs} \dyad{p}{q}_i \dyad{r}{s}_j,
\end{equation*}
where the $\dyad{r}{s}_j$ acts on the $j$-th electron such that:
\begin{multline*}
  \dyad{r}{s}_j\ket{\psi_{p_1}(\vb{x}_1)}\cdots\ket{\psi_{p_j}(\vb{x}_j)}\cdots\ket{\psi_{p_N}(\vb{x}_N)}\\ =
  \begin{cases}
    \ket{\psi_{p_1}(\vb{x}_1)}\cdots\ket{\psi_{r}(\vb{x}_j)}\cdots\ket{\psi_{p_N}(\vb{x}_N)}& \text{if } p_j = s\\
    0 & \text{otherwise}
  \end{cases}
\end{multline*}
Crucially, the $\dyad{r}{s}_j$ does not perform any parity calculations, unlike the $a_p^{\dagger}/a_p$ operator of the second-quantized representation. This simpler operator representation trades-off the larger costs incurred during state initialization.

The first-quantized encoding~\cite{kassalPolynomialtimeQuantumAlgorithm2008,suFaultTolerantQuantumSimulations2021,babbushQuantumSimulationChemistry2019} straightforwardly maps the first-quantized representation using the binary addressing code~\cite{steudtnerFermiontoqubitMappingsVarying2018}.
Each orbital index $p\in\{0,\dots,M-1\}$ is represented by a bitstring of length $\lceil\log_2 M\rceil$.
A Hartree product is thus encoded as:
\begin{align*}
  \mathcal{E}(\Psi) &= \ket{i_1}\ket{i_2}\cdots\ket{i_N},
\end{align*}
where each $\ket{i_j}$ consists of $\lceil\log_2 M\rceil$ qubits, requiring a total of $N\lceil\log_2 M\rceil$ qubits.
To encode $\dyad{r}{s}_j$ in terms of Pauli strings, we decompose it into individual single-qubit operators:
\begin{align*}
  \dyad{r}{s}_j = \bigotimes_{n=1}^{\lceil\log_2 M\rceil} \dyad{r[n]}{s[n]}_{n+(j-1)\lceil\log_2 M\rceil},
\end{align*}
where $r[n]$ represents the $n$-th bit in the binary representation of $r$.
Each single-qubit transition operator can then be expanded into the Pauli basis:
\begin{align*}
  \dyad{0}{0} &= \frac{1}{2}(I+Z) & \dyad{0}{1} &= \frac{1}{2}(X+iY)\\
  \dyad{1}{1} &= \frac{1}{2}(I-Z) & \dyad{1}{0} &= \frac{1}{2}(X-iY).
\end{align*}

\subsubsection{Sorted-list Encoding}
\label{sec:prev_encoding_sl}
The sorted-list encoding represents the second-quantized version of the binary addressing code used in the first-quantized encoding.
\citett{carolanSuccinctFermionData2024} introduced several second-quantized encodings with optimal $\mathcal{O}(N\log M)$ qubit scaling, in which we showcased the specific approach referred to as the sorted-list encoding (detailed in Section 4.2 of their work). Among the variants introduced, this specific encoding offers the simplest encoding along with the most efficient circuit implementations in terms of gate count.
Other encoding schemes increase the complexity of the encoding and circuit implementation to save either the qubit count and/or the circuit depth.
The sorted-list encoding also allows us to create a very compact circuit for the Trotterization implementation, along with efficient conversions between the first-quantized and second-quantized encodings~\cite{kuOptimizingQuantumChemistry2026}.

Consider a Slater determinant expressed in the occupation-number form $\ket{\Psi} = \ket{f_1, f_2,\dots,f_M}$, where $f_p\in\{0,1\}$.
For a system of $N$ electrons, we can alternatively write the Slater determinant as a list of the occupied orbitals indices $\ket{i_1, i_2,\dots,i_N}$ ($f_{i_p}=1$ for all $p=1,\dots,N$), sorted in ascending order ($i_1<i_2<\cdots< i_N$).
The sorted-list encoding concatenates the binary representations of each of these indices:
\begin{equation*}
  \mathcal{E}_{SL}(\ket{i_1, i_2,\dots,i_N}) = \ket{i_1}\ket{i_2}\cdots\ket{i_N}\ket{\infty}\cdots\ket{\infty},
\end{equation*}
where the tensor product is again left implicit.
Similarly with the first-quantized encoding, each orbital index $\ket{i}$ is represented as a binary string of length $\lceil\log_2 M\rceil$.

A sentinel state $\ket{\infty}$ is also introduced to represent ``unoccupied'' registers, used to fill in the remaining slots when fewer than $N$ electrons are present.
By convention, the sentinel state is ordered to be larger than all other valid orbital indices.
Thus, a sorted-list encoding of $N\lceil\log_2 (M+1)\rceil$ qubits can represents all Slater determinants containing less than or equal to $N$ electrons.
This addition of a sentinel state allows the sorted-list encoding to perform particle non-conserving operations as long as enough registers are set aside for the operation.
Further details on the encoding of the creation and annihilation operators can be found in Appendix~\ref{sec:sl_operators}.

\section{Results}
\label{sec:results}
Our comprehensive cost analysis of Trotterization and qubitization techniques for quantum simulation systematically investigates the impact of different encoding schemes and basis sets on both gate cost and qubit count. Here we summarize our results.

Due to the significant differences in algorithm implementations between the MO and plane-wave bases, we present separate discussions of their cost scalings for both the Trotterization and qubitization implementations.
Notably, the plane-wave basis trades its larger basis-set size for a significantly improved gate cost scaling.
For the RPE algorithm with Trotterization, the gate and qubit cost scalings are provided in \autoref{prop:sorted_list_trotter} for the MO basis and \autoref{prop:pw_trotter} for the plane-wave basis.
On the other hand, the cost analysis for the QPE algorithm using qubitization is shown in \autoref{tab:qubitization_mo} for the MO basis and \autoref{prop:qubitization_sorted_list} for the plane-wave basis.

For the Trotterization-based algorithm in the MO basis, we found that the sorted-list encoding obtains a qubit cost advantage in the $N\ll M$ regime.
However, the advantage in gate cost is more nuanced: while it offers better Clifford gate scaling, its Toffoli gate scaling is worse.
Despite this, we still expect an advantage in total quantum resource costs for the sorted-list encoding in the $N/M\to 0$ limit.
In contrast, for the plane-wave basis, the sorted-list encoding maintains parity with the first-quantized encoding in terms of the gate and qubit cost complexities and provides an advantage over the Jordan-Wigner encoding.

For the qubitization-based algorithms in the MO basis, we found similar gate and qubit costs between the sorted-list and the Jordan-Wigner encodings, as the dominant terms are independent of the encoding choice.
Conversely, for qubitization-based algorithms in the plane-wave basis, while the sorted-list and first-quantized encodings achieve the same asymptotic gate and qubit cost scalings, absolute resource requirements still favor the first-quantized encoding.

The two bases discussed also have different applicability for different systems.
The MO basis has been the conventional choice for molecular systems due to its localized nature.
On the other hand, the periodicity of the plane-wave basis is more suitable for periodic systems.
Despite this, the tighter bounds provided by the plane-wave basis may still be preferable over the MO basis for molecular systems.
In such cases, the molecule must be placed in a simulation large enough to minimize interaction with its periodic images.

\begin{table*}
  \centering
  \caption{
    Gate and qubit costs comparison between the sorted-list and Jordan-Wigner encodings for obtaining the ground-state energy of $H$ with precision $\varepsilon$ using the RPE algorithm for the three different Trotter product formulas.
    For the sorted-list encoding, the Clifford gate and the Toffoli gate costs have identical scalings.
    For the partially random approach, $L_{\text{det}}$ and $\lambda_{\text{rand}}$ only refer to the deterministic and random parts, respectively.
    We typically have $L=\mathcal{O}(M^4)$ and $\lambda=\mathcal{O}(M)\sim\mathcal{O}(M^3)$.
  }
  \begin{ruledtabular}
    \begin{tabular}{p{2.8cm}P{5.2cm}P{4.1cm}P{4.1cm}}
      &\multicolumn{1}{c}{Sorted-List}&\multicolumn{2}{c}{Jordan-Wigner}\\
      \cmidrule{2-2}
      \cmidrule{3-4}
      Formula&\multicolumn{1}{c}{Clifford/Toffoli gate}&\multicolumn{1}{c}{Clifford gate}&\multicolumn{1}{c}{Toffoli gate}\\
      \midrule
      Deterministic $p$\textsuperscript{th}order&
      $\mathcal{O}(N\log M[5^{p/2-1}L^{2+1/p}/\varepsilon^{1+1/p}])$&
      $\mathcal{O}(M[5^{p/2-1}L^{2+1/p}/\varepsilon^{1+1/p}])$&
      $\mathcal{O}(5^{p/2-1}L^{2+1/p}/\varepsilon^{1+1/p})$\\
      \addlinespace[2mm]

      Random qDRIFT&
      $\mathcal{O}(N\log M[\lambda^2/\varepsilon^2])$&
      $\mathcal{O}(M[\lambda^2/\varepsilon^2])$&
      $\mathcal{O}(\lambda^2/\varepsilon^2)$\\
      \addlinespace[2mm]

      Partially random&
      $\mathcal{O}(N\log M[5^{p/2-1}L_{\text{det}}^{2+1/p}/\varepsilon^{1+1/p} + \lambda_{\text{rand}}^2/\varepsilon^2])$&
      $\mathcal{O}(M[5^{p/2-1}L_{\text{det}}^{2+1/p}/\varepsilon^{1+1/p} + \lambda_{\text{rand}}^2/\varepsilon^2])$&
      $\mathcal{O}(5^{p/2-1}L_{\text{det}}^{2+1/p}/\varepsilon^{1+1/p} + \lambda_{\text{rand}}^2/\varepsilon^2)$\\
      \addlinespace[1mm]
      \midrule
      Qubit cost&\multicolumn{1}{c}{$\mathcal{O}(N\log M)$}&\multicolumn{2}{c}{$\mathcal{O}(M)$}\\
    \end{tabular}
  \end{ruledtabular}
  \label{tab:trotter_rpe_costs}
\end{table*}

Throughout the paper, we adopt the notation $\tilde{\mathcal{O}}(f) = \mathcal{O}(f\operatorname{polylog}(M,N,\lambda,\varepsilon))$, where $\lambda$ is the Hamiltonian 1-norm, and $\varepsilon$ is the precision parameter.
For our resource estimations, we report non-Clifford gate counts in terms of Toffoli gates, as they are the primary source of non-Clifford gates in the majority of our cost analyses.
One exception to this is Trotter implementation using the Jordan-Wigner encoding, where the sole source of non-Clifford gates comes from the synthesis of the $R_Z(\theta)$ gates to Clifford+T gates.
To maintain a consistent metric of the non-Clifford gate counts, we converted the T gates counts to an equivalent Toffoli counts using a 4:1 ratio~\cite{jonesLowoverheadConstructionsFaulttolerant2013}.

\subsection{Trotterization}
\label{sec:trotter}
Our implementation and cost analysis of the Trotterization circuits uses the single-ancilla phase estimation algorithm called robust phase estimation (RPE)~\cite{guntherPhaseEstimationPartially2026}.
The single ancilla requirement of the RPE algorithm pairs well with the low ancilla overhead of Trotterization.
This algorithm requires the implementation of controlled $e^{-iHt}$ gates, which we approximate using various Trotter formulas.
The Hamiltonian is first decomposed into a linear combination of Hermitian operators $H = \sum_{l=0}^{L-1}h_l H_l$, where each Hermitian $H_l$ is chosen such that $e^{-ih_lH_lt_l}$ can be efficiently implemented as a quantum circuit for arbitrary time steps $t_l$.

The efficiency of implementing these exponential operators, and thus the overall Trotter simulation, is highly dependent on the chosen fermion-to-qubit encoding.
The sorted-list encoding represents the latest attempt in obtaining the optimal $\mathcal{O}(N\log M)$ qubit scaling for simulating a second-quantized particle-conserving Hamiltonians.
With its distinct qubit advantage over the Jordan-Wigner in the $N\ll M$ limit, the objective here is to also provide similar or improved gate complexities.
Historically, this has proven challenging, as earlier encodings that offered qubit savings~\cite{sheeQubitefficientEncodingScheme2022,chengOptimalNumberconservedLinear2025,steudtnerFermiontoqubitMappingsVarying2018,kirbySecondQuantizedFermionicOperators2022,bravyiTaperingQubitsSimulate2017} either incurred severe trade-offs in gate complexities or did not achieve the optimal qubit scaling.

\subsubsection{Trotterization for the MO basis}
\label{sec:trotter_mo}
We begin our analysis with the MO basis, where the $h_{pq}$ and $h_{pqrs}$ coefficients of the Hamiltonians are arbitrary, subject only to the Hermiticity of the Hamiltonian.
Furthermore, we assume the Hamiltonian is decomposed into $L$ Hermitian terms, and its 1-norm is $\lambda=\sum_{l=0}^{L-1}|h_l|$.
\autoref{prop:sorted_list_trotter} formalizes the gate and qubit cost scalings to perform the RPE algorithm that obtains the ground-state energy up to precision $\varepsilon$.

\begin{proposition}
\label{prop:sorted_list_trotter}
The ground-state energy of a Hamiltonian $H$ can be calculated to precision $\varepsilon$ using the RPE algorithm.
For the sorted-list encoding, the gate costs (both Clifford and Toffoli gates) scale as:
\begin{itemize}[nosep]
  \item $\mathcal{O}(N\log M[5^{p/2-1}L^{2+1/p}/\varepsilon^{1+1/p}])$ when using the deterministic $p$\textsuperscript{th} order Trotter formula~\cite{hatanoFindingExponentialProduct2005},
  \item $\mathcal{O}(N\log M[\lambda^2/\varepsilon^2])$ when using the random qDRIFT formula~\cite{campbellRandomCompilerFast2019}, with $\lambda=\sum_{l=0}^{L-1}|h_l|$,
  \item $\mathcal{O}(N\log M[5^{p/2-1}L_{\text{det}}^{2+1/p}/\varepsilon^{1+1/p} + \lambda_{\text{rand}}^2/\varepsilon^2])$ when using a partially random approach~\cite{guntherPhaseEstimationPartially2026}, where the Hamiltonian is partitioned into deterministic ($H_{\text{det}}$) and random ($H_{\text{rand}}$) parts:
    \begin{equation*}
      e^{-iHt} = e^{-iH_{\text{det}}t} e^{-iH_{\text{rand}}t}.
    \end{equation*}
\end{itemize}
For the partially random approach, $L_{\text{det}}$ and $\lambda_{\text{rand}}$ only refer to the deterministic and random parts, respectively.
The qubit costs scales identically across these three strategies at $\mathcal{O}(N\log M)$.
\end{proposition}
\begin{proof}
  See Appendix~\ref{sec:theorem_proofs}.
\end{proof}

\autoref{tab:trotter_rpe_costs} compares the gate and qubit costs of the results in \autoref{prop:sorted_list_trotter} with those of the Jordan-Wigner encoding.
The sorted-list encoding has its qubit requirement scale as $\mathcal{O}(N\log M)$ compared to $\mathcal{O}(M)$ for Jordan-Wigner, providing a significant advantage in the low electron-filling regime ($N\ll M$) prevalent in quantum chemical Hamiltonians.
This qubit saving does comes at a cost of increased Toffoli complexity.
For instance, the time evolution of each Hermitian fermionic operator costs $\mathcal{O}(N\log M)$ Clifford and Toffoli gates in the sorted-list encoding.
In contrast, while the Jordan-Wigner encoding has worse Clifford gate scaling at $\mathcal{O}(M)$, the non-Clifford cost of a Trotter step is exceptionally low, requiring only one $R_Z(\theta)$ gate, synthesized to a small constant number of Clifford+T gates, for every Pauli string.

This discrepancy explains why the sorted-list encoding is much less competitive than the Jordan-Wigner encoding for typical Trotter simulations within chemical accuracy. The magic resources required to implement the $\mathcal{O}(N\log M)$ Toffoli gates are typically more expensive than the $\mathcal{O}(M)$ Clifford gates of the Jordan-Wigner encoding.
However, in the $N/M\to 0$ limit, the sorted-list encoding is expected to regain its advantage due to the logarithmic dependence on $M$ for both space and time resources.
Moreover, recent advancements such as magic state cultivation~\cite{gidneyMagicStateCultivation2024}, which significantly narrows the cost difference between Clifford and non-Clifford gates, is expected to greatly lower the threshold at which the sorted-list encoding is favorable over the Jordan-Wigner encoding.
Note that this trade-off between Clifford and non-Clifford scaling is unique to Trotterization. In the later section, we show that qubitization requires substantial Toffoli resources regardless of the Jordan-Wigner encoding, making the relative overhead of the sorted-list encoding much less significant.

The balanced logarithmic dependence on $M$ is a rare property in fermion-to-qubit encodings, as shown in \autoref{tab:encoding_costs}, with the polylog encoding~\cite{kirbySecondQuantizedFermionicOperators2022} being the only other encoding exhibiting similar dependence on $M$.
Such low $N/M$ ratios in the MO basis may be encountered in scenarios demanding exceptionally high accuracy, such as when studying Van der Waals interactions or hyperfine structures.

It is important to emphasize that $L$ and $\lambda$ scales similarly between the two encodings scheme.
For the Jordan-Wigner encoding, the fermionic operators are decomposed into Pauli strings, yielding $L=\mathcal{O}(M^4)$.
While theoretically possible, the decomposition to Pauli strings is not be feasible for the sorted-list encoding due to an exponential blow-up to $L=\mathcal{O}(M^N)$~\cite{steudtnerFermiontoqubitMappingsVarying2018}.
Instead, circuits implementing the time evolution of Hermitian fermionic operators are directly constructed as detailed in Appendix~\ref{sec:trotter_circuit}.
This approach eliminates the need for decomposition, requiring only the grouping of conjugate operators to ensure Hermiticity (e.g. group $a_p^\dagger a_q$ with $a_q^\dagger a_p$ to form $a_p^\dagger a_q + a_q^\dagger a_p$ and $a_p^\dagger a_q -i a_q^\dagger a_p$).
Consequently, $L$ also scales as $\mathcal{O}(M^4)$ for the sorted-list encoding.
The use of Hermitian fermionic operators for the sorted-list encoding also results in similar $\lambda$ scalings between the two, typically varying between $\mathcal{O}(M)$ to $\mathcal{O}(M^3)$ depending on the specific system~\cite{leeEvenMoreEfficient2021}.

\begin{table}
  \centering
  \caption{
    Clifford, Toffoli, and qubit cost scalings of various second-quantized fermion-to-qubit encodings to perform the time evolution of a single fermionic term (eg. $e^{-i\theta (a_p^\dagger a_q + a_q^\dagger a_p)}$).
  }
  \begin{ruledtabular}
    \begin{tabular}{P{2.7cm}P{3cm}p{2cm}}
      Encoding&Gate Costs&Qubit Cost\\
      \midrule
      Sorted-list \cite{carolanSuccinctFermionData2024}&
      $\mathcal{O}(N\log M)$ Clifford\newline
      $\mathcal{O}(N\log M)$ Toffoli&
      $\mathcal{O}(N\log M)$\\
      \midrule
      Jordan-Wigner \& Parity \cite{jordanUeberPaulischeAequivalenzverbot1928}&
      $\mathcal{O}(M)$ Clifford\newline
      $\mathcal{O}(1)$ Toffoli&
      $\mathcal{O}(M)$\\
      \midrule
      Bravyi-Kitaev \cite{bravyiFermionicQuantumComputation2002}&
      $\mathcal{O}(\log M)$ Clifford\newline
      $\mathcal{O}(1)$ Toffoli&
      $\mathcal{O}(M)$\\
      \midrule
      Qubit Efficient \cite{sheeQubitefficientEncodingScheme2022} \& Random Linear \cite{chengOptimalNumberconservedLinear2025}&
      $\mathcal{O}(M^N)$ Clifford\newline
      $\mathcal{O}(M^N)$ Toffoli&
      $\mathcal{O}(N\log M)$\\
      \midrule
      Segment \cite{steudtnerFermiontoqubitMappingsVarying2018}&
      $\mathcal{O}(N^2)$ Clifford\newline
      $\mathcal{O}(N^2)$ Toffoli&
      $\mathcal{O}(M-M/N)$\\
      \midrule
      Polylog \cite{kirbySecondQuantizedFermionicOperators2022}&
      $\mathcal{O}(N^2\log^5 M)$ Clifford\newline
      $\mathcal{O}(N^2\log^5 M)$ Toffoli&
      $\mathcal{O}(N^2\log^4 M)$\\
    \end{tabular}
  \end{ruledtabular}
  \label{tab:encoding_costs}
\end{table}

\subsubsection{Trotterization for the plane-wave basis}
\label{sec:trotter_pw}
Unlike the MO basis where the $h_{pq}$ and $h_{pqrs}$ coefficients are arbitrary, the plane-wave (along with its dual basis) offers a more structured Hamiltonian, albeit at the trade-off of requiring a much larger basis for equivalent precision.
This allows us to formulate a tighter bound for the gate cost of performing the RPE algorithm, demonstrated in \autoref{prop:pw_trotter}.

\begin{proposition}
\label{prop:pw_trotter}
The ground-state energy of the Hamiltonian $H$ in the plane-wave basis can be calculated to precision $\varepsilon$ using the RPE algorithm, where the gate costs (both Clifford and Toffoli) scale as:
\begin{equation*}
  \mathcal{O}\left( \left( \frac{M^{8/3}}{N^{2/3}} + M^{7/3}N^{2/3} \right)\frac{5^{p/2-1}(M^{1/3}N^{2/3})^{1/p}}{\varepsilon^{1+1/p}} \right)
\end{equation*}
for the second-quantized Jordan-Wigner encoding, and
\begin{equation*}
  \tilde{\mathcal{O}}\left( \left( M^{2/3}N^{4/3} + M^{1/3}N^{8/3} \right)\frac{5^{p/2-1}(M^{1/3}N^{2/3})^{1/p}}{\varepsilon^{1+1/p}} \right)
\end{equation*}
for both the first-quantized encoding and the second-quantized sorted-list encoding.
\end{proposition}
\begin{proof}
  See Appendix~\ref{sec:theorem_proofs}.
\end{proof}

\begin{table*}
  \centering
  \caption{
    Qubit count and Toffoli gate cost of performing an RPE calculation using the deterministic, random, and partially random Trotter formulas~\cite{guntherPhaseEstimationPartially2026} of molecular systems using MO orbitals.
    Due to the high classical costs in determining the partition point between deterministic and random parts, we skipped the cost analysis of the larger systems.
    Qubit counts where the sorted-list encoding has an advantage over the Jordan-Wigner encoding are shown in bold.
  }
  \sisetup{
    table-format = 1.0e+2,
    parse-numbers = true
  }
  \begin{ruledtabular}
    \begin{tabular}{llcSSScSSS}
      &&\multicolumn{4}{c}{Jordan-Wigner}&\multicolumn{4}{c}{Sorted-list}\\
      \cmidrule{3-6}
      \cmidrule{7-10}
      System&Orbitals&Qubits&{Suzuki}&{qDRIFT}&{Partial}&Qubit&{Suzuki}&{qDRIFT}&{Partial}\\
      \midrule

      \ce{H2O}   & 7 (STO-3G)    & 16  & 1.E+09 & 2.E+11 & 2.E+09 & 55           & 3.E+11 & 4.E+14 & 3.E+11 \\
      10 elec    & 19 (6-311G)   & 40  & 5.E+11 & 9.E+12 & 6.E+11 & 77           & 2.E+14 & 2.E+16 & 2.E+14 \\
      Model      & 24 (cc-pvdz)  & 50  & 1.E+12 & 2.E+13 & 2.E+12 & 77           & 6.E+14 & 3.E+16 & 6.E+14 \\
      System     & 58 (cc-pvtz)  & 118 & 6.E+14 & 2.E+15 & 5.E+14 & \textbf{88}  & 2.E+17 & 3.E+18 & 2.E+17 \\
                 & 115 (cc-pvqz) & 232 & 7.E+16 & 9.E+16 & 5.E+16 & \textbf{99}  & 3.E+19 & 1.E+20 & 2.E+19 \\
                 & 201 (cc-pv5z) & 404 & 3.E+18 & 2.E+18 &        & \textbf{110} & 2.E+21 & 4.E+21 &        \\ \midrule
      \ce{CO2}   & 15 (STO-3G)   & 32  & 6.E+10 & 2.E+12 & 1.E+11 & 126          & 7.E+13 & 2.E+16 & 5.E+13 \\
      22 elec    & 39 (6-311G)   & 80  & 3.E+13 & 2.E+14 & 2.E+13 & 172          & 3.E+16 & 8.E+17 & 2.E+16 \\
      Model      & 42 (cc-pvdz)  & 86  & 4.E+13 & 2.E+14 & 4.E+13 & 172          & 4.E+16 & 8.E+17 & 3.E+16 \\
      System     & 90 (cc-pvtz)  & 182 & 7.E+15 & 1.E+16 & 5.E+15 & 195          & 7.E+18 & 4.E+19 & 4.E+18 \\
                 & 165 (cc-pvqz) & 332 & 4.E+17 & 3.E+17 &        & \textbf{218} & 4.E+20 & 1.E+21 &        \\
                 & 273 (cc-pv5z) & 548 & 1.E+19 & 4.E+18 &        & \textbf{241} & 1.E+22 & 2.E+22 &        \\ \midrule
      \ce{EC}    & 34 (STO-3G)   & 70  & 2.E+14 & 3.E+14 & 2.E+13 & 340          & 1.E+17 & 1.E+18 & 3.E+16 \\
      46 elec    & 90 (6-311G)   & 182 & 1.E+17 & 4.E+16 &        & 387          & 5.E+19 & 9.E+19 &        \\
      Battery    & 104 (cc-pvdz) & 210 & 1.E+17 & 4.E+16 &        & 387          & 2.E+20 & 3.E+20 &        \\
      Materials  & 236 (cc-pvtz) & 474 & 3.E+19 & 4.E+18 &        & \textbf{434} & 7.E+22 & 4.E+22 &        \\
                 & 460 (cc-pvqz) & 922 & 2.E+21 & 2.E+20 &        & \textbf{481} & 6.E+24 & 2.E+24 &        \\ \midrule
      \ce{LiPF6} & 44 (STO-3G)   & 90  & 3.E+13 & 6.E+13 & 1.E+14 & 522          & 1.E+18 & 7.E+18 & 3.E+17 \\
      72 elec    & 112 (6-311G)  & 226 & 2.E+16 & 8.E+15 &        & 595          & 5.E+20 & 7.E+20 &        \\
      Battery    & 116 (cc-pvdz) & 234 & 6.E+16 & 2.E+16 &        & 595          & 6.E+20 & 7.E+20 &        \\
      Materials  & 244 (cc-pvtz) & 490 & 2.E+19 & 4.E+18 &        & 668          & 1.E+23 & 7.E+22 &        \\
                 & 444 (cc-pvqz) & 890 & 2.E+21 & 2.E+20 &        & \textbf{741} & 1.E+25 & 3.E+24 &        \\
    \end{tabular}
  \end{ruledtabular}
  \label{tab:costings_trotter_mol_mo}
\end{table*}

Due to the sheer number of plane-waves required compared to the other parameters, $M$ is the most crucial parameter determining the overall cost.
As a result, both the first-quantized and sorted-list encodings are most efficient with a leading order gate cost of $\tilde{\mathcal{O}}(M^{2/3+1/p})$, compared to $\mathcal{O}(M^{8/3+1/p})$ for the Jordan-Wigner encoding. This demonstrates a significant advantage for the $\mathcal{O}(N \log M)$ qubit encodings in this basis.

The gate costs of \autoref{prop:pw_trotter} are also significantly lower compared to the MO basis.
To explain this superior scaling, we partition the Hamiltonian into kinetic $\hat{T}$, external potential $\hat{U}$, and coulomb potential $\hat{V}$ terms: $H=\hat{T} + \hat{U} + \hat{V}$.
The kinetic term $\hat{T}$ is diagonal in the plane-wave basis, while the potential terms $\hat{U} + \hat{V}$ are diagonal in the plane-wave dual basis.
Since diagonal terms are mutually commuting, $\hat{T}$ and $\hat{U}+\hat{V}$ can be implemented exactly without any intrinsic Trotter errors.
Consequently, we effectively have $L=2$ for this Hamiltonian decomposition, represented by $\hat{T}$ and $\hat{U}+\hat{V}$.
Due to this low value of $L$, the deterministic formula is evidently more efficient for this Hamiltonian.

For the Jordan-Wigner encoding, both $\hat{T}$ and $\hat{U}+\hat{V}$ are encoded to Pauli strings of $\mathcal{O}(1)$ weight in their respective bases.
The time evolution of these terms can be implemented with $\mathcal{O}(1)$ Clifford and one $R_Z(\theta)$ gate each.
On the other hand, efficient construction of the first quantized Trotter circuit relies on phase kickback~\cite{kassalPolynomialtimeQuantumAlgorithm2008} to implement the diagonal gates.
A drawback of the this approach is that, unlike the Jordan-Wigner encodings where the Hamiltonian coefficients are calculated classically and incorporated into the rotation angles of the $R_Z(\theta)$ gates, these coefficients needs to be calculated in the quantum circuit for the first-quantized approach, implying higher constant factors despite the favorable scaling.

While the sorted-list encoding implementation can be adapted from either approach, the first-quantized approach is chosen for \autoref{prop:pw_trotter}.
This is because, unlike the massive gate cost reductions observed in the Jordan-Wigner encoding, the sorted-list encoding still maintains the $\mathcal{O}(N\log M)$ Clifford and Toffoli scaling when implementing $\hat{T}$ and $\hat{U}+\hat{V}$ operators.
Conversely, converting the first-quantized implementation to the sorted-list encoding is more straightforward, as the action of $\hat{T}$ and $\hat{U}+\hat{V}$ acts identically between the first-quantized and sorted-list encodings, resulting in identical scaling between the two.

\subsubsection{Costings on real-world systems}
\label{sec:trotter_costings}
In this section, we provide numerical results for the Toffoli and qubit costs of performing RPE calculations for several real-world molecular systems.
We chose to highlight the second-order ($p=2$) Suzuki Trotter, qDRIFT, and the hybrid formula using both the Jordan-Wigner and the sorted-list encoding, following the RPE cost analysis done in a previous work~\cite{guntherPhaseEstimationPartially2026}.
The qubit and Toffoli gate costs are shown in \autoref{tab:costings_trotter_mol_mo}.

In terms of the qubit count, it is evident that the sorted-list encoding shows an advantage in the $N\ll M$ regime, with the break-even point observed around $N/M\approx 0.1$.
However, as previously discussed, the Jordan-Wigner encoding requires only a single $R_Z(\theta)$ gate as its only non-Clifford gate, while the sorted-list encoding requires dense applications of Toffoli gates, in addition to the $R_Z(\theta)$ gate, due to its inherent binary comparison operations.
Consequently, the sorted-list encoding typically requires 2-4 orders of magnitude more non-Clifford resources than its Jordan-Wigner counterpart.
This highlights a clear trade-off between qubit efficiency and Toffoli complexity across different encoding schemes and Trotterization strategies.

The major advantage of the above Trotterization circuit is its conceptual simplicity as the Hamiltonian simulation can implemented as a sequence of time evolution circuits, making it a more realistic choice for early fault-tolerant implementations.
With the exception of the first-quantized and sorted-list implementations in the plane-wave basis, the exponentiation of the Hermitian terms turns the $h_{pq}$ and $h_{pqrs}$ coefficients of the Hamiltonian into rotation angles applied on $R_Z(\theta)$ gates.
In comparison, qubitization circuits uses QROMs to load those Hamiltonian coefficients into the circuit, a significantly more complex operation.

The complexity of the qubitization circuit trades off with more efficient scaling with respect to the precision parameter $\varepsilon$ and Hamiltonian 1-norm $\lambda$.
In terms of the precision parameter, the QPE and RPE algorithms both contribute a $1/\varepsilon$ factor towards the gate cost for phase estimation accuracy. Moreover, the Trotter product formula itself contributes an additional $1/\varepsilon^{1/p}$ for the deterministic formula and $1/\varepsilon$ for the random formula.
This results in the overall $\varepsilon$ scaling found in \autoref{prop:sorted_list_trotter} and \ref{prop:pw_trotter}.
On the other hand, qubitization offers a logarithmic dependence on the error scaling for the simulation primitive, resulting in an overall gate cost scaling of $\tilde{\mathcal{O}}(1/\varepsilon)$ when combined with the QPE algorithm.
Qubitization also offers similar advantages over the Hamiltonian 1-norm $\lambda$, where the gate cost has a linear dependence with $\lambda$, compared to the quadratic dependence for the qDRIFT Trotter formula.

\subsection{Qubitization}
\label{sec:qubitizations}
For qubitization circuits, the Hamiltonian adopts a linear combination of unitaries (LCU) representation:
\begin{equation*}
  H=\sum_{l=0}^{L-1} h_l U_l,
\end{equation*}
where $h_l$ are positive real coefficients ($h_l\in\mathbb{R}^{+}$), and $U_l$ are unitary operators.
The QPE algorithm is then implemented via the Szegedy quantum walk~\cite{babbushEncodingElectronicSpectra2018}
\begin{equation*}
  \mathcal{W} = (2 \operatorname{PREPARE}\dyad{0}\operatorname{PREPARE}^{\dagger}\otimes \mathbb{I} - \mathbb{I})\operatorname{SELECT},
\end{equation*}
where $\operatorname{PREPARE}$ and $\operatorname{SELECT}$ constitute the major gate and qubit costs of the algorithm.
These circuits can be described at a high-level as:
\begin{align*}
  \operatorname{PREPARE}\ket{0} &= \sum_{l=0}^{L-1}\sqrt{\frac{|h_l|}{\lambda}} \ket{l},\\
  \operatorname{SELECT}\ket{l}\ket{\psi} &= \ket{l}U_l\ket{\psi},\label{eqn:majorana_select}
\end{align*}
with $\lambda=\sum_{l=0}^{L-1}|h_l|$ being the 1-norm of the Hamiltonian.
To obtain the ground-state energy of the Hamiltonian with precision $\varepsilon$, the QPE algorithm requires $\mathcal{O}(\lambda/\varepsilon)$ applications of $\mathcal{W}$.

\subsubsection{Qubitization in the MO basis}
\label{sec:qubitizations_mo}

\begin{table*}
  \centering
  \caption{
    Clifford gate, Toffoli gate, and qubit cost scaling for obtaining the ground-state energy of a Hamiltonian of 1-norm $\lambda$ with precision $\varepsilon$ using the QPE algorithm.
    Various rank reduction techniques to mitigate the $\mathcal{O}(M^4)$ complexity of the two-electron terms in the MO basis are included.
    Due to the use of Quantum Read-Only Memories (QROMs) in the QPE algorithm, we provided two QROM implementations~\cite{lowTradingGatesDirty2024}, one that minimizes ancilla qubits (``Minimize Ancilla'') and another that minimize the Toffoli gate cost (``Minimize Toffoli'').
  }
  \begin{ruledtabular}
    \begin{tabular}{l>{\centering}p{2.2cm}>{\centering}p{2.2cm}>{\centering}p{2.2cm}>{\centering}p{2.2cm}l}
      &\multicolumn{2}{c}{Minimize Ancilla}&\multicolumn{2}{c}{Minimize Toffoli}&\\
      \cmidrule{2-3}
      \cmidrule{4-5}
      Method&Jordan-Wigner&Sorted-List&Jordan-Wigner&Sorted-List&Resources\\
      \midrule
      Sparse~\cite{berryQubitizationArbitraryBasis2019}
      &\multicolumn{2}{c}{$\tilde{\mathcal{O}}(\lambda M^4/\varepsilon)$}
      &\multicolumn{2}{c}{$\tilde{\mathcal{O}}(\lambda M^4/\varepsilon)$}
      &Cliffords\\
      &\multicolumn{2}{c}{$\tilde{\mathcal{O}}(\lambda M^4/\varepsilon)$}
      &\multicolumn{2}{c}{$\tilde{\mathcal{O}}(\lambda M^2/\varepsilon)$}
      &Toffolis\\
      &$\tilde{\mathcal{O}}(M)$
      &$\tilde{\mathcal{O}}(N)$
      &\multicolumn{2}{c}{$\tilde{\mathcal{O}}(M^2)$}
      &Qubits\\
      \midrule
      Single-Factorization~\cite{berryQubitizationArbitraryBasis2019}
      &\multicolumn{2}{c}{$\tilde{\mathcal{O}}(\lambda M^3/\varepsilon)$}
      &\multicolumn{2}{c}{$\tilde{\mathcal{O}}(\lambda M^3/\varepsilon)$}
      &Cliffords\\
      &\multicolumn{2}{c}{$\tilde{\mathcal{O}}(\lambda M^3/\varepsilon)$}
      &\multicolumn{2}{c}{$\tilde{\mathcal{O}}(\lambda M^{3/2}/\varepsilon)$}
      &Toffolis\\
      &$\tilde{\mathcal{O}}(M)$
      &$\tilde{\mathcal{O}}(N)$
      &\multicolumn{2}{c}{$\tilde{\mathcal{O}}(M^{3/2})$}
      &Qubits\\
      \midrule
      Double-Factorization~\cite{vonburgQuantumComputingEnhanced2021}
      &\multicolumn{2}{c}{$\tilde{\mathcal{O}}(\lambda M^3/\varepsilon)$}
      &\multicolumn{2}{c}{$\tilde{\mathcal{O}}(\lambda M^3/\varepsilon)$}
      &Cliffords\\
      &\multicolumn{2}{c}{$\tilde{\mathcal{O}}(\lambda M^3/\varepsilon)$}
      &\multicolumn{2}{c}{$\tilde{\mathcal{O}}(\lambda M^{3/2}/\varepsilon)$}
      &Toffolis\\
      &$\tilde{\mathcal{O}}(M)$
      &$\tilde{\mathcal{O}}(N)$
      &\multicolumn{2}{c}{$\tilde{\mathcal{O}}(M^{3/2})$}
      &Qubits\\
      \midrule
      Tensor Hypercontraction~\cite{leeEvenMoreEfficient2021}
      &\multicolumn{2}{c}{$\tilde{\mathcal{O}}(\lambda M^2/\varepsilon)$}
      &\multicolumn{2}{c}{$\tilde{\mathcal{O}}(\lambda M^2/\varepsilon)$}
      &Cliffords\\
      &\multicolumn{2}{c}{$\tilde{\mathcal{O}}(\lambda M^2/\varepsilon)$}
      &$\tilde{\mathcal{O}}(\lambda M/\varepsilon)$&$\tilde{\mathcal{O}}(\lambda MN/\varepsilon)$
      &Toffolis\\
      &$\tilde{\mathcal{O}}(M)$
      &$\tilde{\mathcal{O}}(N)$
      &\multicolumn{2}{c}{$\tilde{\mathcal{O}}(M)$}
      &Qubits\\
    \end{tabular}
  \end{ruledtabular}
  \label{tab:qubitization_mo}
\end{table*}

Unlike the Trotterization case where the Hamiltonian coefficients $h_{pq}$ and $h_{pqrs}$ can be arbitrary, we restrict our Hamiltonian such that $h_{pq}\in\mathbb{R}$ and $h_{pqrs}\in\mathbb{R}$.
Furthermore, we assume that the $h_{pqrs}$ coefficients adopt the $8$-fold symmetry, which is typically obeyed by the Hamiltonians of molecular systems.
In this case, the gate and qubit costs of the QPE algorithm for such Hamiltonians are formalized in \autoref{tab:qubitization_mo}.

A natural unitary decomposition of the Hamiltonian here involves the Majorana operators:
\begin{align*}
  \gamma_{p,\alpha,0} &= a_{p,\alpha}^{\dagger} + a_{p,\alpha} & \gamma_{p,\alpha,1} &= ia_{p,\alpha}^{\dagger}-ia_{p,\alpha},
\end{align*}
with the $\operatorname{SELECT}$ circuit implementing a product of either two or four of them for the one- and two-electron terms, respectively.
The Jordan-Wigner implementation of this involves unitary iteration circuits~\cite{babbushEncodingElectronicSpectra2018} requiring $\mathcal{O}(M)$ gates.
In contrast, the sorted-list implementation, detailed in Appendix~\ref{sec:qubitization_circuit}, has a gate cost scaling of $\mathcal{O}(N\log M)$.
This suggests a clear theoretical advantage for the sorted-list encoding over the Jordan-Wigner in the $N\ll M$ limit.

The implementation of the $\operatorname{PREPARE}$ circuit, however, remains the same between the two encodings.
To mitigate the $L=\mathcal{O}(M^4)$ scaling, various rank-reduction techniques have been proposed to reduce its qubit and gate costs, as listed in \autoref{tab:qubitization_mo}, typically at the cost of expensive classical preprocessing of the Hamiltonian.
Despite these efforts, the gate cost scaling of the $\operatorname{PREPARE}$ circuit still dominates over the $\operatorname{SELECT}$ circuit.
This result in similar overall gate cost scalings between the two encodings when integrated together into the QPE algorithm.
As a result, advanced construction techniques like magic state cultivation~\cite{gidneyMagicStateCultivation2024} is expected to provide similar savings between the Jordan-Wigner and the sorted-list encoding, unlike the Trotter case where the sorted-list encoding receives larger cost reductions.
A notable exception is the tensor hypercontraction approach~\cite{leeEvenMoreEfficient2021}, where a simplified $\operatorname{PREPARE}$ circuit necessitates a more complex $\operatorname{SELECT}$ circuit, for which the Jordan-Wigner encoding fares better than the sorted-list encoding.

Aside from the Jordan-Wigner and the sorted-list encoding, the first-quantized encoding also can be used for qubitization in the MO basis~\cite{georgesQuantumSimulationsChemistry2025}, although it is currently limited to the sparse technique with the exact same asymptotic Clifford, Toffoli and qubit costs as the sorted-list encoding.
This qubitization circuit is implemented by decomposing the $\dyad{r}{s}_j$ operators to Pauli strings and hence, requires extensive preprocessing of the $h_{pq}$ and $h_{pqrs}$ coefficients, resulting in fundamentally different $\operatorname{PREPARE}$ circuits and $\lambda$ values compared to the second-quantized encodings in this paper.
This is in contrast to the sorted-list encoding where the $h_{pq}$ and $h_{pqrs}$ coefficients are identical to the Jordan-Wigner encoding.

\subsubsection{Qubitization in the plane-wave basis}
\label{sec:qubitizations_pw}
The highly structured nature of the plane-wave basis set also significantly benefits the implementation of the qubitization circuits.
We adapted a highly efficient first-quantized implementation of the Szegedy quantum walk operator~\cite{suFaultTolerantQuantumSimulations2021} for our implementation of the QPE algorithm for plane-wave Hamiltonians with the sorted-list encoding.
The gate costs are formalized in \autoref{prop:qubitization_sorted_list}.

\begin{proposition}
  The ground-state energy of a Hamiltonian in the plane-wave basis can be calculated to precision $\varepsilon$ using the QPE algorithm with the second-quantized sorted-list encoding using
  \begin{equation*}
    \tilde{\mathcal{O}}([N^{4/3}M^{2/3} + N^{8/3}M^{1/3}]/\varepsilon)
  \end{equation*}
  gates and $\tilde{\mathcal{O}}(N)$ qubits.
  \label{prop:qubitization_sorted_list}
\end{proposition}
\begin{proof}
  See Appendix~\ref{sec:theorem_proofs}.
\end{proof}

The first-quantized encoding also requires $\tilde{\mathcal{O}}([N^{4/3}M^{2/3} + N^{8/3}M^{1/3}]/\varepsilon)$ gates and $\tilde{\mathcal{O}}(N)$ qubits.
In this case, our sorted-list encoding implementation replaces the $\dyad{r}{s}_j$ transition operators of the first-quantized qubitization circuits, while keeping the remaining circuit implementation unchanged.
As a result, recent optimizations devised for the first-quantized circuits, such as the addition of non-local pseudopotentials~\cite{berryQuantumSimulationRealistic2024}, can be trivially integrated to the sorted-list implementation, while preserving the overall Toffoli and qubit scalings.
Despite this, the sorted-list encoding still requires more Toffoli and qubit resources compared to the first-quantized encoding.
This is primarily attributed the more complex mechanisms required to maintain the ascending order requirement of the sorted-list encoding, detailed in Appendix~\ref{sec:prev_encoding_sl}.
As a result, despite similar gate cost and qubit asymptotic scalings, the first-quantized approach yields a more efficient overall QPE circuit.

\subsubsection{Costings of real-world systems}
\label{sec:qubitizations_costings}

\begin{table*}[!t]
  \centering
  \caption{
    Resource cost comparison for qubitization of various chemical systems.
    We compare the logical qubit and Toffoli counts between the sorted-list encoding and the Jordan-Wigner encoding for (i) to (iv) and the first-quantized encoding for (iv).
    (i) \textbf{FeMoCo}: Benchmarked against SF~\cite{berryQubitizationArbitraryBasis2019}, THC~\cite{leeEvenMoreEfficient2021}, and SOSSA~\cite{lowFastQuantumSimulation2025a}, using SF parameters from \citett{berryQubitizationArbitraryBasis2019}.
    (ii and iii) \textbf{EC and \ce{LiPF6}}: Our Hamiltonians are reconstructed via PySCF~\cite{sunRecentDevelopmentsPySCF2020} with error threshold following \citett{kimFaulttolerantResourceEstimate2022}.
    (iv) \textbf{\ce{LiNiO_2}}: Parameters are adopted from \citett{rubinFaultTolerantQuantumSimulation2023}. We label the $k$-mesh used in the ``System'' column in the square brackets.
    (v) \textbf{\ce{[Fe2S2]^{2-}}}: Hamiltonian and sparsity parameters are obtained from \citet{georgesQuantumSimulationsChemistry2025}.
    For each system, results are presented for QROM configurations that prioritize either minimal Toffoli gates or minimal ancilla qubits~\cite{lowTradingGatesDirty2024}.
  }
  \sisetup{
    table-format = 1.1e+2,
    parse-numbers = true
  }
  \begin{ruledtabular}
    \begin{tabular}{lcllcSlcS}
      &\multirow[b]{2}{1cm}[-1.7mm]{Spatial orbitals}&\multirow[t]{2}{1.2cm}[-1.7mm]{QROM minimizes}&\multicolumn{3}{c}{Previous Works}&\multicolumn{3}{c}{Our Work}\\
      \cmidrule{4-6}\cmidrule{7-9}
      System&&&Source&Qubits&{Toffolis}&Method&Qubits&{Toffolis}\\
      \midrule
      \multicolumn{9}{c}{(i) FeMoCo}\\
      \midrule
      RWSWT~\cite{reiherElucidatingReactionMechanisms2017} & 54 & Toffoli & Berry~\cite{berryQubitizationArbitraryBasis2019} (SF) & 3024 & 1.2E+12 & SF & 835  & 5.8E+10 \\
      54 elec                                              &    & Toffoli & Lee~\cite{leeEvenMoreEfficient2021} (THC)             & 2142 & 5.3E+09 &    &      &               \\
                                                           &    & Toffoli & Low~\cite{lowFastQuantumSimulation2025a} (SOSSA)      & 1132 & 3.4E+08 &    &      &               \\
                                                           &    & Ancilla & Berry~\cite{berryQubitizationArbitraryBasis2019} (SF) & 378  & 2.1E+13 &    & 703  & 6.9E+10 \\
      LLDUC~\cite{liElectronicComplexityGroundstate2019}   & 76 & Toffoli & Berry~\cite{berryQubitizationArbitraryBasis2019} (SF) & 3143 & 1.0E+12 &    & 1368 & 8.1E+10 \\
      113 elec                                             &    & Toffoli & Lee~\cite{leeEvenMoreEfficient2021} (THC)             & 2196 & 3.2E+10 &    &      &               \\
                                                           &    & Toffoli & Low~\cite{lowFastQuantumSimulation2025a} (SOSSA)      & 1454 & 1.0E+09 &    &      &               \\
                                                           &    & Ancilla & Berry~\cite{berryQubitizationArbitraryBasis2019} (SF) & 437  & 2.0E+13 &    & 1349 & 9.1E+10 \\
      \midrule
      \multicolumn{9}{c}{(ii) EC (46 elec)}\\
      \midrule
      STO-3G  & 44  & Toffoli & Kim~\cite{kimFaulttolerantResourceEstimate2022} & 924   & 1.1E+10 & SF & 626  & 3.4E+10 \\
      6-311G  & 90  &         & (DF)                                            & 5772  & 2.2E+11 &    & 900  & 8.2E+11 \\
      cc-pvdz & 104 &         &                                                 & 14963 & 4.6E+11 &    & 1232 & 1.6E+12 \\
      cc-pvtz & 236 &         &                                                 & 33718 & 1.0E+12 &    & 2189 & 4.1E+13 \\
      \midrule
      \multicolumn{9}{c}{(iii) \ce{LiPF6} (72 elec)}\\
      \midrule
      STO-3G  & 44  & Toffoli & Kim~\cite{kimFaulttolerantResourceEstimate2022} & 1146  & 3.6E+10 & SF & 850  & 1.2E+11 \\
      6-311G  & 112 &         & (DF)                                            & 16101 & 6.5E+11 &    & 1459 & 2.5E+12 \\
      cc-pvdz & 116 &         &                                                 & 16667 & 6.0E+11 &    & 1459 & 2.5E+12 \\
      cc-pvtz & 244 &         &                                                 & 34860 & 1.1E+13 &    & 2388 & 4.7E+13 \\
      \midrule
      \multicolumn{9}{c}{(iv) \ce{LiNiO2} (14 elec for R$\overline{3}$m and C2/m, 56 elec for P2/c, and 28 elec for P2\textsubscript{1}/c)}\\
      \midrule
      R$\overline{3}$m [2$\times$2$\times$2]      & 58  & Toffoli & Rubin~\cite{rubinFaultTolerantQuantumSimulation2023} & \num{1.5E+05} & 5.0E+12 & SF & \num{5.9E+03} & 2.9E+13 \\
      R$\overline{3}$m [3$\times$3$\times$3]      & 58  &         & (DF)                                                 & \num{6.0E+05} & 7.3E+13 &    & \num{2.5E+04} & 1.7E+15 \\
      C2/m [2$\times$2$\times$1]                  & 58  &         &                                                      & \num{7.5E+04} & 1.2E+12 &    & \num{2.8E+03} & 7.1E+12 \\
      C2/m [4$\times$4$\times$2]                  & 58  &         &                                                      & \num{6.0E+05} & 9.8E+13 &    & \num{2.7E+04} & 2.0E+15 \\
      P2/c [1$\times$1$\times$1]                  & 232 &         &                                                      & \num{7.6E+04} & 9.7E+11 &    & \num{2.9E+03} & 1.5E+13 \\
      P2/c [2$\times$2$\times$2]                  & 232 &         &                                                      & \num{1.2E+06} & 1.4E+14 &    & \num{2.7E+04} & 3.9E+15 \\
      P2\textsubscript{1}/c [1$\times$2$\times$1] & 116 &         &                                                      & \num{7.5E+04} & 1.3E+12 &    & \num{2.9E+03} & 2.0E+13 \\
      P2\textsubscript{1}/c [2$\times$4$\times$2] & 116 &         &                                                      & \num{1.2E+06} & 1.2E+14 &    & \num{2.7E+04} & 5.4E+15 \\
      \midrule
      \multicolumn{9}{c}{(v) \ce{[Fe2S2]^{2-}} (72 elec)}\\
      \midrule
             & 16  & Ancilla & Georges~\cite{georgesQuantumSimulationsChemistry2025} & 198 & 1.8E+11 & Sparse & 245 & 2.2E+10 \\
             & 32  &         & (Sparse \& 1\textsuperscript{st}                      & 229 & 1.1E+13 &        & 280 & 1.1E+12 \\
             & 64  &         & quant)                                                & 261 & 9.5E+14 &        & 315 & 1.3E+14 \\
             & 128 &         &                                                       & 293 & 4.8E+16 &        & 347 & 7.3E+15 \\
    \end{tabular}
  \end{ruledtabular}
  \label{tab:qubitization_mol_mo_compare}
\end{table*}

\begin{table*}
  \centering
  \caption{
    Costings of the qubitization operators for molecular systems with the plane-wave basis set. We added the first-quantized plane-wave implementation~\cite{suFaultTolerantQuantumSimulations2021} for comparison.
  }
  \sisetup{
    parse-numbers = true
  }
  \begin{ruledtabular}
    \begin{tabular}
      {
        l
        S[table-format=1.1e+2]
        S[table-format=2.0]
        S[table-format=4.0]
        S[table-format=1.1e+2]
        S[table-format=1.1e+2]
        S[table-format=2.0]
        S[table-format=4.0]
        S[table-format=1.1e+2]
        S[table-format=1.1e+2]
      }
      &&\multicolumn{4}{c}{First-Quantized (Previous work~\cite{suFaultTolerantQuantumSimulations2021})}&\multicolumn{4}{c}{Second-Quantized (Our work)}\\
      \cmidrule{3-6}
      \cmidrule{7-10}
      System&{PWs}&{nQPE}&{Qubits}&{Toffolis/Step}&{Total Toffolis}&{nQPE}&{Qubits}&{Toffolis/Step}&{Total Toffolis}\\
      \midrule
      \ce{H2O}   & 4.1E+03 & 20 & 825  & 1.6E+03 & 1.7E+09 & 20 & 934  & 6.1E+03 & 6.4E+09 \\
      $10$ elec  & 3.3E+04 & 21 & 1015 & 1.9E+03 & 4.1E+09 & 21 & 1139 & 7.4E+03 & 1.5E+10 \\
                 & 2.6E+05 & 22 & 1219 & 2.3E+03 & 9.7E+09 & 22 & 1358 & 8.7E+03 & 3.7E+10 \\
                 & 2.1E+06 & 24 & 1440 & 2.7E+03 & 4.5E+10 & 24 & 1594 & 1.0E+04 & 1.7E+11 \\ \midrule
      \ce{CO2}   & 4.1E+03 & 22 & 1021 & 2.3E+03 & 9.7E+09 & 22 & 1166 & 1.1E+04 & 4.4E+10 \\
      $22$ elec  & 3.3E+04 & 23 & 1255 & 2.8E+03 & 2.4E+10 & 23 & 1415 & 1.3E+04 & 1.1E+11 \\
                 & 2.6E+05 & 25 & 1506 & 3.3E+03 & 1.1E+11 & 25 & 1681 & 1.5E+04 & 5.0E+11 \\
                 & 2.1E+06 & 26 & 1768 & 3.9E+03 & 2.6E+11 & 26 & 1958 & 1.7E+04 & 1.2E+12 \\ \midrule
      \ce{EC}    & 4.1E+03 & 25 & 1363 & 3.7E+03 & 1.2E+11 & 25 & 1580 & 1.9E+04 & 6.5E+11 \\
      $46$ elec  & 3.3E+04 & 26 & 1677 & 4.5E+03 & 3.0E+11 & 26 & 1909 & 2.3E+04 & 1.6E+12 \\
                 & 2.6E+05 & 27 & 2005 & 5.3E+03 & 7.1E+11 & 27 & 2252 & 2.8E+04 & 3.7E+12 \\
                 & 2.1E+06 & 28 & 2347 & 6.1E+03 & 1.6E+12 & 28 & 2609 & 3.2E+04 & 8.5E+12 \\ \midrule
      \ce{LiPF6} & 4.1E+03 & 26 & 1701 & 5.1E+03 & 3.4E+11 & 26 & 1996 & 2.9E+04 & 1.9E+12 \\
      $72$ elec  & 3.3E+04 & 27 & 2097 & 6.2E+03 & 8.3E+11 & 27 & 2407 & 3.5E+04 & 4.7E+12 \\
                 & 2.6E+05 & 28 & 2507 & 7.3E+03 & 2.0E+12 & 28 & 2832 & 4.1E+04 & 1.1E+13 \\
                 & 2.1E+06 & 29 & 2931 & 8.5E+03 & 4.5E+12 & 29 & 3271 & 4.7E+04 & 2.5E+13 \\
    \end{tabular}
  \end{ruledtabular}
  \label{tab:qubitization_mol_pw}
\end{table*}

Here, we present numerical results for the Toffoli gate and qubit costs of QPE calculations for several real-world molecular systems.
For our cost analysis in the MO basis, we utilized the sparse and single-factorization form of the Hamiltonian~\cite{berryQubitizationArbitraryBasis2019}, augmented with additional circuit optimizations suggested by \citett{leeEvenMoreEfficient2021}.
We opted not to highlight the sorted-list implementations of the double-factorization~\cite{vonburgQuantumComputingEnhanced2021} and tensor hypercontraction~\cite{leeEvenMoreEfficient2021} methods as the $\mathcal{O}(M)$ sequence of controlled Pauli rotations in their implementations (Section VII.C.2 of the Supporting Information of \citet{vonburgQuantumComputingEnhanced2021}) would translate to $\mathcal{O}(M)$ Majorana operations.
This would result in a less favorable Toffoli scaling of $\mathcal{O}(MN\log M)$ as compared to the $\mathcal{O}(M)$ scaling achieved with the Jordan-Wigner representation.

\autoref{tab:qubitization_mol_mo_compare} presents the calculated qubit and Toffoli costs required for QPE on several model chemical systems with results from previous works for comparison.
Given that the Toffoli cost of the QROM can potentially be reduced through the use of either ``dirty'' or ``clean'' qubits~\cite{lowTradingGatesDirty2024}, we included a cost analysis for two QROM implementations: one that minimizes the ancilla qubits and another that minimizes the Toffoli cost.

Our first comparison focuses on the FeMoCo active spaces defined by \citett{reiherElucidatingReactionMechanisms2017} and \citett{liElectronicComplexityGroundstate2019}, both of which are clearly in the $N\approx M$ regime.
It is evident that the Jordan-Wigner encoding with advanced tensor factorization techniques, including tensor hypercontraction (THC)~\cite{leeEvenMoreEfficient2021} and sum-of-squares spectrum amplification (SOSSA)~\cite{lowFastQuantumSimulation2025a}, exhibits the lowest overall Toffoli costs.
Nevertheless, the sorted-list encoding manages to maintain similar Toffoli cost as an equivalent single-factorized Jordan-Wigner calculation.
To demonstrate a transition to the $N \ll M$ regime, we performed the same cost analysis with battery materials Ethylene Carbonate \ce{C3H4O3} (EC) and \ce{LiPF6}.
For these systems, the sorted-list encoding achieves a qubit advantage over the double-factorized (DF) Jordan-Wigner implementation~\cite{kimFaulttolerantResourceEstimate2022}, while incurring comparable Toffoli costs.
We also included in our cost analysis the periodic material \ce{LiNiO2}, using the method of \citett{rubinFaultTolerantQuantumSimulation2023}.
While plane-wave bases are more ubiquitous for periodic systems, the atom-centered basis used here requires fewer spin orbitals, placing the systems here near the $N\approx M$ regime.
As a result, we can see similar qubit and Toffoli costs between the Jordan-Wigner and the sorted-list encodings.
Finally, we compared the sorted-list encoding with the first-quantized encoding using the \ce{[Fe2S2]^{2-}} system.
As both the sorted-list and the first-quantized encodings are both $\mathcal{O}(N\log M)$-qubit encodings,
they require similar qubit costs, with the sorted-list encoding being only slightly larger.
Despite this, the sorted-list encoding shows lower Toffoli costs than the first-quantized encoding.

We also performed cost analysis on the same systems using the plane-wave basis.
\autoref{tab:qubitization_mol_pw} presents a cost comparison between the first-quantized implementation~\cite{suFaultTolerantQuantumSimulations2021} and our sorted-list implementation.
Despite similar asymptotic cost scalings, we found that the sorted-list encoding requires around 5 times more Toffolis and 10\% more qubit resources compared to the first-quantized encoding.

\section{Discussions}
\label{sec:discussion}

We present the cost analysis of phase estimation algorithms via Trotterization and qubitization utilizing the sorted-list encoding~\cite{carolanSuccinctFermionData2024}.
The sorted-list encoding demonstrates a $\mathcal{O}(N\log M)$ scaling with respect to both the qubit count and the gate cost to implement a fermionic operator, offering a qubit advantage over conventional $\mathcal{O}(M)$ qubit encodings in the $N\ll M$ regime.
While this is an improvement over previous electron-conserving encoding schemes, its trade-offs in the context of phase estimation algorithms had not yet been thoroughly analyzed.
We also obtained numerical estimates for the qubit and Toffoli costs of phase estimation for several chemical systems, comparing them with previous works~\cite{berryQubitizationArbitraryBasis2019,leeEvenMoreEfficient2021,lowFastQuantumSimulation2025a,kimFaulttolerantResourceEstimate2022,rubinFaultTolerantQuantumSimulation2023,georgesQuantumSimulationsChemistry2025}.

For the Trotterization circuits in the MO basis, the sorted-list encoding achieved an obvious qubit advantage when $N \ll M$, with numerical experiments showing the break-even point at $N/M \approx 0.1$.
This, however, comes at a cost of an increased Toffoli gate cost compared to the Jordan-Wigner encoding, at roughly 2-4 orders of magnitude higher for the chemical systems we tested.
Despite this, due to better Clifford gate scaling, our theoretical analysis of the sorted-list encoding predicted an eventual quantum resource advantage of the Jordan-Wigner encoding in the $N/M\to 0$ limit.
  Especially so with recent advancements on magic state cultivation~\cite{gidneyMagicStateCultivation2024}, which greatly reduces the cost disparity between Clifford and non-Clifford gates.
In contrast, the structured nature of the plane-wave basis allowed for a different, more efficient implementation compared to the MO basis.
Consequently, the sorted-list encoding offers both qubit and gate cost advantage for Trotterization circuits in the plane-wave basis.
Due to the similarity between the first-quantized and the sorted-list encoding, we are able to leverage the hybrid quantization scheme~\cite{kuOptimizingQuantumChemistry2026} to obtain similar gate cost scalings between the two.

Regarding qubitization circuits, we obtained similar gate and qubit costs between the Jordan-Wigner and sorted-list encodings for calculations in the MO basis.
This parity arises because the gate and qubit complexities are dominated by the $\operatorname{PREPARE}$ circuit, which is identical for both encodings.
Numerical estimation of the qubitization circuits further demonstrated these comparable gate costs.
Similar gate and qubit asymptotic cost scalings is also achieved between the first-quantized and sorted-list encodings when using the plane-wave basis.
Despite this, maintaining the ascending-order requirement of the sorted-list encoding requires a more complex qubitization circuit.
When estimating these costs on real-world molecular systems, this disparity resulted in around 5 times more Toffoli gate costs and 10\% more qubits compared to the first-quantized encoding.

When performing phase estimation on large molecules, the sublinear gate scaling of the plane-wave qubitization circuits~\cite{suFaultTolerantQuantumSimulations2021} is very attractive, complemented with an $\tilde{\mathcal{O}}(N)$ qubit scaling, which is ideal for the large $M$ inherent to the plane-wave basis.
However, utilizing the plane-wave basis for molecular systems presents several caveats.
The most prominent is the necessity to eliminate interactions between periodic images. Such errors can typically be reduced by increasing the simulation cell size, which incurs larger $M$, or by modifying the Coulomb potential to ignore long-range interactions~\cite{broqvistHybridfunctionalCalculationsPlanewave2009,sundararamanRegularizationCoulombSingularity2013,bylaskaFilonlikeIntegrationStrategy2020}, which typically increases the complexity of the $\operatorname{PREPARE}$ circuits.

Conversely, for the simulation of smaller systems in near-term partially fault-tolerant devices, the QROMs required for qubitization circuits presents implementation challenges in achieving sufficiently low error rates~\cite{phalakOptimizationQuantumReadOnly2022}.
As a result, Trotterization circuits in the MO basis using the Jordan-Wigner encoding may be preferable in this scenario.
The larger scaling is offset by the significantly smaller $M$ required for these molecules.

\section{Conclusion}
\label{sec:conclusion}

For future practitioners with access to a fault-tolerant quantum computer, the sorted-list encoding serves as a compelling, qubit-efficient alternative to the Jordan-Wigner encoding for both Trotterization and qubitization circuits in the $N\ll M$ regime, with the trade-off of minimal to moderate increases in both Clifford and Toffoli gate costs.
Although a direct theoretical comparison is challenging due to fundamental differences in Hamiltonian structures and 1-norms~\cite{georgesQuantumSimulationsChemistry2025}, our numerical benchmarks also show lower Toffoli costs for the sorted-list encoding over the first-quantized encoding when the compact molecular orbital bases are used.
Such $N\ll M$ regime for molecular orbitals bases are often encountered when highly accurate chemical calculations are required, such as in transition or bond breaking states, the study of hyperfine structures, van-der Waals forces, or when spectroscopic levels of accuracy are required.
The $N\ll M$ regime can also occur when less compact, but highly structured basis-sets are used, such as the plane-wave basis.
In this case, we found that both the first-quantized and the sorted-list encoding share the same asymptotic gate and qubit cost scalings.
Despite this, the first-quantized encoding remains preferable for qubitization circuits due to lower constant factors.

Future work will focus on further optimizing the Trotterization and qubitization circuits to reduce their constant overheads.
For qubitization circuits using the MO basis, there is an clear opportunity to further exploit the particle conservation symmetry inherent in chemical Hamiltonians.
Techniques such as the Block-Invariant Symmetry Shift (BLISS)~\cite{loaizaBlockInvariantSymmetryShift2023} could be particularly beneficial here, as they can reduce the 1-norm of Hamiltonians and, consequently, the T gate cost of the qubitization circuits.
Additionally, the applicability of other symmetries, such as $S_z$ symmetry and molecular point group symmetries, within the context of the sorted-list encoding warrants further investigation.
Finally, the efficient implementation of qubitization circuits for double-factorized and tensor hypercontracted Hamiltonians in the sorted-list encoding remains an open research question.

\begin{acknowledgments}
A.H. gratefully acknowledges the sponsorship from National Natural Science Foundation of China (NSFC) (Grant No. 62541160274), City University of Hong Kong (Project No. 7006103), and Hon Hai Research Institute (Project No. 9231594, 9231677).
This work was carried out using the computational facilities, CityU Burgundy, managed and provided by the Computing Services Centre at City University of Hong Kong (\url{https://www.cityu.edu.hk}).
\end{acknowledgments}

\section*{Author Contributions}
The project was conceived by C.K. and Y.C.
Theoretical results were proved by C.K. in discussion with M.H.
Numerical simulations and analysis were performed by C.K. in discussion with A.H.
All authors contributed to the write-up.

\section*{Competing Interests}
The authors declare no competing interests.

\appendix




\makeatletter
\renewcommand{\p@subsection}{\thesection.}
\renewcommand{\p@subsubsection}{\thesection.\thesubsection.}
\makeatother

\section{Theorem Proofs}
\label{sec:theorem_proofs}
\begin{proof}[Proof of Proposition 1]
  For the sorted-list encoding, we decompose the Hamiltonian into a linear combination of the following hermitian operators:
  \begin{gather*}
    a_p^\dagger a_q + a_q^\dagger a_p,\ ia_p^\dagger a_q - ia_q^\dagger a_p,\\
    a_p^\dagger a_q^\dagger a_r a_s + a_s^\dagger a_r^\dagger a_q a_p,\text{ and } ia_p^\dagger a_q^\dagger a_r a_s - ia_s^\dagger a_r^\dagger a_q a_p.
  \end{gather*}
  Thus, we must construct circuit implementations for the time evolution of the above unitaries:
  \begin{equation}
    \begin{gathered}
      e^{i\theta(a_p^\dagger a_q + a_q^\dagger a_p)},\ e^{i\theta(ia_p^\dagger a_q - ia_q^\dagger a_p)},\\
      e^{i\theta(a_p^\dagger a_q^\dagger a_r a_s + a_s^\dagger a_r^\dagger a_q a_p)},\text{ and } e^{i\theta(ia_p^\dagger a_q^\dagger a_r a_s - ia_s^\dagger a_r^\dagger a_q a_p)}.
    \end{gathered}
    \label{eqn:thm1_evolution}
  \end{equation}
  We detail these circuit implementations in Appendix~\ref{sec:trotter_circuit}, with exact gate costs shown in \autoref{tab:building_blocks_trotter_costings}.
  When decomposed to a Clifford+T universal gate set, each of these unitaries in \autoref{eqn:thm1_evolution} requires $\mathcal{O}(N\log M)$ Clifford gates, T gates and qubits for its implementation.

  Next, \autoref{tab:trotter_product_formula_costs} specifies the total number of applications of the unitaries from \autoref{eqn:thm1_evolution} needed for the RPE algorithm to calculate the ground-state energy with precision $\varepsilon$.
  By multiplying the costs from the table by the $\mathcal{O}(N\log M)$ gate costs for each unitary application, we reproduce the overall costs reported in the main text.
\end{proof}

\begin{proof}[Proof of Proposition 2]
  To obtain the ground-state with precision $\varepsilon$, the RPE algorithm~\cite{guntherPhaseEstimationPartially2026} requires $\mathcal{O}(1/\varepsilon\delta)$ applications of the $p$\textsuperscript{th}-order Trotter formula $S_p(\delta)$, which approximates $e^{-iH\delta}$.
  The time step $\delta$ is chosen such that the Trotter error is bounded by $\varepsilon\delta$.
  Previous work has established tight bounds for such Trotter errors~\cite{suNearlyTightTrotterization2021}, which derived:
  \begin{multline*}
    |S_p(\delta) - e^{-iH\delta}| =\\
   \mathcal{O}\left( \left( \frac{M^{2/3}}{N^{2/3}}+M^{1/3}N^{2/3} \right)^p M^{1/3}N^{2/3}\delta^{p+1} \right).
  \end{multline*}
  To bound this error below $\varepsilon\delta$, we find the required $\delta$:
  \begin{equation*}
    \delta \sim \mathcal{O}\left( \varepsilon^{1/p}\left( \frac{M^{2/3}}{N^{2/3}}+M^{1/3}N^{2/3} \right)^{-1}(M^{1/3}N^{2/3})^{-1/p} \right).
  \end{equation*}
  substituting this $\delta$ into $\mathcal{O}(1/\varepsilon\delta)$ yields the total number of $S_p(\delta)$ applications:
  \begin{equation}
    \mathcal{O}\left( \frac{1}{\varepsilon\delta} \right) = \mathcal{O}\left( \left( \frac{M^{2/3}}{N^{2/3}}+M^{1/3}N^{2/3} \right)\frac{(M^{1/3}N^{2/3})^{1/p}}{\varepsilon^{1+1/p}} \right).\label{eqn:thm2_ntrotter}
  \end{equation}

  With the required number of $S_p(\delta)$ applications determined, we are left with obtaining the gate cost to implement $S_{p}(\delta)$ for each of the encodings.
  Unlike the MO basis, we exclusively use the deterministic Trotter formula as the plane-wave Hamiltonian can be decomposed into $L=2$ terms, represented by the kinetic $\hat{T}$ and potential $\hat{U}+\hat{V}$ terms.
  This decomposition is discussed in Appendix~\ref{sec:trotterization_pw}.

  Following a previous Jordan-Wigner implementation~\cite{babbushLowDepthQuantumSimulation2018}, we find a gate cost of $\mathcal{O}(5^{p/2-1}M^2)$ for one application of $S_p(\delta)$.
  We included the $5^{p/2-1}$ factor from \autoref{eqn:suzuki_trotter} for our cost analysis.
  For the first-quantized encoding, one application of $S_p(\delta)$ has a gate cost of $\tilde{\mathcal{O}}(5^{p/2-1}N^2)$~\cite{kassalPolynomialtimeQuantumAlgorithm2008}.
  The gate costs in the main text for both encodings are then reproduced by multiplying these encoding specific costs for one $S_p(\delta)$ application by \autoref{eqn:thm2_ntrotter}.

  For the cost analysis of the sorted-list encoding, we can adapt either the Jordan-Wigner or the first-quantized approaches.
  When adapting from the Jordan-Wigner implementation, each application of $S_p(\delta)$ requires $\mathcal{O}(5^{p/2-1}M^2)$ applications of $e^{i\theta a_p^\dagger a_p}$
  Each of these applications contributes a major cost of $\mathcal{O}(N\log M)$ Clifford and T gates (see \autoref{tab:building_blocks_trotter_costings}).
  Consequently, the Clifford and T gate costs of $S_p(\delta)$ scale as $\mathcal{O}(5^{p/2-1}M^2 N\log M)$, which is worse than the original Jordan-Wigner encoding.

  Conversely, using the first-quantized approach yields identical gate cost scaling.
  Consider the Slater determinant $\ket{\nu_1}\otimes\cdots\otimes\ket{\nu_N}$ in the sorted-list encoding, where $\nu_1,\cdots,\nu_N$ are indices of the occupied orbitals.
  The time evolution of the kinetic term $\hat{T}^{(2)}$ (see Appendix~\ref{sec:trotterization_pw}) can be written as:
  \begin{align*}
    e^{-i\hat{T}^{(2)}t}\ket{\nu_1}\otimes&\cdots\otimes\ket{\nu_N}\\
    &= e^{-it\sum_{\nu=1}^MT_\nu a_\nu^\dagger a_\nu}\ket{\nu_1}\otimes\cdots\otimes\ket{\nu_N}\\
    &= \prod_{\nu=1}^M e^{-it T_\nu a_\nu^\dagger a_\nu}\ket{\nu_1}\otimes\cdots\otimes\ket{\nu_N}\\
    &= \prod_{i=1}^N e^{-iT_{\nu_i}t}\ket{\nu_1}\otimes\cdots\otimes\ket{\nu_N}\\
    &= e^{-i\hat{T}^{(1)}t}\ket{\nu_1}\otimes\cdots\otimes\ket{\nu_N}.
  \end{align*}
  This shows that $e^{-i\hat{T}^{(2)}t} = e^{-i\hat{T}^{(1)}t}$ when applied to a qubit wavefunction in the sorted-list encoding.
  Therefore, we can use the same first-quantized implementation for the sorted-list encoding.
  Through similar arguments, the same can be said for the external $e^{-i\hat{U}^{(2)}t}$ and coulomb potential $e^{-i\hat{V}^{(2)}t}$ terms, maintaining the same cost for $S_p(\delta)$ between the first-quantized and sorted-list encoding.
  We are left with the implementation of the fermionic fast Fourier transform circuits required to switch between the plane-wave for $\hat{T}^{(2)}$ and its dual basis for $\hat{U}^{(2)} + \hat{V}^{(2)}$.
  This can be done efficiently using the hybrid quantization scheme introduced in a previous work~\cite{kuOptimizingQuantumChemistry2026}, where we convert the wavefunction to the first-quantized encoding, performed the QFT in the first-quantization, and converting back to the sorted-list encoding with a gate cost of $\tilde{\mathcal{O}}(N)$.
  This results in identical gate cost scaling for the entire RPE algorithm, as shown in the main text.
\end{proof}

\begin{proof}[Proof of Preposition 3]
  We adapt the construction of the $\operatorname{PREPARE}$ and $\operatorname{SELECT}$ circuits for the sorted-list encoding from a first-quantized implementation~\cite{suFaultTolerantQuantumSimulations2021}.
  The first-quantized implementation is summarized in Appendix~\ref{sec:prev_qubitization_fq} while our adaptation is detailed in Appendix~\ref{sec:qubitization_plane_wave}.
  For the sorted-list encoding, the $\operatorname{PREPARE}$ circuit costs $\tilde{\mathcal{O}}(1)$ gates while the $\operatorname{SELECT}$ circuit costs $\tilde{\mathcal{O}}(N)$.
  To obtain the ground-state with precision $\varepsilon$, the QPE circuit requires $\mathcal{O}(\lambda/\varepsilon)$ applications of $\operatorname{PREPARE}$ and $\operatorname{SELECT}$ circuits.
  Substituting $\lambda = \mathcal{O}(N^{1/3}M^{2/3} + N^{5/3}M^{1/3})$ for the 1-norm of plane-wave Hamiltonians~\cite{suFaultTolerantQuantumSimulations2021}, we derive the total gate cost:
  \begin{equation*}
    \tilde{\mathcal{O}}\left( \frac{\lambda N}{\varepsilon} \right) = \tilde{\mathcal{O}}\left( \frac{N^{4/3}M^{2/3}+N^{8/3}M^{1/3}}{\varepsilon} \right).\qedhere
  \end{equation*}
\end{proof}

\section{Creation and Annihilation Operators for the Sorted-List Encoding}
\label{sec:sl_operators}

\begin{table}[t]
  \centering
  \caption{
    Costings of the constant and indexed variants of the $=p$ and $<p$ circuits using the implementation of~\cite{berryQubitizationArbitraryBasis2019}.
    $n_0(p)$ and $n_1(p)$ represents the number of `0' and `1' bits, respectively, in the binary representation of $p$.
  }
  \begin{ruledtabular}
    \begin{tabular}{lrll}
      Circuit&&Constant&Indexed\\
      \midrule
      $=p$                     & $C^{\log M}X$ & $1$           & $1$          \\
      \autoref{fig:eq_circuit} & $CX$          & $0$           & $2\ltw M$    \\
                               & $X$           & $2n_0(p)$     & $2\ltw M$    \\ \midrule
      $<p$                     & $C^2X$        & $\ltw M - 1$  & $\ltw M$     \\
      \autoref{fig:lt_circuit} & $CX$          & $6\ltw M - 4$ & $6\ltw M -5$ \\
                               & $X$           & $2n_1(p)$     & $0$          \\
    \end{tabular}
  \end{ruledtabular}
  \label{tab:equal_less_costings}
\end{table}

\begin{figure}[H]
  \centering
  \hspace*{\fill}
  \subfloat[\label{fig:eq_const_circuit}]{\includegraphics{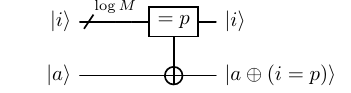}}
  \hspace*{\fill}
  \subfloat[\label{fig:eq_index_circuit}]{\includegraphics{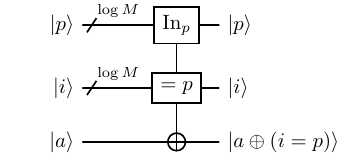}}
  \hspace*{\fill}
  \caption{The $=p$ circuit~\cite{carolanSuccinctFermionData2024}. This circuit flips ancilla $\ket{a}$ when the value of $i$ is equal to (a) a constant $p$ or (b) an indexed value $p$.}
  \label{fig:eq_circuit}
\end{figure}

\begin{figure}[H]
  \centering
  \hspace*{\fill}
  \subfloat[\label{fig:lt_const_circuit}]{\includegraphics{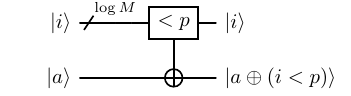}}
  \hspace*{\fill}
  \subfloat[\label{fig:lt_index_circuit}]{\includegraphics{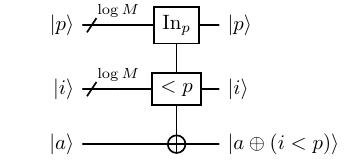}}
  \hspace*{\fill}
  \caption{
    The $<p$ circuit~\cite{carolanSuccinctFermionData2024}. This circuit flips ancilla $\ket{a}$ when the value of $i$ is less than (a) a constant $p$ or (b) an indexed value $p$.
    While our work only uses the $<p$ gate, the original work includes the $\le p$ and $>p$ variant which can be implemented similarly~\cite{berryQubitizationArbitraryBasis2019}.
  }
  \label{fig:lt_circuit}
\end{figure}

Before elaborating how fermionic operations are encoded, we first introduce several building block gates frequently used in the sorted-list encoding.
The $=p$ gate (\autoref{fig:eq_const_circuit}) accepts one register as input and flips a target ancilla qubit if that register is equal to a constant value $p$.
Similarly, the $<p$ gate (\autoref{fig:lt_const_circuit}) accepts one register as input and flips a target ancilla qubit if that register is equal to a constant value $p$.
Both the $=p$ and $<p$ gates has indexed variants, shown in \autoref{fig:eq_index_circuit} and~\ref{fig:lt_index_circuit}, respectively, for use in the qubitization circuits, where an additional register is added to accept a superposition of values for $p$.
The $=p$ gate can be implemented using a multicontrolled Toffoli gate ($C^{\log M}X$ gate) while the implementation for the $<p$ gate can be found in Appendix H of a previous work~\cite{berryQubitizationArbitraryBasis2019}.
The gate cost for both the constant and indexed variant of $=p$ and $<p$ are shown in \autoref{tab:equal_less_costings}.
We also have the bubble gate $U_p$ (\autoref{fig:bubble_ori_circuit}) which accepts two registers as input and swaps the two registers if one of them is equal to $p$ and the other is larger than $p$.
The original implementation of $U_p$ uses $\ltw M$ controlled swap gates, and four $=p$ and $>p$ gates, where the $>p$ gate can be implemented similarly as the $<p$ gate~\cite{berryQubitizationArbitraryBasis2019}.
For the purposes of our implementation, we introduced two implementations of the $U_p$ gate specially designed for the Trotterization and qubitization circuits, respectively, detailed in Appendix~\ref{sec:prev_encoding_sl_opt}.
Finally, the $a\leftrightarrow b$ (\autoref{fig:swap_circuit}) gate outputs $b$ when the input is $a$ and outputs $a$ when the input is $b$, where both $a$ and $b$ are constant values. Otherwise the input remains unchanged.

Fermionic operations on this encoding are first decomposed to Majorana operators.
The Majorana operators are then further decomposed to $\operatorname{sgn-rank}$ and $\operatorname{bit-flip}$ operators.
\begin{align*}
  a_p^\dagger &= \frac{1}{2}(\gamma_{p,0}-i\gamma_{p,1}),\\
  a_p &= \frac{1}{2}(\gamma_{p,0}+i\gamma_{p,1}),\\
  \mathcal{E}(\gamma_{p,0}) &= \operatorname{bit-flip}(p)\operatorname{sgn-rank}(p-1),\\
  \mathcal{E}(\gamma_{p,1}) &= i\operatorname{bit-flip}(p)\operatorname{sgn-rank}(p).
\end{align*}
The $\operatorname{sgn-rank}(p)$ operator calculates the parity of the number of occupied states for orbital indices less than or equal to $p$. It then multiplies the wavefunction by $-1$ for odd parity and $+1$ for even parity. On the other hand, the $\operatorname{bit-flip}(p)$ operator flips the occupation of orbital $p$.
The circuit implementations of the $\operatorname{sgn-rank}$ and $\operatorname{bit-flip}$ operators can be found in \autoref{fig:sgnrank_circuit} and~\ref{fig:bitflip_circuit}, respectively.
Essentially, the $\operatorname{sgn-rank}(j)$ circuit applies a $Z$ gate for every register containing an orbital index less than or equal to $j$.
On the other hand, $\operatorname{bit-flip}(j)$ first checks if $j$ is occupied.
If $j$ is occupied it replaces $\ket{j}$ with $\ket{\infty}$, otherwise it replaces $\ket{\infty}$ with $\ket{j}$.
After replacing the target register with the appropriate value, that register is sorted back to the correct ascending order.
As the Majorana operators changes the number of occupied orbitals, the encoding needs to set aside more registers than the total number of electrons.
More specifically, the one-electron terms will require 2 extra registers in the encoding than the total number of electrons, while the two-electron terms will require 4 extra registers.
Such registers are filled with the $\ket{\infty}$ state so that the swap circuit in \autoref{fig:bitflip_circuit} can be done.

\subsection{Additional Optimizations}
\label{sec:prev_encoding_sl_opt}
\begin{figure}[H]
  \centering
  \includegraphics{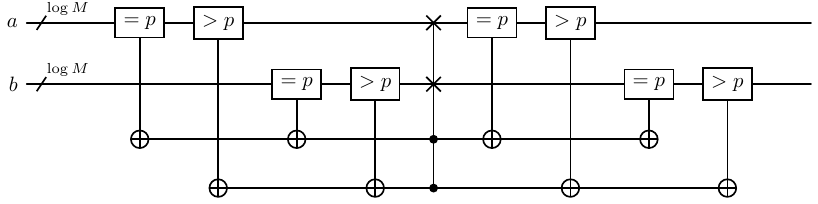}
  \caption{The bubble circuit $U_p$~\cite{carolanSuccinctFermionData2024}. This circuit swaps the two registers $a$ and $b$ when one of them is equal to $p$ and the other is larger than $p$.}
  \label{fig:bubble_ori_circuit}
\end{figure}

\begin{figure}[H]
  \centering
  \includegraphics{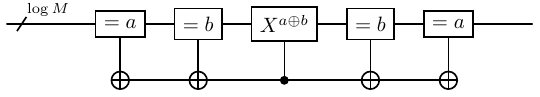}
  \caption{The $a\leftrightarrow b$ circuit~\cite{carolanSuccinctFermionData2024}. This circuit outputs $b$ when the input is $a$ and outputs $a$ when the input is $b$. Otherwise, the input is unchanged. The $X^{a\oplus b}$ is composed of multiple CNOT gate according to the binary representation of $a\oplus b$.}
  \label{fig:swap_circuit}
\end{figure}

\begin{figure}[H]
  \centering
  \includegraphics{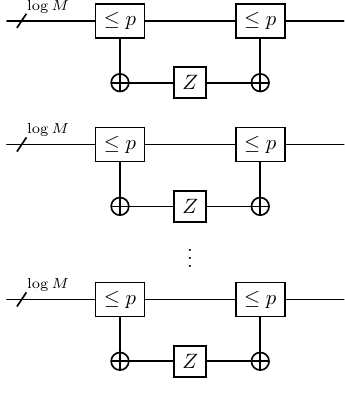}
  \caption{The $\operatorname{sgn-rank}(p)$ circuit~\cite{carolanSuccinctFermionData2024}. The $\le p$ circuit is shown on \autoref{fig:lt_const_circuit}. Quantum wires not extending to both ends of the circuit are ancilla qubits.}
  \label{fig:sgnrank_circuit}
\end{figure}

\begin{figure}[H]
  \centering
  \includegraphics[width=0.8\linewidth]{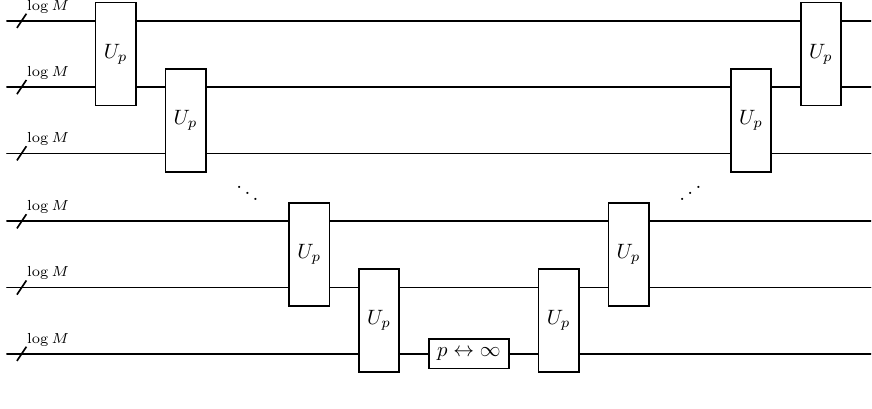}
  \caption{The $\operatorname{bit-flip}(p)$ circuit~\cite{carolanSuccinctFermionData2024}. The bubble circuit $U_p$ is shown on \autoref{fig:bubble_ori_circuit}, while the swap circuit $p\leftrightarrow\infty$ is shown in \autoref{fig:swap_circuit}.}
  \label{fig:bitflip_circuit}
\end{figure}

\begin{figure}[H]
  \centering
  \includegraphics{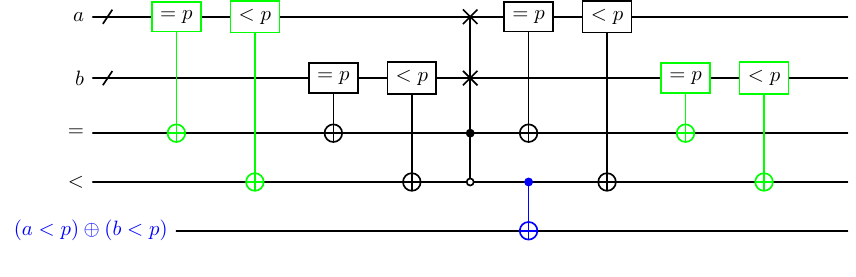}
  \caption{The modified bubble circuit from \autoref{fig:bubble_ori_circuit}.
    Here $<p$ is used instead of $>p$ so that parity information can also be optionally tapped as shown in blue.
    Finally, as this circuit is implemented sequentially to all consecutive pairs of registers, the $=$ and $<$ ancillae can be reused.
    As a result, the $=p$ and $<p$ at qubits $a$ on the left and the $=p$ and $<p$ at qubits $b$ on the right side shown in green can be omitted for the $U'_p$ circuits in the middle as demonstrated in \autoref{fig:bubble_cascade}.
    This circuit has an input of size $2\ltw M$ qubits and requires $2\ltw M + 4$ ancillae.
  }
  \label{fig:bubble_circuit}
\end{figure}

\begin{figure}[H]
  \centering
  \includegraphics[width=\linewidth]{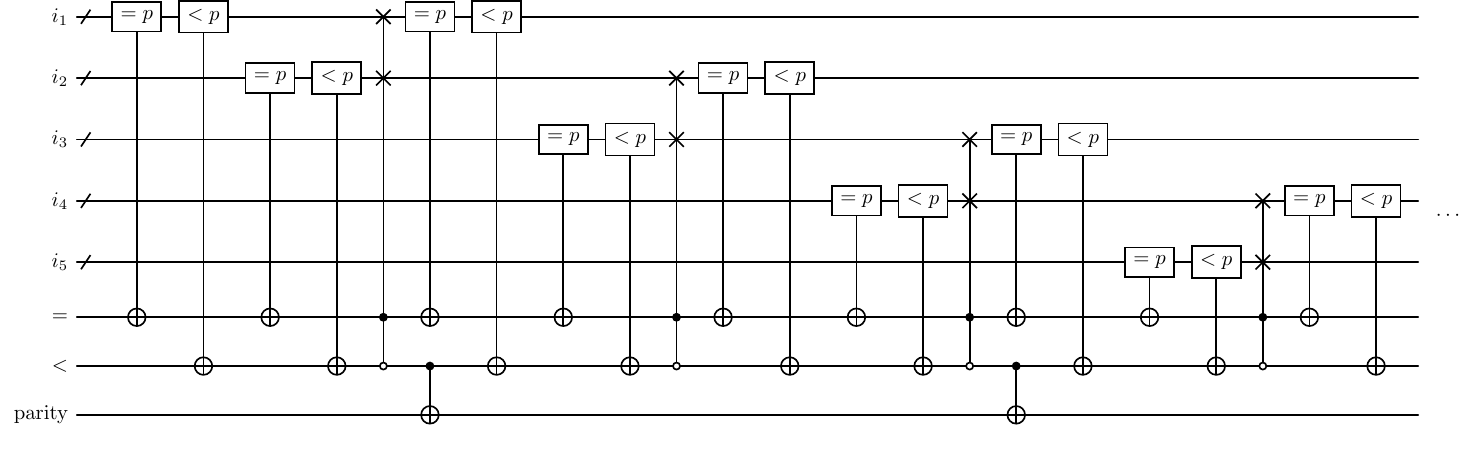}
  \caption{This circuit shows the cascade of the modified bubble circuit $U'_p$ using the same ancilla for the $=$ and $<$ calculation. As a result, some of the $=p$ and $<=p$ gates can be omitted. The CNOT gate is also added to every other bubble circuit such that the parity qubit contains $(i_1<p)\oplus(i_2<p)\oplus\cdots\oplus (i_N<p)$.}
  \label{fig:bubble_cascade}
\end{figure}

Here, we show two optimizations we made on the bubble circuit (\autoref{fig:bubble_ori_circuit}) denoted as $U'_p$ and $U''_p$.
The $U'_p$ gate is shown in \autoref{fig:bubble_circuit}.
In this case, we modified the circuit to use the $<p$ gate instead of the $>p$ gate.
This only results in an open control for the swap gate as no two registers has the same value in the sorted-list encoding as long as $p\ne \infty$.
This allows parity information of the two inputs to be optionally extracted into an ancilla qubit for parity calculations. Next, as the $U_p$ circuit is applied sequentially to consecutive pairs of registers, the ancillae are reused so that some circuit elements can be omitted, as shown in \autoref{fig:bubble_circuit} and~\ref{fig:bubble_cascade}.

\begin{figure}[H]
  \centering
  \includegraphics{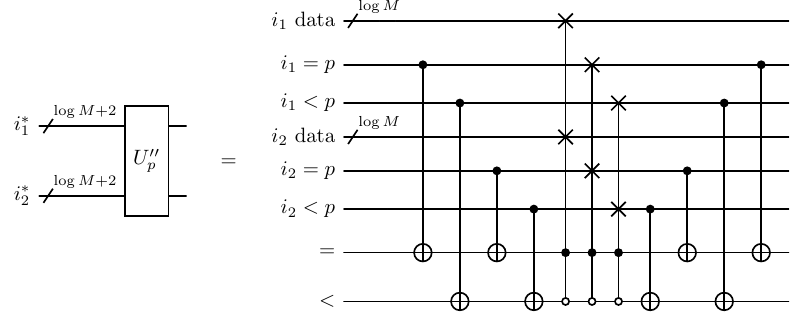}
  \caption{
    The circuit implementing $U_p''$, where the $=p$ and $<p$ gates are precalculated beforehand.
    The registers $\psi_n^*$ contain $\ltw M$ qubits storing the binary representation of the orbital indices (shown as $\psi_n$ data), along with 2 additional ancilla qubits to store the precalculated $=p$ and $<q$ data (shown as $\psi_n=p$ and $\psi_n<p$, respectively).
  }
  \label{fig:bubble_circuit2}
\end{figure}

On the other hand, the $U''_p$ gate precalculates $=p$ and $<p$ information of the registers in the beginning. This optimized bubble gate $U''_p$ halves the number $=p$ and $<p$ gates, with the trade-off of increasing the CSWAP gate by 2 for every application of the $U''_p$ gate, as shown in \autoref{fig:bubble_circuit2}. As a result, the use of this circuit would add two additional ancillae to store $=p$ and $<p$ information for every register of the sorted-list encoding.

\setonecolumn

\settwocolumn

\section{Hamiltonian Simulation}
\label{sec:hamiltonian_simulation}
Numerous quantum algorithms have been proposed to calculate the electronic structure of quantum chemical systems.
This includes Variational Quantum Eigensolver (VQE)~\cite{peruzzoVariationalEigenvalueSolver2014,tillyVariationalQuantumEigensolver2022,grimsleyAdaptiveVariationalAlgorithm2019,yordanovQubitexcitationbasedAdaptiveVariational2021}, quantum imaginary time evolution algorithms~\cite{mottaDeterminingEigenstatesThermal2020,chenQuantumImaginarytimeControl2023}, quantum Monte Carlo methods~\cite{hugginsUnbiasingFermionicQuantum2022}, Quantum Phase Estimation (QPE)~\cite{kitaevQuantumMeasurementsAbelian1995,nielsenQuantumComputationQuantum2012}, among others.
In our paper, we focus on the implementation of the Quantum Phase Estimation (QPE) algorithm, which is one of the fundamental quantum algorithms to calculate the eigenvalues of a unitary circuit.
As typical quantum chemical Hamiltonians are not unitary, it cannot be directly used with the QPE algorithm.
Instead, a proxy operator, implementable in quantum circuits, is constructed such that the Hamiltonian's eigenvalues can be obtained from the eigenvalues of the proxy operator.

\subsection{Robust Phase Estimation with Trotterization}
\label{sec:trotterization}
\begin{figure}
  \centering
  \includegraphics{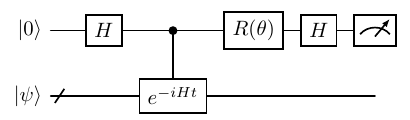}
  \caption{
    The Hadamard test circuit to obtain $\ev{e^{-iHt}}{\psi}$.
    The angle $\theta$ of the phase gate $R(\theta)$ is set to $0$ for $\Re \ev{e^{-iHt}}{\psi}$ and $\pi/2$ for $\Im \ev{e^{-iHt}}{\psi}$.
  }
  \label{fig:rpe_circuit}
\end{figure}

One example of the proxy operator is the Hamiltonian evolution $e^{-iHt}$, which is often implemented via Trotterization~\cite{berryEfficientQuantumAlgorithms2006,hatanoFindingExponentialProduct2005}.
The eigenvalues of the evolution, $E_k(e^{-iHt})$, are related to the eigenvalues of the Hamiltonian, $E_k(H)$, by the following equation:
\begin{equation*}
  E_k(e^{-iHt}) = \exp\left( -iE_k(H)t \right).
\end{equation*}
Robust Phase Estimation (RPE)~\cite{belliardoAchievingHeisenbergScaling2020,higginsDemonstratingHeisenberglimitedUnambiguous2009,kimmelRobustCalibrationUniversal2015,niLowdepthAlgorithmsQuantum2023,guntherPhaseEstimationPartially2026} is particularly well-suited for use with Trotter formulas due to the low number of ancilla qubits required.
The RPE algorithm performs phase estimation with only one ancilla qubit, typically via Hadamard tests, as depicted in \autoref{fig:rpe_circuit}.
Given an error parameter $\varepsilon_{\text{qpe}}$, we perform the Hadamard test to obtain the expectation value of $e^{-iHt}$ at different values of $t$:
\begin{align*}
  g(t) &= \ev{e^{-iHt}}{\Psi}; \quad t = 2^1, \dots, 2^{n_{\text{qpe}}}\\
       &= \sum_{k}\left|\langle\Psi|\Psi_k\rangle\right|^2 e^{-iE_kt}.
\end{align*}
By interpreting $g(t)$ as a time signal, the ground-state energy can be extracted as the lowest frequency component via signal processing techniques~\cite{guntherPhaseEstimationPartially2026}.
To obtain the ground-state energy with precision $\varepsilon$, we require $\mathcal{O}(1/\varepsilon\delta)$ applications of the Trotter formula $S(\delta)$ that approximates the Hamiltonian evolution $e^{-iH\delta}$.
In this case, the Trotter time step $\delta$ is chosen to ensure the Trotter error is bounded by $\varepsilon \delta$.

Implementing $S(\delta)$ involves decomposing the Hamiltonian into a linear combination of Hermitian operators $H_l$, such that each individual $e^{-iw_lH_l\delta_l}$ can be implemented as a quantum circuit:
\begin{equation}
  H = \sum_{l=0}^{L-1}w_l H_l.\label{eqn:hamiltonian_lcu}
\end{equation}
For standard encodings such as the Jordan-Wigner, Parity, and Bravyi-Kitaev~\cite{jordanUeberPaulischeAequivalenzverbot1928,bravyiFermionicQuantumComputation2002}, the individual Hermitian terms are typically chosen to be Pauli strings.

The time evolution $e^{-iH\delta}$ can then be approximated as a product of individual $e^{-iw_lH_l\delta_l}$ terms, with multiple variations of the exponential product formulas available offering varying levels of accuracy.
One such formula is the deterministic $p$\textsuperscript{th}-order Suzuki formula~\cite{hatanoFindingExponentialProduct2005} shown below:
\begin{equation}
  \begin{aligned}
    e^{-iH\delta} &\approx S^{(\text{det})}_p(\delta),\\
    S^{(\text{det})}_1(\delta) &= \prod_{l=0}^{L-1}e^{-iw_lH_l\delta},\\
    S^{(\text{det})}_2(\delta) &= \prod_{l=0}^{L-1}e^{-iw_lH_l\delta/2}{\prod_{l=L-1}^{0*}}e^{-iw_lH_l\delta/2},\\
    S^{(\text{det})}_{p+2}(\delta) &= (S^{(\text{det})}_{p}(s_p\delta))^2S^{(\text{det})}_p((1-4s_p)\delta)\\
    &\phantom{=}\times(S^{(\text{det})}_{p}(s_p\delta))^2,
  \end{aligned}
  \label{eqn:suzuki_trotter}
\end{equation}
where the product $\prod^*$ is taken in reverse order and $s_p = (4 - \sqrt[p+1]{4})^{-1}$.
Another approach is the random qDRIFT formula~\cite{campbellRandomCompilerFast2019}:
\begin{equation*}
  e^{-iH\delta} \approx S_N^{(\text{rand})}(\delta) = \prod_{l=1}^N e^{-i\lambda H_{j} \delta/N}.
\end{equation*}
In this case, the index $j$ is chosen randomly using a probability distribution $p_l = h_l / \lambda$ and $N$ is chosen based on the precision parameter $\varepsilon$.
Finally, the partially random formula partitions the Hamiltonian into deterministic and random implementations:
\begin{equation*}
  e^{-iHt} \approx e^{-iH_{\text{deter}}t}e^{-iH_{\text{rand}}t}.
\end{equation*}
\autoref{tab:trotter_product_formula_costs} presents the total cost of performing the RPE calculation to obtain the ground-state of a Hamiltonian with precision $\varepsilon$ using these three Trotter formulas.

\begin{table}
  \centering
  \caption{
    The number of applications of the time evolution of the elementary Hermitians operators $e^{-iw_lH_lt_l}$ required to obtain the ground-state with precision $\varepsilon$ using the various Trotter product formulas.
    These costs are obtained from a previous work~\cite{guntherPhaseEstimationPartially2026}.
    Here, $L$ is the number of Hermitian terms, $\lambda$ is the 1-norm of the Hamiltonian.
  }
  \begin{ruledtabular}
    \begin{tabular}{ll}
      Formula&Cost\\
      \midrule
      Deterministic $p$\textsuperscript{th}-order~\cite{hatanoFindingExponentialProduct2005}&$\mathcal{O}(5^{p/2-1}L^{2+1/p}/\varepsilon^{1+1/p})$\\
      Random qDRIFT~\cite{campbellRandomCompilerFast2019}&$\mathcal{O}(\lambda^2/\varepsilon^2)$\\
      Partially random~\cite{guntherPhaseEstimationPartially2026}&$\mathcal{O}(5^{p/2-1}L^{2+1/p}/\varepsilon^{1+1/p} + \lambda^2/\varepsilon^2)$\\
    \end{tabular}
  \end{ruledtabular}
  \label{tab:trotter_product_formula_costs}
\end{table}

\subsubsection{Trotterization for the Plane-Wave Basis}
\label{sec:trotterization_pw}
As mentioned in the main text, a lower gate cost can be achieved with the plane-wave basis due to the inherent structure of the plane-wave Hamiltonian.
We define $\nu$ as the indices for the plane-wave (momentum) basis, and $p$ as indices for the plane-wave dual (position) basis.
These two bases are related by the Fourier transform:
\begin{equation*}
  \ket{\nu} = \frac{1}{\sqrt{M}}\sum_p \ket{p} e^{-ik_{\nu}\cdot r_p}.
\end{equation*}
The same relationship holds in the second quantization for creation and annihilation operators, linking the plane-wave basis operators ($a_\nu^\dagger/a_p$) and the plane-wave dual basis operators ($c_p^\dagger/c_p$):
\begin{align*}
  a_{\nu}^{\dagger} &= \frac{1}{\sqrt{M}}\sum_p c_p^\dagger e^{-ik_{\nu}\cdot r_p}.
\end{align*}
Conversion between the two bases can be performed using the quantum Fourier transform (QFT) circuit~\cite{coppersmithApproximateFourierTransform2002,halesImprovedQuantumFourier2000} for the first quantization, and the fermionic fast Fourier transform (FFFT)~\cite{babbushEncodingElectronicSpectra2018} for the second quantization.

The Hamiltonian can then be written as:
\begin{equation}
  H = \operatorname{FT}^{\dagger}\hat{T}\operatorname{FT} +\ \hat{U} + \hat{V},\label{eqn:h_trotter_pw}\\
\end{equation}
where $\hat{T}$, $\hat{U}$, and $\hat{V}$ represents the kinetic, external, and coulomb operators, respectively, expressed in first and second quantization as follows:
\begin{align*}
  \hat{T}^{(1)} &= \sum_i^N\sum_{\nu}^MT_{\nu} \dyad{\nu}_i,& \hat{T}^{(2)} &= \sum_{\nu}^M T_{\nu} a_{\nu}^\dagger a_{\nu},\\
  \hat{U}^{(1)} &= \sum_i^N\sum_{p}^M U_{p} \dyad{p}_i,& \hat{U}^{(2)} &= \sum_{p}^M U_{p}n_p,\\
  \hat{V}^{(1)} &= \sum_{i\ne j}^N \sum_{p\ne q}^M V_{pq} \dyad{p}_i\dyad{q}_j,& \hat{V}^{(2)} &= \sum_{p\ne q}^M V_{pq}n_p n_q,
\end{align*}
where $n_p = c_p^\dagger c_p$ denotes the number operator in the dual basis.
Furthermore, the operator $\operatorname{FT}$ represents applying QFT for all the electronic degrees of freedom for the first quantization, or the FFFT operator~\cite{babbushEncodingElectronicSpectra2018} for the second quantization.

The terms $\dyad{\nu}_i$ are diagonal in the plane-wave basis, while $\dyad{p}_i$ and $\dyad{p}_i\dyad{q}_j$ are diagonal in the plane-wave dual basis.
The same property holds for their second-quantized counterparts, allowing all the kinetic operators to mutually commute in the plane-wave basis and all the potential operators to mutually commute in the plane-wave dual basis.
This mutually commuting property is highly advantageous for Trotterization, as the time evolution of $\hat{T}$ and $\hat{U}+\hat{V}$ can be implemented exactly, without any intrinsic Trotter errors.
\begin{align*}
  e^{-i\hat{T}^{(1)}\delta} &= \prod_{i}^N e^{-i\hat{T}_i^{(1)}\delta} = \prod_{i}^N e^{-i\delta\sum_{\nu}^M T_{\nu}\dyad{\nu}_i},\\
  e^{-i(\hat{U}^{(1)}+\hat{V}^{(1)})\delta} &= \prod_{i}^Ne^{-i\hat{U}_i^{(1)}\delta}\prod_{i\ne j}^Ne^{-i\hat{V}_{ij}^{(1)}\delta}\\
  &= \prod_{i}^Ne^{-i\delta\sum_{p}^MU_{p}\dyad{p}_i}\\
  &\phantom{=}\times\prod_{i\ne j}^Ne^{-i\delta\sum_{pq}^MV_{pq}\dyad{p}_i\dyad{q}_j},\\
  e^{-i\hat{T}^{(2)}\delta} &= \prod_{\nu}^Me^{-i\delta T_{\nu}a_{\nu}^{\dagger}a_{\nu}},\\
  e^{-i(\hat{U}^{(2)}+\hat{V}^{(2)})\delta} &= \prod_{p}^Me^{-i\delta U_{p}n_p}\prod_{p\ne q}^Me^{-i\delta V_{pq}n_pn_q}.
\end{align*}
Thus, we effectively have $L=2$ for the Hamiltonian decomposition, comprising $\hat{T}$ and $\hat{U}+\hat{V}$.
Due to the low value of $L$, it is evident that the deterministic formula is more efficient for this Hamiltonian.
As an example, the second-order ($p=2$) Suzuki-Trotter product formula can be written as:
\begin{equation*}
  S_2(\delta) = e^{-i(\hat{U}+\hat{V})\delta/2}\operatorname{FT}^{\dagger}e^{-i\hat{T}\delta}\operatorname{FT}e^{-i(\hat{U}+\hat{V})\delta/2}.
\end{equation*}

\subsection{Szegedy Quantum Walk with Qubitization}
\label{sec:qubitization}
A more recent method of constructing the proxy operator for phase estimation algorithms is through qubitization.
In this approach, the Szegedy walk operator~\cite{babbushEncodingElectronicSpectra2018}, implementable as a quantum circuit, serves as the proxy unitary.
The construction of this walk operator can be summarized as follows.
Unlike the Trotterization case, the Hamiltonian terms are divided into a linear combination of unitaries, where all the coefficients $w_l$ are positive (with any sign or complex phase incorporated into the unitary itself):
\begin{equation*}
  H = \sum_{l=0}^{L-1} w_l U_l\quad w_l\in\mathbb{R}^+.
\end{equation*}
Those terms are then incorporated into the Szegedy walk operator using the $\operatorname{PREPARE}$ and $\operatorname{SELECT}$ circuits, defined in the high-level as:
\begin{align}
  \operatorname{PREPARE}\ket{0}^{\otimes \ltw L} &= \sum_{l=0}^{L-1} \sqrt{\frac{|w_l|}{\lambda}}\ket{l}^{\otimes \ltw L},\label{eqn:prepare}\\
  \operatorname{SELECT}\ket{l}\otimes \ket{\psi} &= \ket{l}\otimes U_l\ket{\psi}.\label{eqn:select}
\end{align}
The PREPARE circuit prepares the coefficient magnitudes $|w_l|$ on $\ltw L$ ancilla qubits, where $\ket{l}^{\otimes \ltw L}$ is the binary representation of the index $l$, and $\lambda = \sum_{l=0}^{L-1} |w_l|$.
On the other hand, the $\operatorname{SELECT}$ circuit applies the unitaries $U_l$ based on the indices $l$ encoded on the ancilla qubits.
At this point, applying a composition of these two operators yields a normalized representation of the Hamiltonian:
\begin{equation}
  \operatorname{PREPARE}^\dagger\cdot\operatorname{SELECT}\cdot\operatorname{PREPARE}\ket{0}\otimes\ket{\psi} = \ket{0}\otimes \frac{H}{\lambda}\ket{\psi}.\label{eqn:lcu_prepare_select}
\end{equation}
However, \autoref{eqn:lcu_prepare_select} cannot be directly used for the QPE algorithm because $H$ is not unitary, as previously noted.
The Szegedy walk operator~\cite{babbushEncodingElectronicSpectra2018} is constructed instead:
\begin{equation}
  \mathcal{W} = (2\operatorname{PREPARE}\dyad{0}{0}\operatorname{PREPARE}^\dagger\otimes \mathbb{I} - \mathbb{I}\otimes\mathbb{I})\cdot\operatorname{SELECT}.\label{eqn:szegedy_walk}
\end{equation}
The eigenvalues of the walk operator, $E_k(\mathcal{W})$, and the Hamiltonian, $E_k(H)$, are then related by:
\begin{equation}
  E_k(\mathcal{W}) = \pm \arccos(E_k(H) / \lambda).\label{eqn:szegedy_walk_spectrum}
\end{equation}
Consequently, one can obtain $E_k(H)$ by performing the QPE algorithm on $\mathcal{W}$.
For this, we opt to use the optimal QPE algorithm, which minimizes the number of $\mathcal{W}$ applications~\cite{luisOptimumPhaseshiftEstimation1996,babbushEncodingElectronicSpectra2018}, shown in Figure 2 of~\cite{babbushEncodingElectronicSpectra2018}.

Various implementations exist for the $\operatorname{PREPARE}$ and $\operatorname{SELECT}$ circuits.
Implementations for second-quantized Hamiltonians in the MO basis, using the 8-fold symmetry of the coulomb terms, include methods like single-factorization~\cite{berryQubitizationArbitraryBasis2019}, double-factorization~\cite{vonburgQuantumComputingEnhanced2021}, and tensor hypercontraction~\cite{leeEvenMoreEfficient2021}.
These techniques lead to circuit implementations with varying T gate complexities, where lower scaling often comes with the trade-off of more expensive classical preprocessing.
Alternatively, implementations that exploit the inherent structure of the plane-wave basis set typically utilizes the first-quantized encoding~\cite{babbushQuantumSimulationChemistry2019,suFaultTolerantQuantumSimulations2021}.

The following section highlighted the single-factorization method~\cite{berryQubitizationArbitraryBasis2019}, used for the simulation of the second-quantized Hamiltonian in the MO basis-set, along with the first-quantized plane-wave algorithm~\cite{suFaultTolerantQuantumSimulations2021,babbushQuantumSimulationChemistry2019}.
This serves as an introduction to both methods as they will be incorporated into our circuit formulations of the sorted-list encoding.

\subsubsection{Single-Factorization}
\label{sec:prev_qubitization_sf}

\begin{table*}
  \centering
  \caption{Legend for the qubit label of \autoref{eqn:sf_prepare}.}
  \begin{ruledtabular}
    \begin{tabular}{lP{15.8cm}}
      Qubit Label&Description\\
      \midrule
      $a$&Indices of the eigenvalues and eigenvectors of the matrix $W$. $\ket{0}_a$ is reserved for the one-electron term.\\
      $b$ and $c$&Sign of the $T'_{pq}$ for the one-electron term, and the sign of the $g_{pq}^{(l)}$ and $g_{rs}^{(l)}$, respectively, for the two-electron term.\\
      $d$ and $e$&Orbital indices and spin for the Majorana operators.\\
    \end{tabular}
  \end{ruledtabular}
  \label{tab:sq_prepare_legend}
\end{table*}

The single-factorization algorithm~\cite{berryQubitizationArbitraryBasis2019} focuses on reducing the complexity of implementing the two-electron terms in a quantum circuit.
Although it is possible to construct spin-unrestricted implementations of this algorithm, only the spin-restricted version is discussed in previous works~\cite{berryQubitizationArbitraryBasis2019,vonburgQuantumComputingEnhanced2021,leeEvenMoreEfficient2021}.
Furthermore, all the $h_{pq}$ and $h_{pqrs}$ coefficients must be real and that the $h_{pqrs}$ term must have 8-fold symmetry:
\begin{equation}
  \begin{aligned}
    h_{pq}&\in \mathbb{R},\ h_{pqrs}\in \mathbb{R},\\
    h_{pq} &=  h_{qp},\\
    h_{pqrs} &= h_{prqs} = h_{sqrp} = h_{srqp},\\
             &=  h_{qpsr} = h_{qspr} = h_{rpsq} = h_{rspq}.
  \end{aligned}
  \label{eqn:sf_constraint}
\end{equation}
As a result, the more compact molecular orbital basis sets are typically used instead of the plane-wave basis set. Despite this, other basis sets such as the Daubechies wavelets~\cite{hongAccurateEfficientQuantum2022} can also be used as long as the constraints of \autoref{eqn:sf_constraint} are satisfied.

Here, we show a summary of the single-factorization algorithm~\cite{berryQubitizationArbitraryBasis2019} along with the optimizations made in another work~\cite{leeEvenMoreEfficient2021}.
While they directly converted the fermionic operators to Pauli operators using the Jordan-Wigner encoding, we opt to convert the fermionic operators to Majorana operators instead.
First, an alternative second-quantized representation of the Hamiltonian, shown in \autoref{eqn:hamiltonian_sf_1}, is used:
\begin{multline}
  H_{\text{SF}} = \sum_{\sigma\in\left\{ \uparrow,\downarrow \right\}}\sum_{p,q=1}^{M/2}T_{pq}a_{p,\sigma}^{\dagger}a_{q,\sigma}\\
                +\frac{1}{2}\sum_{\alpha,\beta\in\left\{ \uparrow,\downarrow \right\}}\sum_{p,q,r,s=1}^{M/2}V_{pqrs}a_{p,\alpha}^{\dagger}a_{q,\alpha}a_{r,\beta}^{\dagger}a_{s,\beta}.\label{eqn:hamiltonian_sf_1}
\end{multline}
Next, we define the operator $Q_{pq\sigma}$ in terms of Majorana operators:
\begin{align*}
  Q_{pq\sigma} &= i\gamma_{p,0,\sigma}\gamma_{q,1,\sigma},\\
  a_{p,\sigma}^{\dagger}a_{q,\sigma} + a_{q,\sigma}^{\dagger}a_{p,\sigma} &= i\frac{\gamma_{p,0,\sigma}\gamma_{q,1,\sigma}}{2} + i\frac{\gamma_{q,0,\sigma}\gamma_{p,1,\sigma}}{2},\\
  a_{p,\sigma}^{\dagger}a_{p,\sigma} &= \frac{\mathbb{I}}{2} + i\frac{\gamma_{p,0,\sigma}\gamma_{p,1,\sigma}}{2}.
\end{align*}
Due to the symmetries of \autoref{eqn:sf_constraint}, we can then express $H_{\text{SF}}$ in terms of $Q_{pq\sigma}$.
Note that the constant $\mathbb{I}/2$ term in the decomposition of $a_{p,\sigma}^{\dagger}a_{p,\sigma}$ is not implemented by the $Q_{pq\sigma}$ operator and therefore must be explicitly accounted for, resulting in $T'_{pq}$ and a constant term:
\begin{align*}
  H_{\text{SF}} &= H_{\text{const}}\mathbb{I} + \frac{1}{2}\sum_{\sigma\in\left\{ \uparrow,\downarrow \right\}}\sum_{p,q=1}^{M/2}T'_{pq}Q_{pq\sigma}\\
                &\phantom{=}+ \frac{1}{8}\sum_{\alpha,\beta\in\left\{ \uparrow,\downarrow \right\}}\sum_{p,q,r,s=1}^{M/2}V_{pqrs}Q_{pq\alpha}Q_{rs\beta},\\
  H_{\text{const}} &= \sum_{p=1}^{M/2}T_{pp} + \frac{1}{2}\sum_{pr=1}^{M/2}V_{pprr},\\
  T'_{pq} &= T_{pq} + \sum_{r=1}^{M/2}V_{pqrr}.
\end{align*}

Next, the $M/2\times M/2\times M/2\times M/2$ tensor $V_{pqrs}$ is converted to an $M^2/4\times M^2/4$ matrix $W$, with the rows containing the composite index $pq$ and the columns containing the composite index $rs$ ($W_{pM/2+q,rM/2+s} = V_{pqrs}$).
This matrix is then diagonalized to obtain eigenvalues $w_l$ and eigenvectors $g^{(l)}_{pq}$. As a result,
\begin{multline*}
  \frac{1}{8}\sum_{\alpha,\beta} \sum_{p,q,r,s=1}^{M/2}V_{pqrs}Q_{pq\alpha}Q_{rs\beta}\\
  = \sum_{l=1}^{L}w_l \left( \sum_{\sigma\in\left\{ \uparrow,\downarrow \right\}}\sum_{p,q=1}^{M/2}g_{pq}^{(l)}Q_{pq\sigma} \right)^2.
\end{multline*}
In this case, only the largest $L \ll M^2/4$ eigenvalues of $W$ is included, chosen such that this differs from the original Hamiltonian by an acceptable error, with empirical observations showing that $L\sim\mathcal{O}(M\log M)$~\cite{pengHighlyEfficientScalable2017}.
After the $T'_{pq}$, $w_l$, $g_{pq}^{(l)}$ coefficients are calculated, the $\operatorname{PREPARE}$ circuit is implemented as follows:
\begin{align}
  &\operatorname{PREPARE} \rightarrow \ket{0}_a\sum_{pq\sigma}\sqrt{\frac{|T'_{pq}|}{2\lambda_{\text{SF}}}}\ket{\tau_{pq}^{(0)}}_b\ket{0}_c\ket{p;q;\sigma}_d\ket{0}_e\notag\\
  &\hspace{10mm}+ \sum_l\sqrt{\frac{w_l}{\lambda_{\text{SF}}}}\ket{l}_a\sum_{pqrs\alpha\beta}\sqrt{|g_{pq}^{(l)}g_{rs}^{(l)}|}\ket{\tau_{pq}^{(l)}}_b\ket{\tau_{rs}^{(l)}}_c\notag\\
  &\hspace{10mm}\times\ket{p;q;\alpha}_d\ket{r;s;\beta}_e.\label{eqn:sf_prepare}
\end{align}
The preparation of the coefficients is done using the coefficient oracle developed in~\cite{babbushEncodingElectronicSpectra2018}, with the QROM implemented using the techniques that trade dirty or clean qubits to reduce its T gate cost~\cite{lowTradingGatesDirty2024}. On the other hand, the SELECT circuit is implemented as two circuits below. $\operatorname{SELECT}_1$ is applied for any value of $\ket{l}_a$, while $\operatorname{SELECT}_2$ is conditionally applied when $\ket{l}_a$ is nonzero:
\begin{align*}
  \operatorname{SELECT}_1\ket{\tau}_b\ket{p;q;\alpha}_d\ket{\psi}
  &= (-1)^{\tau}i\times\ket{\tau}_b\ket{p;q;\alpha}_d \nonumber\\
  &\phantom{=}\otimes \gamma_{p,0,\alpha}\gamma_{q,1,\alpha}\ket{\psi},\\
  \operatorname{SELECT}_2\ket{\tau}_c\ket{r;s;\beta}_e\ket{\psi}
  &= (-1)^{\tau}i\times\ket{\tau}_c\ket{r;s;\beta}_e \nonumber\\
  &\phantom{=}\otimes \gamma_{r,0,\beta}\gamma_{s,1,\beta}\ket{\psi}.
\end{align*}
One final issue to account for is the conversion from the spatial-orbital indices to the spin-orbital indices. This is done using a simple controlled addition circuit~\cite{cuccaroNewQuantumRipplecarry2004}.
\autoref{tab:sq_prepare_legend} details the indices $a$ to $e$ of the PREPARE circuit.

\subsubsection{First-Quantized Encoding}
\label{sec:prev_qubitization_fq}

\begin{table*}
  \centering
  \caption{Legend for the qubit label of \autoref{eqn:fq_pw_prepare}.}
  \begin{ruledtabular}
    \begin{tabular}{lP{15.8cm}}
      Qubit Label&Description\\
      \midrule
      $a$&$\ket{0}_a$ tells the SELECT circuit to apply $\hat{T}^{(1)}$\\
         &$\ket{1}_a$ tells the SELECT circuit to apply $\hat{U}^{(1)}+\hat{V}^{(1)}$\\
      $b$&Used to cancel out the unitary application when $(p+\nu)\notin G$ and when $(q-\nu)\notin G$ for $\hat{U}^{(1)}$ and $\hat{V}^{(1)}$ (see \autoref{eqn:fq_hamiltonian_pw_u} and~\ref{eqn:fq_hamiltonian_pw_v})\\
      $c$&Specifies if $i=j$ for the case of $\hat{V}^{(1)}$ (see \autoref{eqn:fq_hamiltonian_pw_v})\\
      $d$ and $e$&The electron indices $i$ and $j$, respectively\\
      $w$&$\ket{0}_w$, $\ket{1}_w$, $\ket{2}_w$ specifies the Cartesian axes $x,y,$ and $z$, respectively, used for the SELECT circuit to apply $\hat{T}^{(1)}$\\
      $g$ and $h$& Used for the application of $\|k_p\|^2$ in $\hat{T}^{(1)}$ (see \autoref{eqn:fq_hamiltonian_pw_t})\\
      $i$&$\ket{1}_a\ket{0}_i$ tells the SELECT circuit to apply $\hat{U}^{(1)}$\\
         &$\ket{1}_a\ket{1}_i$ tells the SELECT circuit to apply $\hat{V}^{(1)}$\\
      $j$ and $k$&Creates the $1/\|k_{\nu}\|^2$ coefficients for $\hat{U}^{(1)}$ and $\hat{V}^{(1)}$ (see \autoref{eqn:fq_hamiltonian_pw_u} and~\ref{eqn:fq_hamiltonian_pw_v})\\
      $l$&The ion indices $I$ for $\hat{U}^{(1)}$ (see \autoref{eqn:fq_hamiltonian_pw_u})\\
      $m$ and $n$&Instead of performing arithmetic operations on the $i$\textsuperscript{th} and $j$\textsuperscript{th} registers, respectively, the $\operatorname{SELECT}$ circuit swaps those registers with these temporary ancillae, controlled on qubit $d$ and $e$. The arithmetic operations $\ket{p_i+\nu}_m$ and $\ket{q_j-\nu}_n$ are then performed here and the resulting values are swapped back to the original register.\\
    \end{tabular}
  \end{ruledtabular}
  \label{tab:fq_pw_prepare_legend}
\end{table*}

To leverage the good structure of the plane-wave basis, we divide the Hamiltonian into the kinetic $\hat{T}$, external potential $\hat{U}$, and coulomb potential $\hat{V}$ terms.
In the first-quantization, those can be written as:
{
  \begin{gather}
    H^{(1)} = \hat{T}^{(1)} + \hat{U}^{(1)} + \hat{V}^{(1)},\label{eqn:fq_hamiltonian_pw}
  \end{gather}
  \begin{gather}
    \hat{T}^{(1)} = \sum_{i=1}^{N}\left(\sum_{p\in G}\frac{\|k_p\|^2}{2}\dyad{p}{p}_i\right),\label{eqn:fq_hamiltonian_pw_t}\\
    \hat{U}^{(1)} = -\frac{4\pi}{\Omega}\sum_{i=1}^{N}\sum_{\substack{\nu\in G\\\nu\ne 0}}\left( \sum_{\substack{p\in G\\(p-\nu)\in G}} \sum_{I=1}^{N_{at}}Z_I\frac{e^{-ik_{\nu}\cdot \vb{X}_I}}{\|k_{\nu}\|^2} \dyad{p-\nu}{p}_i \right),\label{eqn:fq_hamiltonian_pw_u}\\
    \hat{V}^{(1)} = \frac{2\pi}{\Omega}\sum_{\substack{i,j=1\\i\ne j}}^{N}\sum_{\substack{\nu\in G\\\nu\ne 0}}\left( \sum_{\substack{p,q\in G\\(p+\nu)\in G\\(q-\nu)\in G}} \frac{1}{\|k_{\nu}\|^2} \dyad{p+\nu}{p}_i \dyad{q-\nu}{q}_j\right).\label{eqn:fq_hamiltonian_pw_v}
  \end{gather}
}
The $\hat{T}^{(1)}$, $\hat{U}^{(1)}$, $\hat{V}^{(1)}$ terms can be decomposed into a linear combination of unitary terms (after normalization), shown as brackets in the equation above.
The action of such unitaries (with the normalization omitted for the sake of clarity) on an arbitrary Hartree product can be described by the following equations.
{
  \allowdisplaybreaks[4]
  \begin{widetext}
    \begin{align*}
      \ket{a}_1\cdots\ket{p}_i\cdots\ket{z}_N &\xRightarrow{T\text{ unitary}} \frac{\|k_p\|^2}{2}(\ket{a}_1\cdots\ket{p}_i\cdots\ket{z}_N),\\
      \ket{a}_1\cdots\ket{p}_i\cdots\ket{z}_N &\xRightarrow{U\text{ unitary}} \left( \sum_{I=1}^{N_{at}}Z_I \frac{e^{-ik_{\nu}\cdot \vb{X}_I}}{\|k_{\nu}\|^2} \right)(\ket{a}_1\cdots\ket{p-\nu}_i\cdots\ket{z}_N),\\
      \ket{a}_1\cdots\ket{p}_i\cdots\ket{q}_j\cdots\ket{z}_N &\xRightarrow{V\text{ unitary}} \frac{1}{\|k_{\nu}\|^2}(\ket{a}_1\cdots\ket{p+\nu}_i\cdots\ket{q-\nu}_j\cdots\ket{z}_N).
    \end{align*}
    Using the qubit labels $a$ to $l$ as explained in \autoref{tab:fq_pw_prepare_legend}, the unitary formulations above results in the following form of the $\operatorname{PREPARE}$ and $\operatorname{SELECT}$ circuits:
    \begin{align}
      \operatorname{PREPARE} &\rightarrow (\cos\theta \ket{0}+\sin\theta\ket{1})_a\ket{+}_b\frac{1}{\sqrt{N}}\left( \sqrt{N-1}\ket{0}_c\sum_{i\ne j=1}^{N}\ket{i}_d\ket{j}_e + \ket{1}_c\sum_{j=1}^{N}\ket{j}_d\ket{j}_e \right)\notag\\
                             &\phantom{=}\times\left( \frac{1}{\sqrt{3}}\sum_{w=0}^2\ket{w}_f \right)\left( \frac{1}{2^{n_p-1}-1}\sum_{r,s=0}^{n_p-2}2^{(r+s)/2}\ket{r}_g\ket{s}_h \right)\left(\sqrt{\frac{\lambda_U}{\lambda_U+\lambda_V}}\ket{0}_i + \sqrt{\frac{\lambda_V}{\lambda_U+\lambda_V}}\ket{1}_i \right)\notag\\
                             &\phantom{=}\times\left( \sqrt{\frac{p_\nu}{\lambda_\nu}}\ket{0}_j\sum_{\nu\in G_0}\frac{1}{\|\nu\|}\ket{\nu}_k + \sqrt{1-p_{\nu}}\ket{1}_j\ket{\nu^{\bot}}_k \right)\left( \frac{1}{\sqrt{\sum_IZ_I}}\sum_{I=1}^{N_{at}}\sqrt{Z_I}\ket{I}_l \right),\label{eqn:fq_pw_prepare}
    \end{align}
    \begin{align}
      \operatorname{SELECT}_{\hat{T}^{(1)}} \ket{b}_b\ket{j}_e\ket{w}_f\ket{r}_g\ket{s}_h\ket{q_j}_n &= (-1)^{b(q_{w,r}q_{w,s}\oplus 1)}\ket{b}_b\ket{j}_e\ket{w}_f\ket{r}_g\ket{s}_h\ket{q_j}_n,\notag\\
      \operatorname{SELECT}_{\hat{U}^{(1)}} \ket{b}_b\ket{j}_e\ket{0}_i\ket{\nu}_k\ket{I}_l\ket{q_j}_n &= -e^{-ik_{\nu}\cdot \vb{X}_I}(-1)^{b[(q-v)\notin G]}\ket{b}_b\ket{j}_e\ket{0}_i\ket{\nu}_k\ket{I}_l\ket{q_j-\nu}_n,\notag\\
      \operatorname{SELECT}_{\hat{V}^{(1)}} \ket{b}_b\ket{i}_d\ket{j}_e\ket{1}_i\ket{\nu}_k\ket{p_i}_m\ket{q_j}_n &= (-1)^{b([(p+\nu)\notin G]\vee[(q-\nu)\notin G])}\ket{b}_b\ket{i}_d\ket{j}_e\ket{1}_i\ket{\nu}_k\ket{p_i+\nu}_m\ket{q_j-\nu}_n. \label{eqn:fq_pw_select}
    \end{align}
  \end{widetext}
}

Here, we separate the circuit based on the application of $\hat{T}^{(1)}$, $\hat{U}^{(1)}$, and $\hat{V}^{(1)}$.
The indices outside the ket corresponds to ancillas required for the qubitization circuits as explained in \autoref{tab:fq_pw_prepare_legend}. In this case, ancillas $a$ to $l$ are ancillas generated by the PREPARE circuit of \autoref{eqn:fq_pw_prepare} while ancillas $m$ and $n$ are temporary registers to hold the value of registers $i$ and $j$ during the operation of the SELECT circuit.

\section{Trotterization Implementation for the Sorted-List Encoding}
\label{sec:trotter_circuit}
\subsection{Circuit Implementations}
\label{sec:trotter_circuit_impl}

For our Trotterization implementation of the sorted-list encoding, the Hamiltonian evolution is decomposed into the following unitaries:
\begin{gather}
  \exp(i\theta(a_p^\dagger a_q + a_q^\dagger a_p)),\label{eqn:unitary_one_real}\\
  \exp(i\theta(ia_p^\dagger a_q - ia_q^\dagger a_p)),\label{eqn:unitary_one_imag}\\
  \exp(i\theta(a_p^\dagger a_q^\dagger a_r a_s + a_s^\dagger a_r^\dagger a_q a_p)),\label{eqn:unitary_two_real}\\
  \exp(i\theta(ia_p^\dagger a_q^\dagger a_r a_s - ia_s^\dagger a_r^\dagger a_q a_p)).\label{eqn:unitary_two_imag}
\end{gather}
Notice that $(a_p^\dagger a_q \pm a_q^\dagger a_p)^2$ can be written as:
\begin{align*}
  (a_p^\dagger a_q \pm a_q^\dagger a_p)^2
  &= (a_p^\dagger a_q)^2 + (a_q^\dagger a_p)^2 \\
  &\phantom{=}\ \pm a_p^\dagger a_q a_q^\dagger a_p \pm a_q^\dagger a_p a_p^\dagger a_q\\
  &= \pm (a_p^\dagger a_q a_q^\dagger a_p + a_q^\dagger a_p a_p^\dagger a_q),
\end{align*}
which means:
\begin{align*}
  (a_p^\dagger a_q + a_q^\dagger a_p)^2\ket{\psi} &=
  \begin{cases}
    \ket{\psi}& \text{ if } a_p^\dagger a_q\ket{\psi} \ne 0\\
    \ket{\psi}& \text{ if } a_q^\dagger a_p\ket{\psi} \ne 0\\
    0&\text{ otherwise}
  \end{cases},\\
  (i(a_p^\dagger a_q - a_q^\dagger a_p))^2\ket{\psi} &=
  \begin{cases}
    \ket{\psi}& \text{ if } a_p^\dagger a_q\ket{\psi} \ne 0\\
    \ket{\psi}& \text{ if } a_q^\dagger a_p\ket{\psi} \ne 0\\
    0&\text{ otherwise}
  \end{cases}.
\end{align*}
Thus, given any $\ket{\psi}$ such that either $a_p^\dagger a_q\ket{\psi} \ne 0$ or $a_q^\dagger a_p\ket{\psi} \ne 0$ (either orbital $p$ is occupied, or orbital $q$ is occupied, but not both), we can expand the exponent of \autoref{eqn:unitary_one_real} into the Taylor series:
\begin{equation*}
  \exp(i\theta(a_p^\dagger a_q + a_q^\dagger a_p)) = (\cos\theta + i\sin\theta (a_p^\dagger a_q + a_q^\dagger a_p))\ket{\psi}.
\end{equation*}
If $\ket{\psi}$ does not satisfy the previous condition, it reverts to the identity operator. Through similar arguments, we can obtain similar identities for \autoref{eqn:unitary_one_imag} to~\ref{eqn:unitary_two_imag}.
{
  \allowdisplaybreaks[4]
  \begin{align*}
    &\exp(i\theta(a_p^\dagger a_q + a_q^\dagger a_p))\ket{\psi}\\
    &= \begin{cases}
      (\cos\theta + i\sin\theta (a_p^\dagger a_q))\ket{\psi}&\text{ if } a_p^\dagger a_q\ket{\psi} \ne 0\\
      (\cos\theta + i\sin\theta (a_q^\dagger a_p))\ket{\psi}&\text{ if } a_q^\dagger a_p\ket{\psi} \ne 0\\
      \ket{\psi}&\text{ otherwise}
    \end{cases}\\
    &\exp(i\theta(ia_p^\dagger a_q - ia_q^\dagger a_p))\ket{\psi}\\
    &= \begin{cases}
      (\cos\theta + i\sin\theta (i a_p^\dagger a_q))\ket{\psi}&\text{ if } a_p^\dagger a_q\ket{\psi} \ne 0\\
      (\cos\theta + i\sin\theta (-i a_q^\dagger a_p))\ket{\psi}&\text{ if } a_q^\dagger a_p\ket{\psi} \ne 0\\
      \ket{\psi}&\text{ otherwise}
    \end{cases}\\
    &\exp(i\theta(a_p^\dagger a_q^\dagger a_r a_s + a_s^\dagger a_r^\dagger a_q a_p))\ket{\psi}\\
    &= \begin{cases}
      (\cos\theta + i\sin\theta (a_p^\dagger a_q^\dagger a_r a_s))\ket{\psi}&\text{ if } a_p^\dagger a_q^\dagger a_r a_s\ket{\psi} \ne 0\\
      (\cos\theta + i\sin\theta (a_s^\dagger a_r^\dagger a_q a_p))\ket{\psi}&\text{ if } a_s^\dagger a_r^\dagger a_q a_p\ket{\psi} \ne 0\\
      \ket{\psi}&\text{ otherwise}
    \end{cases}\\
    &\exp(i\theta(ia_p^\dagger a_q^\dagger a_r a_s - ia_s^\dagger a_r^\dagger a_q a_p))\ket{\psi}\\
    &= \begin{cases}
      (\cos\theta + i\sin\theta (i a_p^\dagger a_q^\dagger a_r a_s))\ket{\psi}&\text{ if } a_p^\dagger a_q^\dagger a_r a_s\ket{\psi} \ne 0\\
      (\cos\theta + i\sin\theta (-i a_s^\dagger a_r^\dagger a_q a_p))\ket{\psi}&\text{ if } a_s^\dagger a_r^\dagger a_q a_p\ket{\psi} \ne 0\\
      \ket{\psi}&\text{ otherwise}
    \end{cases}
  \end{align*}
}

\begin{figure*}
  \centering
  \includegraphics[width=0.95\linewidth]{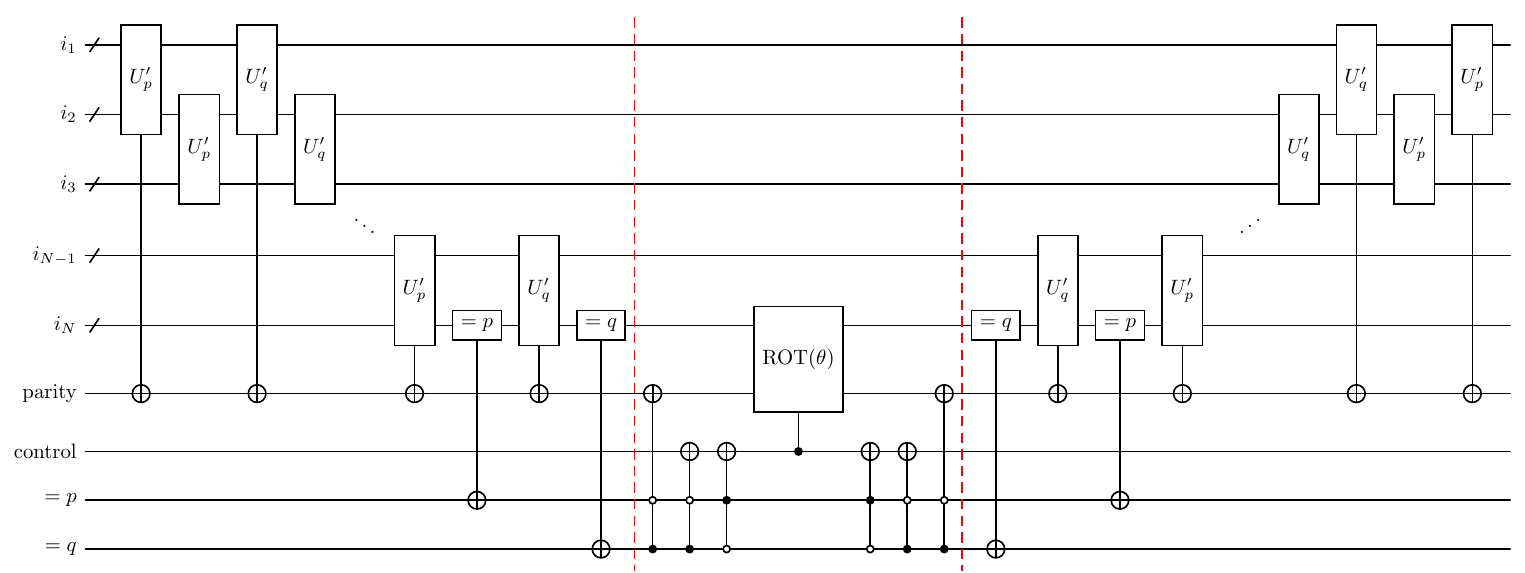}
  \caption{
    The circuit implementing $\exp(i\theta (a_p^\dagger a_q+a_q^\dagger a_p))$ and $\exp(i\theta (ia_p^\dagger a_q - ia_q^\dagger a_p))$ for $p > q$.
    For $\exp(i\theta (a_p^\dagger a_q+a_q^\dagger a_p))$, $\operatorname{ROT}(\theta)$ is a controlled Pauli evolution gate $\exp(i\theta (X^{p\oplus q}_{\psi_N}\otimes Z_{parity}))$, where $X$ gates are placed in the `1' bits of the binary representation of $p\oplus q$, and a $Z$ gate is placed on the parity ancilla.
    The $\operatorname{ROT}(\theta)$ for $\exp(i\theta (ia_p^\dagger a_q - ia_q^\dagger a_p))$ is implemented similarly, with one of the $X$ replaced by a $Y$.
    Both can be implemented efficiently using the techniques in~\cite{yordanovEfficientQuantumCircuits2020}.
    Note that the $=p$ gate is implicitly done in the $U'_p$ cascade.
    This is achieved by using the $=$ ancilla in $U'_p$ of \autoref{fig:bubble_circuit} as the $=p$ ancilla here and omitting the $=p$ gate in the final $U'_p$ gate of $\psi_{N-1}$ and $\psi_N$.
    The $=q$ gate is also done implicitly with the $U'_q$ cascade.
    The parity calculation is performed by obtaining parity information from the $U'_p$ and $U'_q$ gates as shown in \autoref{fig:bubble_circuit} such that the parity ancilla contains the expression of \autoref{eqn:one_elec_rot_parity} just after the $p$ and $q$ indices are bubbled down.
    The red dashed line separates the circuit into the $\operatorname{BUBBLE}$ operation, the $\operatorname{ROTATE}$ operation, and the $\operatorname{BUBBLE}^\dagger$ operation from left to right.
    In total, this circuit acts on $N\ltw M$ data qubits and requires $N(\ltw M + 1) + 6$ ancillae.
  }
  \label{fig:one_elec_rot}
\end{figure*}

\begin{figure*}
  \centering
  \includegraphics[width=\textwidth]{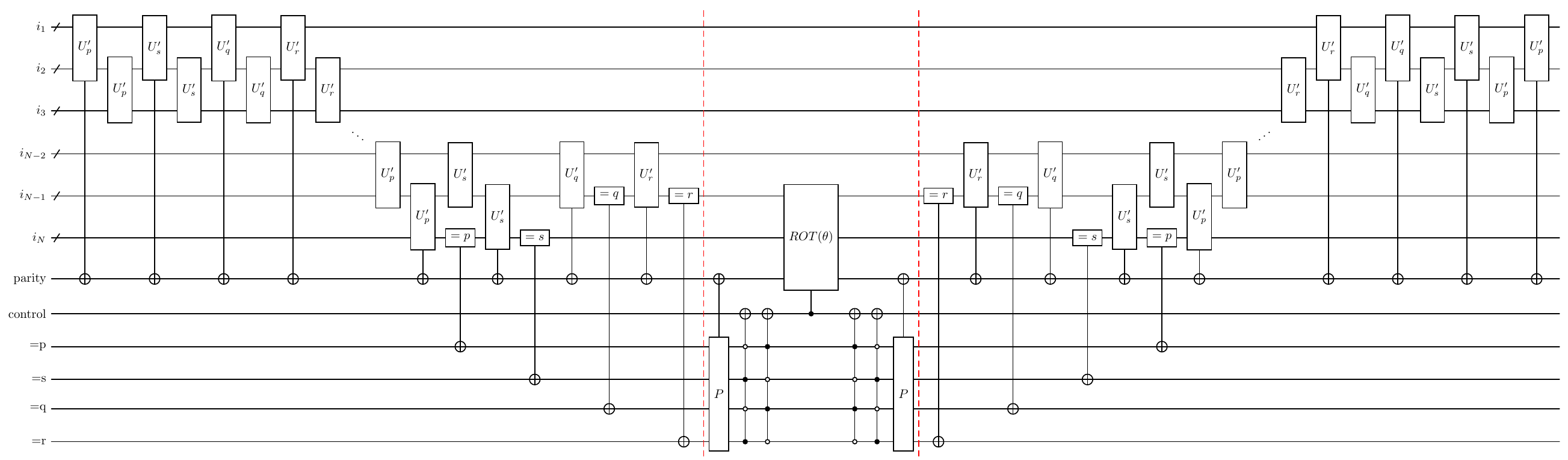}
  \caption{
    The circuit implementing $\exp(i\theta (a_p^\dagger a_q^\dagger a_r a_s +a_s^\dagger a_r^\dagger a_q a_p))$ and $\exp(i\theta (ia_p^\dagger a_q^\dagger a_r a_s -ia_s^\dagger a_r^\dagger a_q a_p))$.
    The gate $P$ flips the parity qubit according to \autoref{eqn:two_elec_rot_parity2}. For $\exp(i\theta (a_p^\dagger a_q^\dagger a_r a_s +a_s^\dagger a_r^\dagger a_q a_p))$, $\operatorname{ROT}(\theta)$ is a controlled Pauli evolution gate $\exp(i\theta X^{q\oplus r}_{\psi_{N-1}}X^{p\oplus s}_{\psi_N}\otimes Z_{\text{parity}})$, where $X$ gates are placed in the `1' bits of the binary representation of $q\oplus r$ and $p\oplus s$, and a $Z$ gate is placed on the parity ancilla.
    The $\operatorname{ROT}(\theta)$ for $\exp(i\theta (ia_p^\dagger a_q^\dagger a_r a_s - ia_s^\dagger a_r^\dagger a_q a_p))$ is implemented similarly, with one of the $X$ replaced by a $Y$.
    Both can be implemented efficiently using the techniques in~\cite{yordanovEfficientQuantumCircuits2020}.
    The parity calculation is done by obtaining parity information from the $U'_p$, $U'_q$, $U'_r$, $U'_s$ gates as shown in \autoref{fig:bubble_circuit} such that the parity ancilla contains the expression of \autoref{eqn:two_elec_rot_parity} just after the $p$, $q$, $r$, and $s$ indices are bubbled down.
    The red dashed line separates the circuit into the $\operatorname{BUBBLE}$ operation, the $\operatorname{ROTATE}$ operation, and the $\operatorname{BUBBLE}^\dagger$ operation from left to right.
    In total, this circuit acts on $N\ltw M$ data qubits and requires $N(\ltw M + 1) + 10$ ancillae.
  }
  \label{fig:two_elec_rot}
\end{figure*}

\begin{figure}
  \centering
  \includegraphics[width=\linewidth]{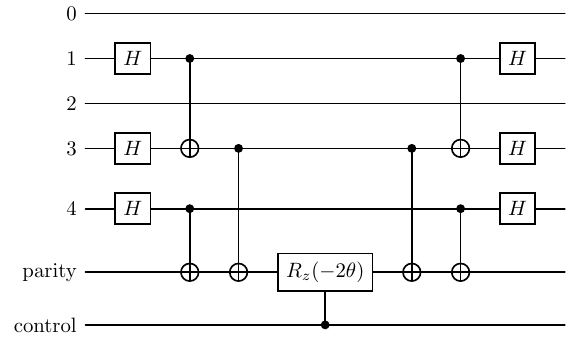}
  \caption{The controlled rotation circuit implementing $\exp(i\theta X_{p\oplus q}\otimes Z_{\text{parity}})$, where $p\oplus q$ has binary representation of $11010$. We used the CNOT cascade techniques discussed in~\cite{yordanovEfficientQuantumCircuits2020}.}
  \label{fig:controlled_rotation}
\end{figure}

The circuit implementing both \autoref{eqn:unitary_one_real} and~\ref{eqn:unitary_one_imag} is shown in \autoref{fig:one_elec_rot}.
Looking at the right-hand side of \autoref{eqn:unitary_one_real} and~\ref{eqn:unitary_one_imag}, the circuit would apply a Pauli evolution if $p$ or $q$ is occupied, but not both. Otherwise, it will perform the identity operation.
This is achieved by first using the $U'_p$ and $U'_q$ gates of \autoref{fig:bubble_circuit} to bubble down both the $p$ and $q$ orbitals to the last register.
Parity calculations are also carried out by tapping the ancilla qubits of some of the $U'_p$ and $U'_q$ gates.
At this point, register $\psi_N$ would contain either $p$ or $q$, if orbital $p$ or $q$ is occupied, respectively.
The $=p$ and $=q$ gates are then used to check whether orbital $p$ or $q$ are occupied.
If only one of orbital $p$ or $q$ is occupied, we would apply an evolution gate swapping between the binary representation of $p$ and $q$ on register $\psi_N$, as shown in \autoref{eqn:rot_eqn_one}.
This controlled evolution gate also includes another ancilla that calculates the parity, which we will elaborate further towards the end of this section.
\autoref{fig:controlled_rotation} shows an example of the evolution gate applied on register $\psi_N$ and the parity ancilla, written as:
\begin{equation}
  \operatorname{ROT}(\theta) = \exp(i\theta (X^{p\oplus q}_{\psi_N}\otimes Z_{parity})).\label{eqn:rot_eqn_one}
\end{equation}
The circuit implementing \autoref{eqn:unitary_one_imag} is done almost identically, with one of the $X$ gates in \autoref{eqn:rot_eqn_one} replaced with the $Y$ gate.

The circuit implementing the two electron unitaries of \autoref{eqn:unitary_two_real} and~\ref{eqn:unitary_two_imag} is shown in \autoref{fig:two_elec_rot}.
This is done similarly with the one-electron case, just that the $p$ and $s$ indices are bubbled down to register $\psi_N$, while the $q$ and $r$ indices are bubbled down to register $\psi_{N-1}$.
Parity calculations are again carried out by tapping the ancilla qubits of some of the bubble gates.
The final adjustments to parity are made using the $P$ gate which is a CNOT gate controlled by one of the $=p$, $=q$, $=r$, or $=s$ ancillae, depending on the exact value of the indices.
Finally, we have the following controlled rotation gate applied on the second last register, the last register, and the parity ancilla, written as:
\begin{equation}
  \operatorname{ROT}(\theta) = \exp(i\theta (X^{q\oplus r}_{\psi_{N-1}}\otimes X^{p\oplus s}_{\psi_N}\otimes Z_{parity})).\label{eqn:rot_eqn_two}
\end{equation}
This Pauli rotation gate is applied when both $p$ and $q$ are occupied and both $r$ and $s$ are unoccupied or when both $r$ and $s$ are occupied and both $p$ and $q$ are unoccupied as shown with the Toffoli gates targeting the ancilla qubit labeled ``control''.

\begin{figure}
  \centering
  \includegraphics[width=0.6\linewidth]{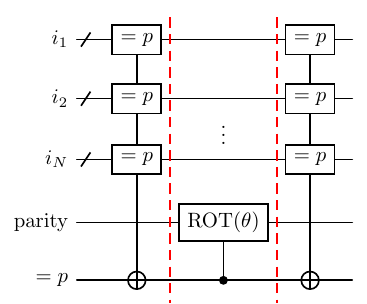}
  \caption{This circuit implements $\exp(i\theta a_p^\dagger a_p)$.
    The red dashed line separates the circuit into the $\operatorname{BUBBLE}$ operation, the $\operatorname{ROTATE}$ operation, and the $\operatorname{BUBBLE}^\dagger$ operation from left to right.
    In total, this circuit acts on $N\ltw M$ qubits and requires $N(\ltw M-1) + 2$ ancillae.
  }
  \label{fig:one_elec_rot_same}
\end{figure}

\begin{figure}
  \centering
  \includegraphics[width=0.8\linewidth]{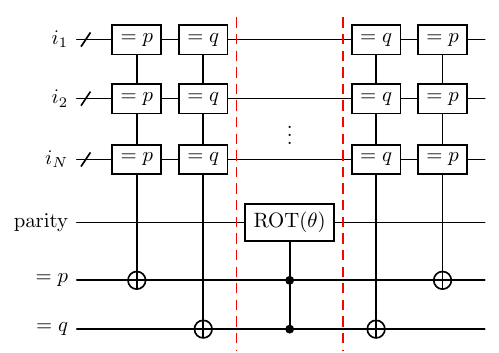}
  \caption{This circuit implements $\exp(i\theta a_p^\dagger a_q^\dagger a_q a_p)$.
    The red dashed line separates the circuit into the $\operatorname{BUBBLE}$ operation, the $\operatorname{ROTATE}$ operation, and the $\operatorname{BUBBLE}^\dagger$ operation from left to right.
    In total, this circuit acts on $N\ltw M$ qubits and requires $N(\ltw M-1) + 3$ ancillae.
  }
  \label{fig:two_elec_rot_same_2}
\end{figure}

\begin{figure*}
  \centering
  \includegraphics[width=\linewidth]{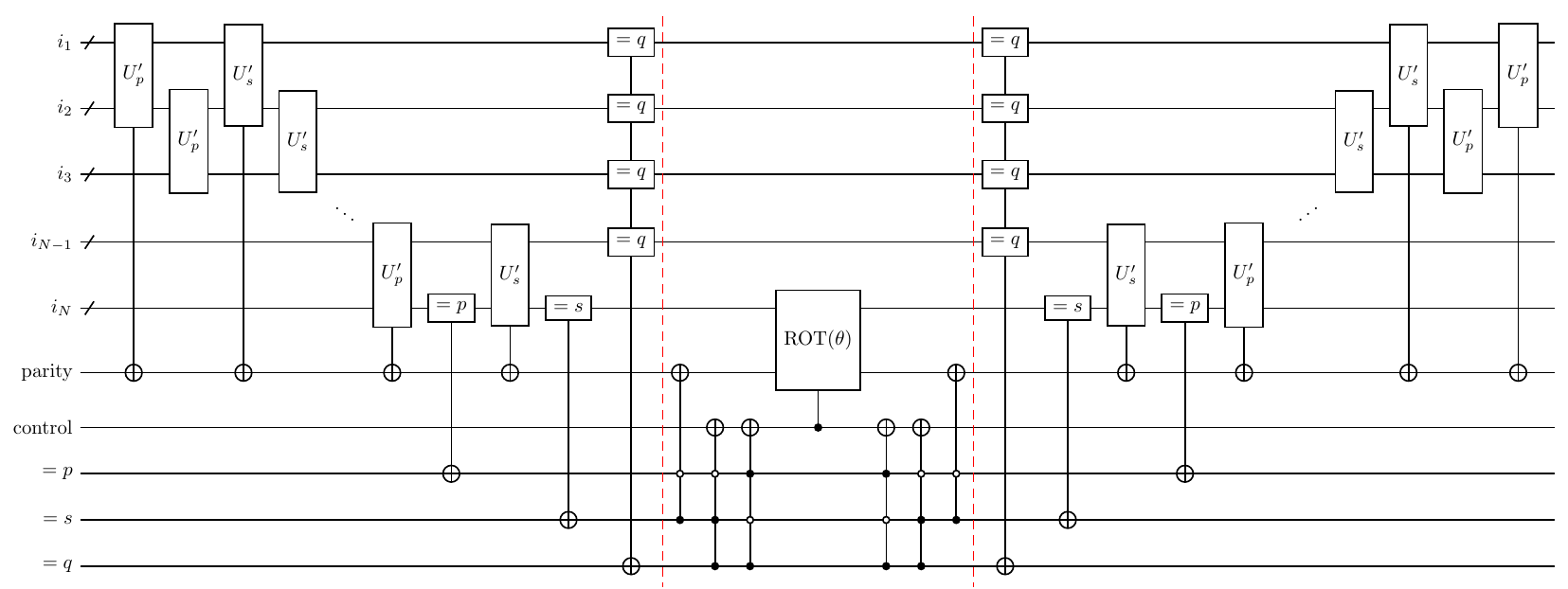}
  \caption{This circuit implements $\exp(i\theta (a_p^\dagger a_q^\dagger a_q a_s + a_s^\dagger a_q^\dagger a_q a_p))$ and $\exp(i\theta (ia_p^\dagger a_q^\dagger a_q a_s - ia_s^\dagger a_q^\dagger a_q a_p))$.
    The red dashed line separates the circuit into the $\operatorname{BUBBLE}$ operation, the $\operatorname{ROTATE}$ operation, and the $\operatorname{BUBBLE}^\dagger$ operation from left to right.
    In total, this circuit acts on $N\ltw M$ qubits and requires $N(\ltw M+1) + 7$ ancillae.
  }
  \label{fig:two_elec_rot_same_1}
\end{figure*}

The circuits of \autoref{fig:one_elec_rot} and~\ref{fig:two_elec_rot} can be slightly simplified when there are common indices, mainly:
\begin{multline*}
  \exp\left(i\theta a_p^\dagger a_p\right),\ \exp\left(i\theta a_p^\dagger a_q^\dagger a_q a_p\right),\ \text{and}\\
  \exp\left(i\theta(ia_p^\dagger a_q^\dagger a_q a_s - ia_s^\dagger a_q^\dagger a_q a_p)\right).
\end{multline*}
Hence, we show specialized circuits for the following in \autoref{fig:one_elec_rot_same}-\ref{fig:two_elec_rot_same_1}.
The structure of the circuits of \autoref{fig:one_elec_rot} and~\ref{fig:two_elec_rot} results in one more opportunity for optimization.
Suppose that we have two successive unitary operations with common indices as follows:
\begin{equation*}
  e^{i\theta_2(a_p^\dagger a_r + a_r^\dagger a_p)} e^{i\theta_1(a_p^\dagger a_q + a_q^\dagger a_p)}.
\end{equation*}
Due to the common orbital index $p$, the $U'_p$ gate on the right side of the $e^{i\theta_1(a_p^\dagger a_q + a_q^\dagger a_p)}$ circuit cancels out with the $U'_p$ gate on the left side of the $e^{i\theta_2(a_p^\dagger a_q + a_q^\dagger a_p)}$. As a result, the two $U'_p$ gates can be omitted from the circuit.

One last thing to consider is the parity calculation. The parity calculation can be explained by first looking into the Jordan-Wigner encoding of the fermionic operators~\cite{steudtnerFermiontoqubitMappingsVarying2018}
{
  \allowdisplaybreaks[4]
  \begin{align}
    a_p^\dagger a_q &= X_pX_q\left(\frac{I+Z_p}{2}\right)\left(\frac{I-Z_q}{2}\right)\Theta_{pq}\notag\\
                    &\phantom{=}\times\left( \prod_{i=0}^{p-1} Z_i \right)\left( \prod_{i=0}^{q-1} Z_i \right),\label{eqn:one_elec_parity}\\
    a_p^\dagger a_q^\dagger &a_r a_s\notag\\
                            &= X_pX_qX_rX_s\left( \frac{I+Z_p}{2} \right)\left( \frac{I+Z_q}{2} \right)\notag\\
                                    &\phantom{=}\times\left( \frac{I-Z_r}{2} \right)\left( \frac{I-Z_s}{2} \right)\Theta_{pq}\Theta_{pr}\Theta_{ps}\Theta_{qr}\Theta_{qs}\Theta_{rs}\notag\\
                                    &\phantom{=}\times\left( \prod_{i=0}^{p-1} Z_i \right)\left( \prod_{i=0}^{q-1} Z_i \right)\left( \prod_{i=0}^{r-1} Z_i \right)\left( \prod_{i=0}^{s-1} Z_i \right),\label{eqn:two_elec_parity}
  \end{align}
}
where we have $\Theta_{ij}=+1$ if $i \le j$ and $\Theta_{ij}=-1$ if $i > j$.
The equation above can be divided into the parity, update, and projection terms~\cite{steudtnerFermiontoqubitMappingsVarying2018}.
The $X$ gates are the update terms, the $(I\pm Z)/2$ gates are the projection terms, and finally the $\Theta_{ij}$ and the $\prod^{p}_{i=0}Z_i$ are the parity terms. Since the $\prod^{p}_{i=0}Z_i$ terms are the same between $a_p^\dagger a_q$ and $a_q^\dagger a_p$ and between $a_p^\dagger a_q^\dagger a_r a_s$ and $a_s^\dagger a_r^\dagger a_q a_p$, those terms can be implemented as follows:
{
  \begin{align}
    &\left( \prod_{i=0}^{p-1} Z_i \right)\left( \prod_{i=0}^{q-1} Z_i \right)\notag\\
    &\Rightarrow (\psi_1 < p)\oplus\cdots\oplus(\psi_N < p)\oplus(\psi_1 < q)\oplus\cdots\oplus(\psi_N < q),\label{eqn:one_elec_rot_parity}\\
    &\left( \prod_{i=0}^{p-1} Z_i \right)\left( \prod_{i=0}^{q-1} Z_i \right)\left( \prod_{i=0}^{r-1} Z_i \right)\left( \prod_{i=0}^{s-1} Z_i \right)\notag\\
    &\Rightarrow (\psi_1 < p)\oplus\cdots\oplus(\psi_N < p)\oplus(\psi_1 < q)\oplus\cdots\oplus(\psi_N < q)\nonumber\\
    &\ \oplus(\psi_1 < r)\oplus\cdots\oplus(\psi_N < r)\oplus(\psi_1 < s)\oplus\cdots\oplus(\psi_N < s).\label{eqn:two_elec_rot_parity}
  \end{align}
}
Such information can be tapped from the ancilla of the bubble gate $U'$ as shown in \autoref{fig:bubble_circuit}.
On the other hand, we also have the $\Theta$ terms that differ between $a_p^\dagger a_q$ and $a_q^\dagger a_p$ and between $a_p^\dagger a_q^\dagger a_r a_s$ and $a_s^\dagger a_r^\dagger a_q a_p$, respectively. Those are
\begin{equation}
  \Theta_{pq} \quad \text{and}\quad\Theta_{pq}\Theta_{pr}\Theta_{ps}\Theta_{qr}\Theta_{qs}\Theta_{rs}\label{eqn:two_elec_rot_parity2},
\end{equation}
respectively. As a result, those terms are implemented as a CNOT gate with one of the $=p,=q,=r$, or $=s$ ancillae as the control and the parity ancilla as the target as shown in \autoref{fig:one_elec_rot} and the $P$ gate in \autoref{fig:two_elec_rot}.

\begin{table}
  \centering
  \caption{
    Costings of the building blocks of the Trotterization circuits.
    $n_0(p)$ and $n_1(p)$ represents the number of `0' and `1' bits, respectively, in the binary representation of $p$.
    The circuit $\operatorname{Bubble}_p^N$ shown in \autoref{fig:trotter_bubble} is composed of $(2N-1)$ $[=p]$ gates, $2N$ $[<p]$ gates, $(2N-2)$ $X$ gates, $(N-1)\ltw M$ CSWAP gates, and $\frac{1}{2}N+\frac{1}{2}(N\mod 2)$ $CX$ gates.
  }
  \begin{ruledtabular}
    \begin{tabular}{lrP{5.2cm}}
      Circuit&\multicolumn{2}{l}{Costings}\\
      \midrule
      $\operatorname{Rot}^1_{p}$      & $CRZ$         & $1$                                  \\
      \autoref{fig:trotter_rot1_p}    &               &                                      \\ \midrule
      $\operatorname{Rot}^2_{pq}$     & $C^2X$        & $1$                                  \\
      \autoref{fig:trotter_rot2_pq}   & $CRZ$         & $1$                                  \\ \midrule
      $\operatorname{Rot}^1_{pq}$     & $C^2X$        & $4$                                  \\
      \autoref{fig:trotter_rot1_pq}   & $CX$          & $2n_1(p\oplus q)+2$                  \\
                                      & $X$           & $8$                                  \\
                                      & $H$           & $2n_1(p\oplus q)$                    \\
                                      & $CRZ$         & $1$                                  \\ \midrule
      $\operatorname{Rot}^2_{pqs}$    & $C^3X$        & $4$                                  \\
      \autoref{fig:trotter_rot2_pqs}  & $CX$          & $2n_1(p\oplus s)+2$                  \\
                                      & $X$           & $8$                                  \\
                                      & $H$           & $2n_1(p\oplus s)$                    \\
                                      & $CRZ$         & $1$                                  \\ \midrule
      $\operatorname{Rot}^2_{pqrs}$   & $C^4X$        & $4$                                  \\
      \autoref{fig:trotter_rot2_pqrs} & $CX$          & $2[n_1(p\oplus s)+n_1(q\oplus r)]+4$ \\
                                      & $X$           & $16$                                 \\
                                      & $H$           & $2[n_1(p\oplus s)+n_1(q\oplus r)]$   \\
                                      & $CRZ$         & $1$                                  \\ \midrule
      $\operatorname{Bubble}_p^N$     & $C^{\log M}X$ & $2N-1$                               \\
      \autoref{fig:trotter_bubble}    & CSWAP         & $N\ltw M-N$                          \\
                                      & $C^2X$        & $2N\ltw M-2N$                        \\
                                      & $CX$          & $12N\ltw M-7.5N+0.5(N\mod 2)$        \\
                                      & $X$           & $(4N-2)n_0(p) + 4Nn_1(p) + 2N - 2$   \\
    \end{tabular}
  \end{ruledtabular}
  \label{tab:trotter_building_blocks}
\end{table}

\subsection{Cost Analysis}
\label{sec:trotter_circuit_exact}
\begin{figure*}
  \centering

  \begin{minipage}[b]{0.48\textwidth}
    \centering
    \subfloat[\label{fig:trotter_rot1_p}]{\includegraphics{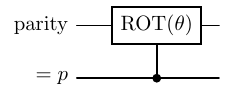}}
  \end{minipage}
  \hfill
  \begin{minipage}[b]{0.48\textwidth}
    \centering
    \subfloat[\label{fig:trotter_rot2_pq}]{\includegraphics{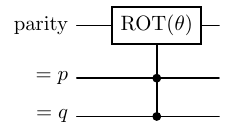}}
  \end{minipage}

  \vspace{5mm}

  \begin{minipage}[b]{0.48\textwidth}
    \centering
    \subfloat[\label{fig:trotter_rot1_pq}]{\includegraphics{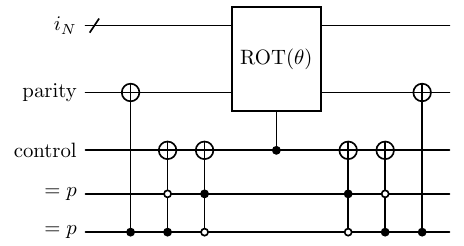}}
  \end{minipage}
  \hfill
  \begin{minipage}[b]{0.48\textwidth}
    \centering
    \subfloat[\label{fig:trotter_rot2_pqs}]{\includegraphics{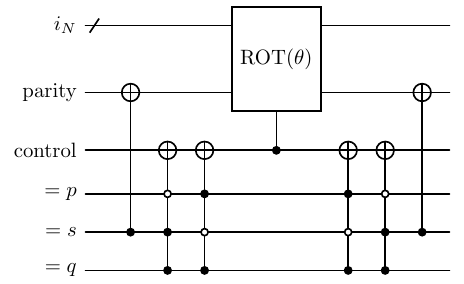}}
  \end{minipage}

  \vspace{5mm}

  \begin{minipage}[b]{0.48\textwidth}
    \centering
    \subfloat[\label{fig:trotter_rot2_pqrs}]{\includegraphics{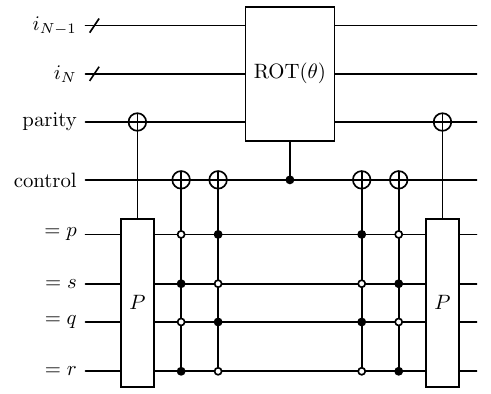}}
  \end{minipage}
  \hfill
  \begin{minipage}[b]{0.48\textwidth}
    \centering
    \subfloat[\label{fig:trotter_bubble}]{\includegraphics{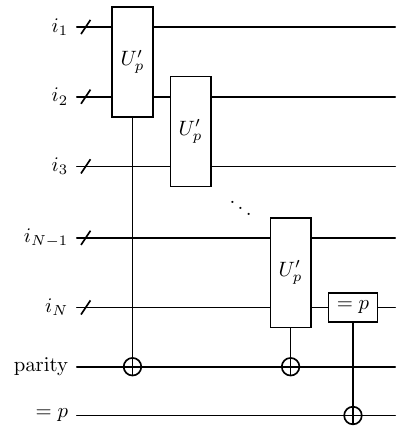}}
  \end{minipage}
  \caption{Building blocks of the Trotterization circuits with costings shown in \autoref{tab:trotter_building_blocks}}
  \label{fig:trotter_building_blocks}
\end{figure*}

\autoref{tab:trotter_building_blocks} shows the costings of the building blocks of the Trotterization circuits, with those building blocks displayed in \autoref{fig:trotter_building_blocks}, while the costings of the Trotterization circuits are shown in \autoref{tab:building_blocks_trotter_costings}.

\begin{table*}
  \centering
  \caption{
    Costings of the Trotterization circuits.
    $n_0(p)$ and $n_1(p)$ represents the number of `0' and `1' bits, respectively, in the binary representation of $p$.
  }
  \begin{ruledtabular}
    \begin{tabular}{l>{\raggedleft}p{1.5cm}P{12cm}}
      Circuit&\multicolumn{2}{l}{Costings}\\
      \midrule
      \multirow[t]{8}{3.8cm}{$e^{i\theta(a_p^\dagger a_q + a_q^\dagger a_p)}$ and $e^{i\theta(ia_p^\dagger a_q - ia_q^\dagger a_p)}$ (\autoref{fig:one_elec_rot})}
      & Blocks & $2[\operatorname{Bubble}^N_p] + 2[\operatorname{Bubble}^N_q] + [\operatorname{Rot}^1_{pq}]$ \\
      \cmidrule{2-3}
      & $C^{\log M}X$ & $8N-4$                                               \\
      & CSWAP         & $4N\ltw M-4N$                                        \\
      & $C^2X$        & $8N\ltw M - 8N + 4$                                  \\
      & $CX$          & $48N\ltw M - 30N + 2(N\mod 2) + 2 + 2n_1(p\oplus q)$ \\
      & $X$           & $(8N-4)[n_0(p)+n_0(q)]+8N[n_1(p)+n_1(q)]+8N$         \\
      & $H$           & $2n_1(p\oplus q)$                                    \\
      & $CRZ$         & $1$                                                  \\
      \midrule
      \multirow[t]{10}{3.8cm}{$e^{i\theta (a_p^\dagger a_q^\dagger a_r a_s + a_s^\dagger a_r^\dagger a_q a_p)}$ and $e^{i\theta (ia_p^\dagger a_q^\dagger a_r a_s - ia_s^\dagger a_r^\dagger a_q a_p)}$ (\autoref{fig:two_elec_rot})}
      & Blocks & $2[\operatorname{Bubble}^N_p] + 2[\operatorname{Bubble}^N_s] + 2[\operatorname{Bubble}^{N-1}_q] + 2[\operatorname{Bubble}^{N-1}_r] + [\operatorname{Rot}^2_{pqrs}]$ \\
      \cmidrule{2-3}
      & $C^{\log M}X$ & $16N-6$                                                      \\
      & $C^4X$        & $4$                                                          \\
      & CSWAP         & $8N\ltw M - 8N - 4\ltw M + 4$                                \\
      & $C^2X$        & $16N\ltw M - 16N - 8\ltw M +8$                               \\
      & $CX$          & $96N\ltw M-60N-48\ltw M+36+2[n_1(p\oplus s)+n_1(q\oplus r)]$ \\
      & $X$           & $16N-8+(8N-4)[n_0(p)+n_0(s)]+8N[n_1(p)+n_1(s)]+(8N-12)[n_0(q)+n_0(r)]+(8N-8)[n_1(q)+n_1(r)]$\\
      & $H$           & $2[n_1(p\oplus s)+n_1(q\oplus r)]$                           \\
      & $CRZ$         & $1$                                                          \\
      \midrule
      \multirow[t]{4}{3.8cm}{$e^{i\theta a_p^\dagger a_p}$ (\autoref{fig:one_elec_rot_same})}
      & Blocks & $2N[=p] + [\operatorname{Rot}^1_{p}]$ \\
      \cmidrule{2-3}
      & $C^{\log M}X$   & $2N$                                  \\
      & $CRZ$           & $1$                                   \\
      & $X$             & $2Nn_0(p)$                            \\
      \midrule
      \multirow[t]{5}{3.8cm}{$e^{i\theta a_p^\dagger a_q^\dagger a_q a_p}$ (\autoref{fig:two_elec_rot_same_2})}
      & Blocks & $2N[=p] + 2N[=q] + [\operatorname{Rot}^2_{pq}]$ \\
      \cmidrule{2-3}
      & $C^{\log M}X$ & $4N$                \\
      & $C^2X$        & $2$                 \\
      & $X$           & $4N[n_0(p)+n_0(q)]$ \\
      & $CRZ$         & $1$                 \\
      \midrule
      \multirow[t]{9}{3.8cm}{$e^{i\theta (a_p^\dagger a_q^\dagger a_q a_s + a_s^\dagger a_q^\dagger a_q a_p)}$ and $e^{i\theta (ia_p^\dagger a_q^\dagger a_q a_s - ia_s^\dagger a_q^\dagger a_q a_p)}$ (\autoref{fig:two_elec_rot_same_1})}
      & Blocks & $2[\operatorname{Bubble}^N_p] + 2[\operatorname{Bubble}^N_s] + 2(N-1)[=q] + [\operatorname{Rot}^2_{pqs}]$ \\
      \cmidrule{2-3}
      & $C^{\log M}X$ & $10N-6$                                      \\
      & $C^3X$        & $4$                                          \\
      & CSWAP         & $4N\ltw M - 4N$                              \\
      & $C^2X$        & $8N\ltw M - 8N$                              \\
      & $CX$          & $48N\ltw M-30N-2(N\mod 2)+2+2n_1(p\oplus s)$ \\
      & $X$           & $(8N-4)[n_0(p)+n_0(s)]+8N[n_1(p)+n_1(s)]+8N+(4N-4)n_0(q)$\\
      & $H$           & $2n_1(p\oplus s)$                            \\
      & $CRZ$         & $1$                                          \\
    \end{tabular}
  \end{ruledtabular}
  \label{tab:building_blocks_trotter_costings}
\end{table*}

\subsection{Simulation Methods for Trotterization cost}
\label{sec:simulation_details}
As mentioned in the main manuscript, tight theoretical error bounds are challenging to obtain.
As a result, we used experimental error bounds obtained from a previous work~\cite{guntherPhaseEstimationPartially2026} that uses the RPE algorithm.
We used Equation~E4 in the appendix of that work~\cite{guntherPhaseEstimationPartially2026} for the deterministic Trotter formulas, reproduced here below with the variables explained in \autoref{tab:trotter_legend}:
\begin{equation}
  G = 5\pi N_{\text{stage}}L_{\text{det}}G_{\text{avg}}\sqrt{\frac{(p+1)^{1+1/p}}{p}}\frac{C_{\text{gs}}^{1/p}}{\varepsilon^{1+1/p}}.\label{eqn:trotter_deter}
\end{equation}
Of particular note is the constant $C_{\text{gs}}$ which is obtained from empirical calculations.
For our calculations, we used Figure 8 of~\cite{guntherPhaseEstimationPartially2026} to obtain the following formula with respect to the 1-norm of the Hamiltonian $\lambda$:
\begin{equation*}
  C_{\text{gs}} = \num{3.470E-05}\lambda^{2.081}.
\end{equation*}

On the other hand, the cost for the randomized Trotter formula is as follows:
\begin{equation}
  G = G_{\text{avg}} \frac{16.3\lambda^2}{\varepsilon^2},\label{eqn:trotter_rand}
\end{equation}
with the variables explained in \autoref{tab:trotter_legend}.
Finally, the partially randomized formula has the following costs:
\begin{multline}
  G = 30G_{\text{avg}}N_{\text{stage}} L_{\text{det}} e^{2/\kappa}\frac{0.1\pi}{\delta\sqrt{\varepsilon^2-C_{\text{gs}}^2\delta^4}}\\
  + \frac{280}{9}G_{\text{avg}}\kappa e^{2/\kappa}\frac{(0.1\pi\lambda_{\text{rand}})^2}{\varepsilon^2 - C_{\text{gs}}^2\delta^4},\label{eqn:trotter_partial}
\end{multline}
where the variable $\kappa$ is minimized analytically using a quadratic formula, and the variable $\delta$ is minimized numerically.

\begin{table*}
  \centering
  \caption{Legend for the parameters of \autoref{eqn:trotter_deter}-\ref{eqn:trotter_partial}.}
  \begin{ruledtabular}
    \begin{tabular}{lP{16cm}}
      Parameters&Description\\
      \midrule
      $\varepsilon$&Target precision for the energy, set to chemical accuracy (\num{1.6e-3} Ha)\\
      $G$&Total gate costs\\
      \midrule
      $L_{\text{det}}$&Number of deterministic Trotter terms\\
      $\delta$&Trotter step size\\
      $p$& Order of the Trotter formula, set to $2$ (Suzuki-Trotter) for our calculations\\
      $N_{\text{stage}}$&$2\times 5^{p/2-1}$, set to $2$ (Suzuki-Trotter) for our calculations\\
      $C_{\text{gs}}$&Trotter error constant with respect to the ground state energy\\
      $G_{\text{avg}}$&Average gate costs to implement one unitary\\
      \midrule
      $\kappa$&Scaling factor for the number of evolution terms for the randomized Trotter formula\\
      $\lambda_{\text{rand}}$&$1$-norm of terms treated randomly for the partially random Trotter formula\\
    \end{tabular}
  \end{ruledtabular}
  \label{tab:trotter_legend}
\end{table*}

\section{Qubitization Implementation for the Sorted-List Encoding}
\label{sec:qubitization_circuit}
\subsection{Circuit Implementations}
\label{sec:qubitization_circuit_impl}
Here, we show the circuit implementation and cost analysis of the $\operatorname{SELECT}$ circuit.
While a Majorana-free SELECT circuit can be constructed with similar gate cost scaling, we instead opt to implement Majorana operators for our qubitization circuit due to its simpler construction and lower T gate cost.
The Majorana-free version of this SELECT circuit can be found in Appendix~\ref{sec:qubitization_circuit_majorana_free}.
As we are applying Majorana operators for our circuit, the encoding needs to have at least $N+4$ registers to accommodate the two-electron terms.
Additionally, we make use of the $U''_p$ optimizations of the bubble gate detailed in Appendix~\ref{sec:prev_encoding_sl_opt}.
This requires 2 additional ancilla qubits for every register to hold temporary values during each Majorana operations.

\begin{figure}
  \centering
  \includegraphics{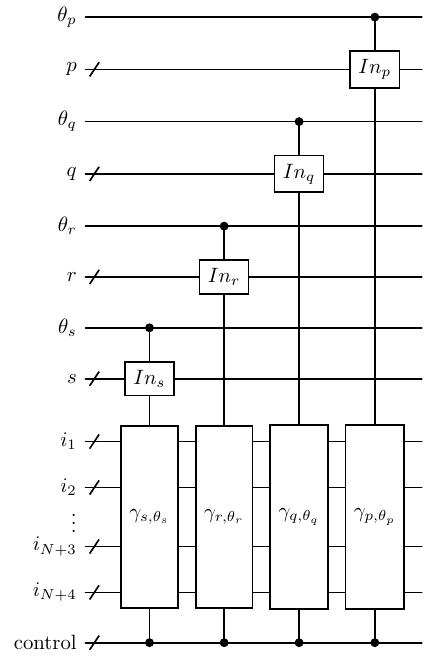}
  \caption{Circuit implementing $\operatorname{SELECT}$, made up of four blocks of the $\gamma_{p,\theta_p}$ circuits, with the individual blocks defined in \autoref{fig:select_majorana}.}
  \label{fig:select_a}
\end{figure}

For reference, we use the following form of the Majorana operators:
\begin{align*}
  \gamma_{p,0} &= a_p^{\dagger} + a_p, & a_p^\dagger &= \frac{\gamma_{p,0}-i\gamma_{p,1}}{2},\\
  \gamma_{p,1} &= i(a_p^\dagger - a_p), & a_p &= \frac{\gamma_{p,0}+i\gamma_{p,1}}{2}.
\end{align*}
The SELECT circuit then applies four Majorana operators based on the indices of the ancilla qubits:
\begin{align*}
  \operatorname{SELECT}&\ket{\theta_p;p;\theta_q;q;0;0;0;0}\ket{\psi}\nonumber\\
                         &= \ket{\theta_p;p;\theta_q;q;0;0;0;0}\gamma_{p,\theta_p}\gamma_{q,\theta_q}\ket{\psi},\nonumber\\
  \operatorname{SELECT}&\ket{\theta_p;p;\theta_q;q;\theta_r;r;\theta_s;s}\ket{\psi}\nonumber\\
                       &= \ket{\theta_p;p;\theta_q;q;\theta_r;r;\theta_s;s} \gamma_{p,\theta_p}\gamma_{q,\theta_q}\gamma_{r,\theta_r}\gamma_{s,\theta_s}\ket{\psi}.
\end{align*}
To distinguish between the one- and two-electron terms, we can either use controlled variants of the $\gamma_{r,\theta_r}\gamma_{s,\theta_s}$ terms or by setting $\theta_r = \theta_s = 0$ and $r = s$ when applying the one-electron terms.
\autoref{fig:select_a} shows the circuit that implements $\operatorname{SELECT}$, with the circuit of the individual Majorana operator $\gamma_{p,\theta_p}$ shown in \autoref{fig:select_majorana}.
Like in the Trotterization case, the $\mathcal{O}(N\log M)$ scaling is maintained with respect to the qubit count (including the ancillae) and gate count (including non-Toffoli gates).
Furthermore, the conversion from fermionic to Majorana operators can be trivially implemented using Hadamard and S gates. This is illustrated in \autoref{fig:fermion_to_majorana_both} and the following equation:
\begin{align}
  \sum_p \dyad{p}\otimes a_p^{\dagger} &\Rightarrow \sum_{\theta\in\{0,1\}}\sum_p\frac{(-i)^\theta}{2}\dyad{\theta}\otimes\dyad{p}\otimes \gamma_{p,\theta}\label{eqn:fermion_to_majorana_dagger}\\
  \sum_p \dyad{p}\otimes a_p &\Rightarrow \sum_{\theta\in\{0,1\}}\sum_p\frac{i^\theta}{2}\dyad{\theta}\otimes\dyad{p}\otimes \gamma_{p,\theta}.\label{eqn:fermion_to_majorana}
\end{align}
Note that the $\theta$ above needs to be reflected for the qubitization circuits in addition to the orbital index $p$.

\subsection{Cost Analysis}
\label{sec:qubitization_circuit_cost}
\begin{figure*}
  \centering
  \includegraphics[width=\textwidth]{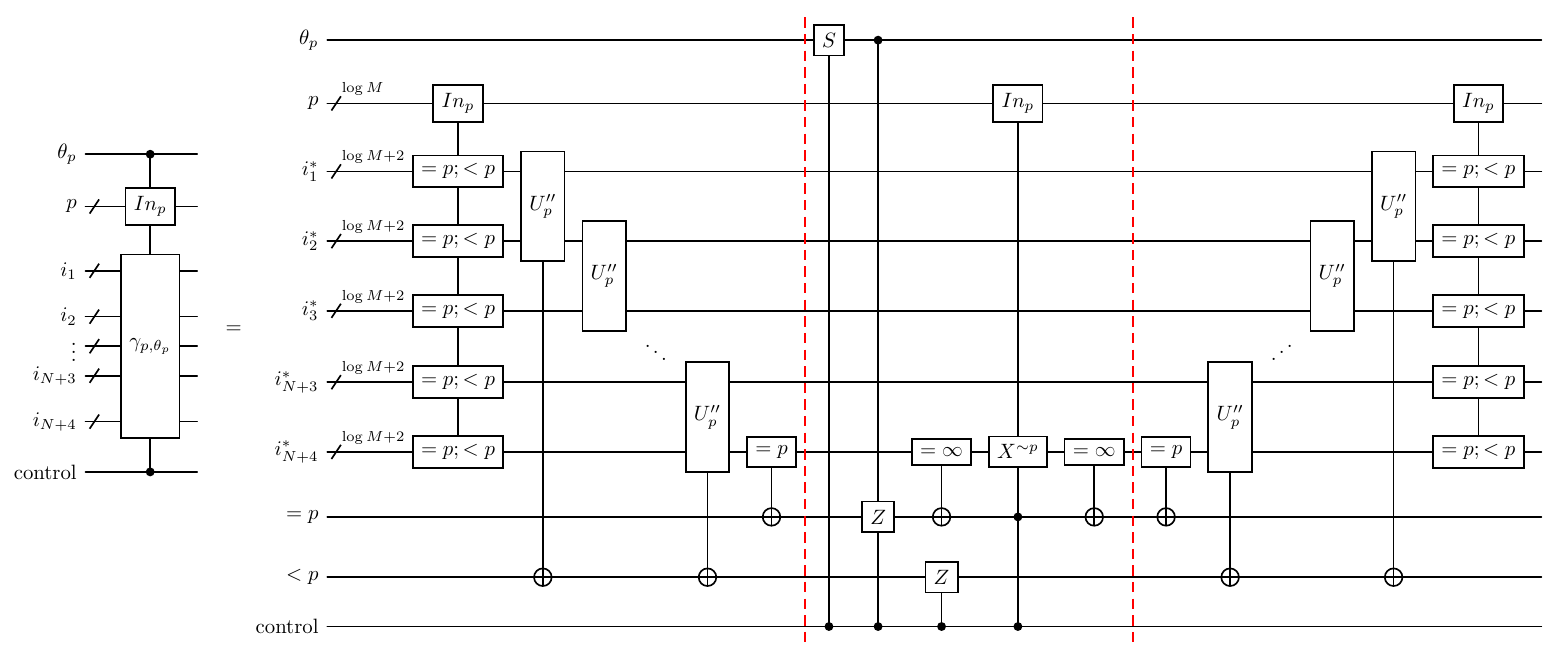}
  \caption{
    The circuit implementing $\gamma_{p,\theta_p}$.
    The bubble circuits $U''_p$ uses precalculated $=p$ and $<p$ information to reduce gate count and circuit depth.
    As a result, the registers $\psi_n^*$ contain $\ltw M$ qubits storing the binary representation of the orbital indices, along with 2 additional ancilla qubits to store $=p$ and $<q$ data generated by the $=p;<p$ gates, totaling $\ltw M + 2$ qubits.
    The implementation of $U_p''$ is shown in \autoref{fig:bubble_circuit2}.
    The dotted red line divides the circuit into three parts: $\operatorname{BUBBLE}$, $\operatorname{SWAP}$ and $\operatorname{BUBBLE}^\dagger$.
  }
  \label{fig:select_majorana}
\end{figure*}

\begin{figure*}
  \centering
  \hspace*{\fill}
  \subfloat[\label{fig:fermion_to_majorana_dagger}]{\includegraphics{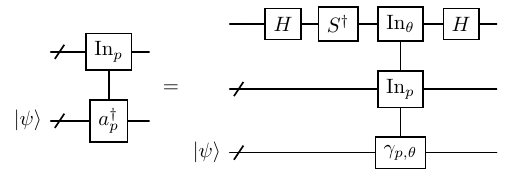}}
  \hspace*{\fill}
  \subfloat[\label{fig:fermion_to_majorana}]{\includegraphics{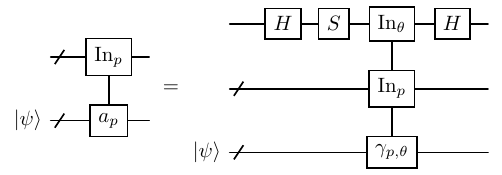}}
  \hspace*{\fill}

  \caption{
      The circuit to convert fermionic operators to Majorana operators. In this case, the $\theta$ for $\gamma_{p,\theta}$ is generated using Hadamard and S gates.
  }
  \label{fig:fermion_to_majorana_both}
\end{figure*}

Similar to the Trotterization circuits, we divide the $\gamma_{p,\theta_p}$ circuit into three parts as shown with the dotted red line of \autoref{fig:select_majorana}. The exact costings of such terms are shown in \autoref{tab:select_costings} along with the costings of the $\gamma_{p,\theta_p}$.

\begin{table}
  \centering
  \caption{Costings of the gates of the $\gamma_{p,\theta_p}$ circuit of \autoref{fig:select_majorana}. Here, $M$ is the number of orbitals and $N$ is the total number of electrons.}
  \begin{ruledtabular}
    \begin{tabular}{p{1.4cm}>{\raggedleft}p{1.5cm}P{5.4cm}}
      Circuit& \multicolumn{2}{l}{Costings}\\
      \midrule
      $\operatorname{BUBBLE}$  & $C^{\log M}X$ & $N+4$                                                       \\
                               & CSWAP         & $(N+4)\ltw M - \ltw M + 2(N+4)-2$                           \\
                               & $C^2X$        & $(N+4)\ltw M + 2(N+4)-2$                                    \\
                               & $CX$          & $8(N+4)\ltw M - \frac{1}{2}(N+4) + \frac{1}{2}(N\mod2) - 5$ \\
                               & $X$           & $\ltw M$                                                    \\
                               \midrule
      $\operatorname{SWAP}$    & $C^{\log M}X$ & $2$                                                         \\
                               & $C^2X$        & $\ltw M+3$                                                  \\
                               & $CX$          & $1$                                                         \\
                               & $H$           & $4$                                                         \\
                               & $S$           & $1$                                                         \\
                               \midrule
      $\gamma_{p,\theta_p}$ & $C^{\log M}X$ & $2(N+4)+2$                                                  \\
                            & CSWAP         & $2(N+4)\ltw M +4(N+4)- 2\ltw M-4$                           \\
                            & $C^2X$        & $2(N+4)\ltw M +4(N+4) +\ltw M - 1$                          \\
                            & $CX$          & $16(N+4)\ltw M - (N+4) + (N\mod2) - 9$                     \\
                            & $X$           & $2\ltw M$                                                   \\
                            & $H$           & $4$                                                         \\
                            & $S$           & $1$                                                         \\
    \end{tabular}
  \end{ruledtabular}
  \label{tab:select_costings}
\end{table}

\subsection{Plane-wave implementation}
\label{sec:qubitization_plane_wave}
\begin{figure}[t]
  \centering
  \includegraphics[width=\linewidth]{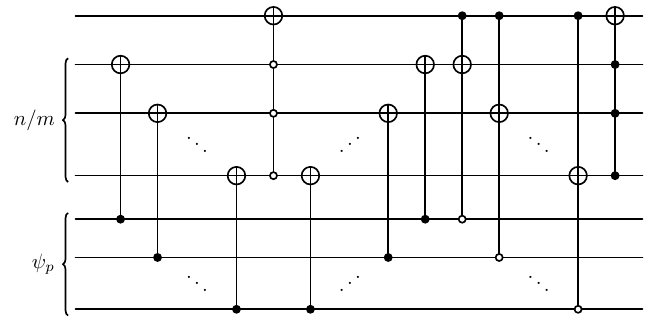}
  \caption{The iterative element implementing \autoref{eqn:zero_element}. Here we iterate $\psi_p$ with $p=1,\dots,N+4$.}
  \label{fig:zero_element}
\end{figure}

\begin{figure*}
  \centering
  \includegraphics{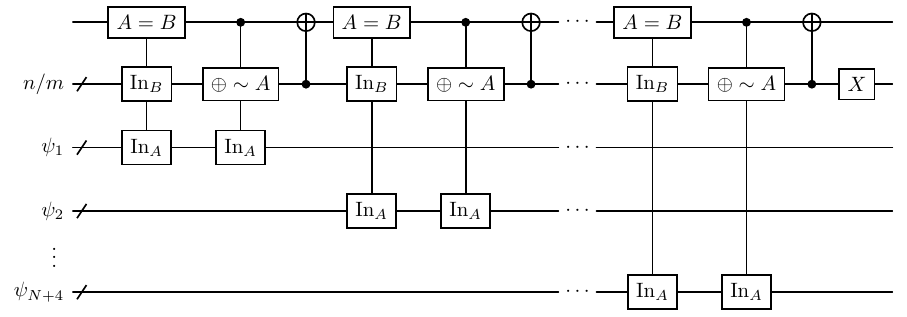}
  \caption{The circuit implementation of the ZERO operation. The $\oplus\sim A$ gate binary adds the complement of the binary representation of $A$, while the CNOT gates represents a multi-controlled Toffoli gate which applies a NOT gate on the ancilla qubit when qubit $n$/$m$ equals $\ket{\infty}$. Finally, the $X$ gate at the end converts the $\ket{\infty}$ state of qubit $n$/$m$ back to $\ket{0}$.}
  \label{fig:zero_operation}
\end{figure*}

  The $\operatorname{PREPARE}$ and $\operatorname{SELECT}$ circuits are adapted from the first-quantized plane-wave qubitization circuit~\cite{suFaultTolerantQuantumSimulations2021} summarized in Appendix~\ref{sec:prev_qubitization_fq}, along with the the state-dependent oracles of \citet{liuBlockEncodingLow2025}.
In the second-quantization, the Hamiltonian can be written as:
{
  \begin{align}
    H^{(2)} &= \hat{T}^{(2)} + \hat{U}^{(2)} + \hat{V}^{(2)},\label{eqn:sq_hamiltonian_pw}\\
    \hat{T}^{(2)} &= \sum_{p\in G}\frac{\|k_p\|^2}{2}a_p^\dagger a_p,\label{eqn:sq_hamiltonian_pw_t}\\
    \hat{U}^{(2)} &= -\frac{4\pi}{\Omega}\sum_{\substack{\nu\in G\\\nu\ne 0}}\sum_{\substack{p\in G\\(p-\nu)\in G}} \left(\sum_{I=1}^{N_{at}}Z_I\frac{-e^{ik_{\nu}\cdot \vb{X}_I}}{\|k_{\nu}\|^2}\right) a_{p-\nu}^{\dagger}a_p,\label{eqn:sq_hamiltonian_pw_u}\\
    \hat{V}^{(2)} &= \frac{2\pi}{\Omega}\sum_{\substack{\nu\in G\\\nu\ne 0}} \sum_{\substack{p,q\in G\\(p+\nu)\in G\\(q-\nu)\in G}} \frac{1}{\|k_{\nu}\|^2} a_{p+\nu}^{\dagger}a_{q-\nu}^{\dagger} a_q a_p.\label{eqn:sq_hamiltonian_pw_v}
  \end{align}
}
Here, the $\operatorname{PREPARE}$ circuit is identical with the first-quantized version. On the other hand, the $\operatorname{SELECT}$ circuit has the following form below:
{
  \allowdisplaybreaks[4]
  \begin{widetext}
    \begin{align*}
      \operatorname{SELECT}_{\hat{T}^{(2)}}&: \ket{b}_b\ket{j}_e\ket{w}_f\ket{r}_g\ket{s}_h\ket{0}_n\ket{\psi}\nonumber\\
                                           &\xRightarrow{\text{COPY}} \ket{b}_b\ket{j}_e\ket{w}_f\ket{r}_g\ket{s}_h\ket{q_j}_n\ket{\psi}\\
                                           &\xRightarrow{\text{APPLY}} \ket{b}_b\ket{j}_e\ket{w}_f\ket{r}_g\ket{s}_h\ket{q_j}_n\ket{a_{q_j}\psi}\\
                                           &\xRightarrow{\text{FQ}} (-1)^{b(q_{w,r}q_{w,s}\oplus 1)}\ket{b}_b\ket{j}_e\ket{w}_f\ket{r}_g\ket{s}_h\ket{q_j}_n\ket{a_{q_j}\psi}\\
                                           &\xRightarrow{\text{APPLY}} (-1)^{b(q_{w,r}q_{w,s}\oplus 1)}\ket{b}_b\ket{j}_e\ket{w}_f\ket{r}_g\ket{s}_h\ket{q_j}_n\ket{a_{q_j}^\dagger a_{q_j}\psi}\\
                                           &\xRightarrow{\text{ZERO}} (-1)^{b(q_{w,r}q_{w,s}\oplus 1)}\ket{b}_b\ket{j}_e\ket{w}_f\ket{r}_g\ket{s}_h\ket{0}_n\ket{a_{q_j}^\dagger a_{q_j}\psi}\\
      \operatorname{SELECT}_{\hat{U}^{(2)}}&: \ket{b}_b\ket{j}_e\ket{0}_i\ket{\nu}_k\ket{I}_l\ket{0}_n\ket{\psi}\nonumber\\
                                           &\xRightarrow{\text{COPY}} \ket{b}_b\ket{j}_e\ket{0}_i\ket{\nu}_k\ket{I}_l\ket{q_j}_n\ket{\psi}\\
                                           &\xRightarrow{\text{APPLY}} \ket{b}_b\ket{j}_e\ket{0}_i\ket{\nu}_k\ket{I}_l\ket{q_j}_n\ket{a_{q_j}\psi}\\
                                           &\xRightarrow{\text{FQ}} -e^{-ik_{\nu}\cdot \vb{X}_I}(-1)^{b[(q-v)\notin G]} \ket{b}_b\ket{j}_e\ket{0}_i\ket{\nu}_k\ket{I}_l\ket{q_j-\nu}_n\ket{a_{q_j}\psi}\\
                                           &\xRightarrow{\text{APPLY}} -e^{-ik_{\nu}\cdot \vb{X}_I}(-1)^{b[(q-v)\notin G]} \ket{b}_b\ket{j}_e\ket{0}_i\ket{\nu}_k\ket{I}_l\ket{q_j-\nu}_n\ket{a_{q_j-\nu}^\dagger a_{q_j}\psi}\\
                                           &\xRightarrow{\text{ZERO}} -e^{-ik_{\nu}\cdot \vb{X}_I}(-1)^{b[(q-v)\notin G]} \ket{b}_b\ket{j}_e\ket{0}_i\ket{\nu}_k\ket{I}_l\ket{0}_n\ket{a_{q_j-\nu}^\dagger a_{q_j}\psi}\\
      \operatorname{SELECT}_{\hat{V}^{(2)}}&: \ket{b}_b\ket{i}_d\ket{j}_e\ket{1}_i\ket{\nu}_k\ket{0}_m\ket{0}_n\ket{\psi}\nonumber\\
                                           &\xRightarrow{\text{COPY}} \ket{b}_b\ket{i}_d\ket{j}_e\ket{1}_i\ket{\nu}_k\ket{p_i}_m\ket{q_j}_n\ket{\psi}\\
                                           &\xRightarrow{\text{APPLY}} \ket{b}_b\ket{i}_d\ket{j}_e\ket{1}_i\ket{\nu}_k\ket{p_i}_m\ket{q_j}_n\ket{a_{q_j}a_{p_i}\psi}\\
                                           &\xRightarrow{\text{FQ}} (-1)^{b([(p+\nu)\notin G]\vee[(q-\nu)\notin G])} \ket{b}_b\ket{i}_d\ket{j}_e\ket{1}_i\ket{\nu}_k\ket{p_i}_m\ket{q_j}_n\ket{a_{q_j}a_{p_i}\psi}\\
                                           &\xRightarrow{\text{APPLY}} (-1)^{b([(p+\nu)\notin G]\vee[(q-\nu)\notin G])} \ket{b}_b\ket{i}_d\ket{j}_e\ket{1}_i\ket{\nu}_k\ket{p_i}_m\ket{q_j}_n\ket{a_{p_i+\nu}^{\dagger}a_{q_j-\nu}^{\dagger}a_{q_j}a_{p_i}\psi}\\
                                           &\xRightarrow{\text{ZERO}} (-1)^{b([(p+\nu)\notin G]\vee[(q-\nu)\notin G])} \ket{b}_b\ket{i}_d\ket{j}_e\ket{1}_i\ket{\nu}_k\ket{0}_m\ket{0}_n\ket{a_{p_i+\nu}^{\dagger}a_{q_j-\nu}^{\dagger}a_{q_j}a_{p_i}\psi}
    \end{align*}
  \end{widetext}
}
First, the COPY operation copies the contents (using controlled CNOT gates) of the $j$\textsuperscript{th} register of $\ket{\psi}$ to ancilla $n$, where the electron index $j$ is stored in ancilla $e$.
For $\operatorname{SELECT}_{\hat{V}^{(2)}}$, we additionally copy the contents of the $i$\textsuperscript{th} register of $\ket{\psi}$  to ancilla $m$, where the electron index $j$ is stored in ancilla $d$.
This is similarly done for the first-quantized qubitization circuits (see Equation~72 of~\cite{suFaultTolerantQuantumSimulations2021}), except that a swap is performed for the first-quantized case.
Next, the APPLY operation performs the fermionic operation, as done in \autoref{eqn:fermion_to_majorana_dagger} and~\ref{eqn:fermion_to_majorana}, using ancilla $m$ and $n$ as indices.
The FQ operation applies the corresponding first-quantized $\operatorname{SELECT}_{\hat{T}^{(1)}}$, $\operatorname{SELECT}_{\hat{U}^{(1)}}$, and $\operatorname{SELECT}_{\hat{V}^{(1)}}$.
Finally, the ZERO operation, shown in \autoref{fig:zero_operation}, resets the value of ancilla $m$ and $q$ back to $\ket{0}$ by iterating through $\ket{\psi}$.
This is possible as $\ket{a_p^\dagger a_q\psi}$ should either have orbital $p$ occupied or $\ket{a_p^\dagger a_q\psi}$ is zero, the same with $\ket{a_p^\dagger a_q^\dagger a_r a_s\psi}$ where we either have both orbitals $p$ and $q$ occupied or $\ket{a_p^\dagger a_q^\dagger a_r a_s\psi}$ is zero.
As a result, we would iterate the following operation for all the registers of $\ket{\psi}$, implemented in \autoref{fig:zero_element}:
\begin{equation}
  \ket{p}\ket{q} \Rightarrow
 \begin{cases}
   \ket{\infty}\ket{q} &\text{ if } p = q\\
   \ket{p}\ket{q} &\text{ otherwise}
 \end{cases}.
 \label{eqn:zero_element}
\end{equation}
In this case, $p$ represents ancilla $m$ and $n$, while $q$ represents the individual registers of $\ket{\psi}$. After iterating through all the registers of $\ket{\psi}$, we apply $X$ gates to convert $\ket{\infty}_n$ back to $\ket{0}_n$ (along with ancilla $m$).

\subsection{Cost Estimates on Model Systems}
\label{sec:qubitization_circuit_estimate}
\autoref{tab:qubitization_mol_mo_sf_1} and \ref{tab:qubitization_mol_mo_sf_2} shows the costings of the qubitization circuits for molecular systems with the MO basis-set.

\subsection{Majorana-free Implementations}
\label{sec:qubitization_circuit_majorana_free}

\begin{figure*}
  \centering
  \includegraphics[width=0.9\linewidth]{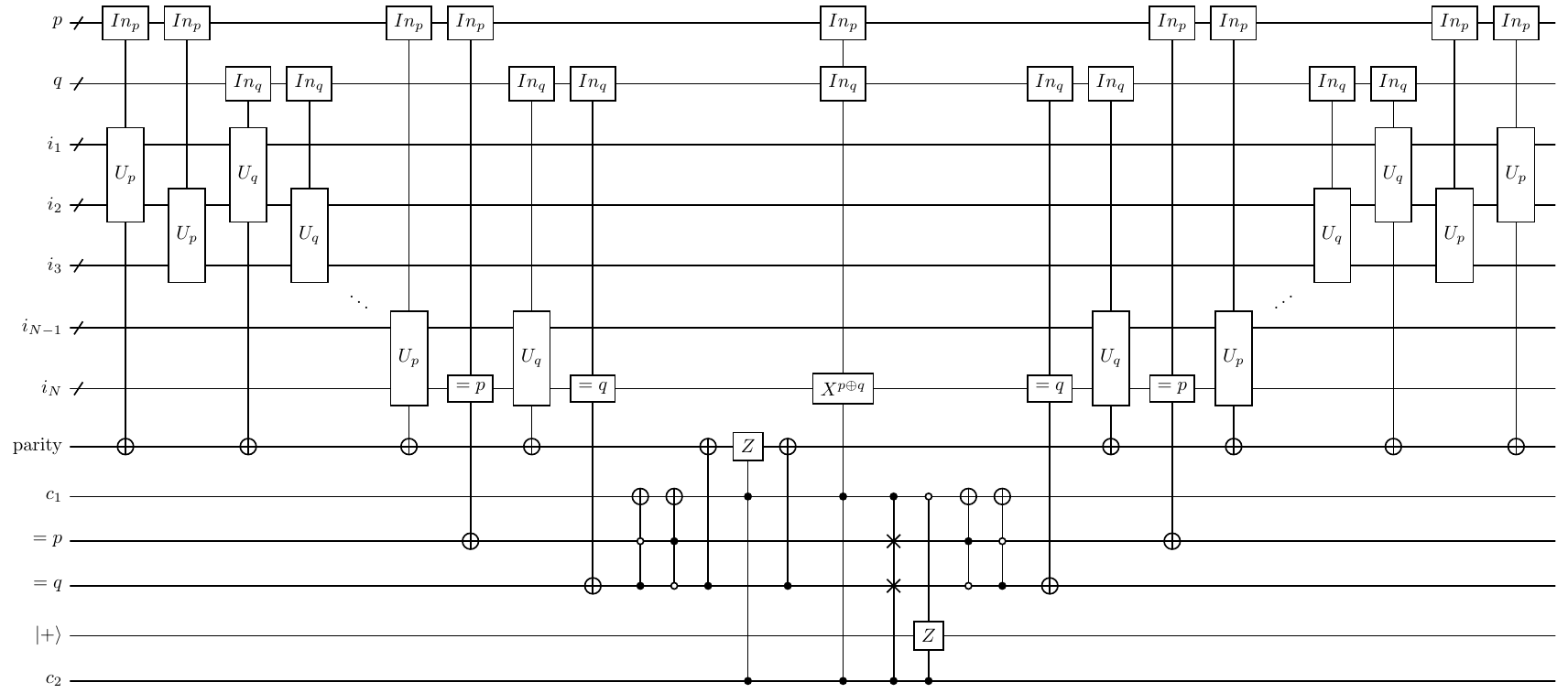}
  \caption{Circuit to implement $a_p^\dagger a_q + a_q^\dagger a_p$. The circuit can be trivially modified to obtain $ia_p^\dagger a_q -ia_q^\dagger a_p$.}
  \label{fig:select_one_majorana_free}
\end{figure*}

\begin{figure*}
  \centering
  \includegraphics[width=\linewidth]{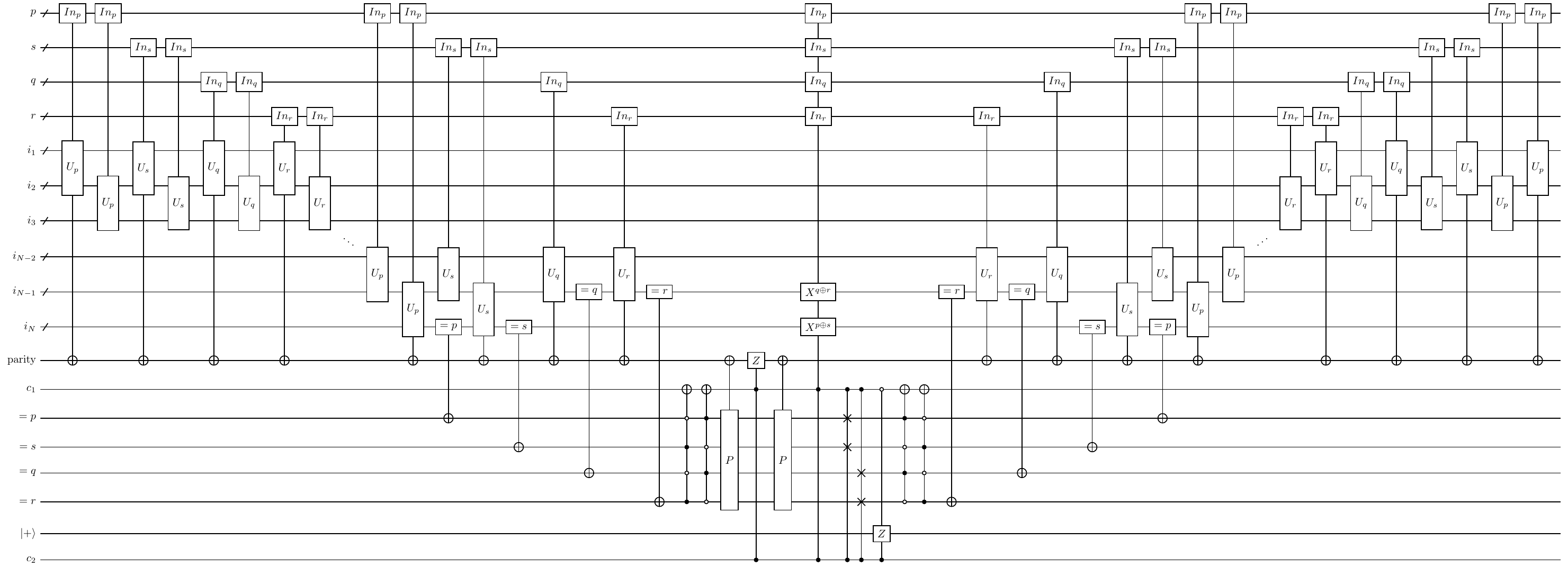}
  \caption{Circuit to implement $a_p^\dagger a_q^\dagger a_r a_s + a_s^\dagger a_r^\dagger a_q a_p$. The circuit can be trivially modified to obtain $ia_p^\dagger a_q^\dagger a_r a_s - ia_s^\dagger a_r^\dagger a_q a_p$.}
  \label{fig:select_two_majorana_free}
\end{figure*}

The mechanisms of the Majorana-free select circuits are similar to the Majorana-free Trotterization circuit.
Given an arbitrary Slater determinant $\ket{\psi}$, the Hermitian operator $a_p^\dagger a_q + a_q^\dagger a_p$ should swap the occupation of $p$ and $q$ if either $p$ is occupied or $q$ is occupied, but not both, while applying the appropriate phase depending on the parity of the wavefunction. Otherwise, $\ket{\psi}$ should vanish.
The former requirement is implemented using similar techniques as the Majorana-free Trotterization circuit, while the latter requirement uses an ancilla qubit in the $\ket{+}$ state as done in~\cite{suFaultTolerantQuantumSimulations2021}. If the requirements are not satisfied, a $Z$ gate is applied in this $\ket{+}$ state, followed by uncomputation. This is done similarly for the other Hermitian operators $ia_p^\dagger a_q - ia_q^\dagger a_p$, $a_p^\dagger a_q^\dagger a_r a_s + a_s^\dagger a_r^\dagger a_q a_p$, and $ia_p^\dagger a_q^\dagger a_r a_s - ia_s^\dagger a_r^\dagger a_q a_p$. As a result, the PREPARE circuit will simply need to prepare an additional ancilla qubit in the $\ket{+}$ state.
\begin{align*}
  \ket{+}_b&\ket{p}\ket{q}\ket{\psi}\\
           &\rightarrow (-1)^{b[(p\in \psi)\oplus (q\in \psi) \oplus 1]}\ket{+}_a\ket{p}\ket{q}\ket{\psi}\\
           &\rightarrow \begin{cases}
             \ket{+}_b\ket{p}\ket{q}a_p^\dagger a_q\ket{\psi}&\text{if } (p\in\psi) \oplus (q\in\psi)\\
             \ket{-}_b\ket{p}\ket{q}a_p^\dagger a_q\ket{\psi}&\text{otherwise}
           \end{cases}
\end{align*}

The circuit to implement $a_p^\dagger a_q + a_q^\dagger a_p$ is shown in \autoref{fig:select_one_majorana_free} while the circuit to implement $a_p^\dagger a_q^\dagger a_r a_s + a_s^\dagger a_r^\dagger a_q a_p$ is shown in \autoref{fig:select_two_majorana_free}. Those circuits can be trivially modified to obtain implementations for $ia_p^\dagger a_q -ia_q^\dagger a_p$ and $ia_p^\dagger a_q^\dagger a_r a_s - ia_s^\dagger a_r^\dagger a_q a_p$. Moreover, the use of the $\ket{+}$ operator can be extended to implement the non-Hermitian variants $a_p^\dagger a_q$ and $a_p^\dagger a_q^\dagger a_r a_s$.

\begin{table*}
  \centering
  \small
  \caption{
    Costings of the select operator along with the qubitization operator for molecular systems with MO basis sets using $\operatorname{PREPARE}_{\text{SF}}$.
    Here, QROM with dirty ancilla assistance is used to minimize the number of ancilla qubits.
    The parameters of the FeMoCo calculations are obtained from~\cite{berryQubitizationArbitraryBasis2019}.
  }
  \sisetup{
    table-format = 1.1e+2,
    parse-numbers = true
  }
  \begin{ruledtabular}
    \begin{tabular}{llcccScScSS}
      &&&\multirow[b]{2}{1cm}[-1.7mm]{QPE steps}&\multicolumn{2}{c}{SELECT}&\multicolumn{2}{c}{PREPARE}&\multicolumn{3}{c}{TOTAL}\\
      \cmidrule{5-6}
      \cmidrule{7-8}
      \cmidrule{9-11}
      System&Basis set&Orbs&&Qubits&{Toffolis/Step}&Qubits&{Toffolis/Step}&Qubits&{Toffolis/Step}&{Total Toffolis}\\
      \midrule
      FeMoCo     & RWSWT (54 elec)~\cite{reiherElucidatingReactionMechanisms2017} & 54  & 23 & 525  & 1.1E+04 & 178 & 5.2E+03 & 703  & 1.6E+04 & 6.9E+10 \\
      (from~\cite{berryQubitizationArbitraryBasis2019})
                 & LLDUC (113 elec)~\cite{liElectronicComplexityGroundstate2019}  & 76  & 22 & 1173 & 2.5E+04 & 176 & 6.7E+03 & 1349 & 3.2E+04 & 9.1E+10 \\
      \ce{H2O}   & STO-3G  & 7   & 17 & 87  & 1.7E+03 & 126 & 8.3E+02 & 213  & 2.5E+03 & 2.8E+08 \\
      10 elec    & 6-311G  & 19  & 20 & 115 & 2.3E+03 & 156 & 2.1E+03 & 271  & 4.5E+03 & 4.2E+09 \\
      Model      & cc-pvdz & 24  & 20 & 115 & 2.3E+03 & 160 & 2.7E+03 & 275  & 5.0E+03 & 6.5E+09 \\
      system     & cc-pvtz & 58  & 24 & 129 & 2.7E+03 & 191 & 6.4E+03 & 320  & 9.1E+03 & 1.6E+11 \\
                 & cc-pvqz & 115 & 27 & 143 & 3.0E+03 & 215 & 1.5E+04 & 358  & 1.8E+04 & 2.7E+12 \\
                 & cc-pv5z & 201 & 30 & 157 & 3.4E+03 & 237 & 3.4E+04 & 394  & 3.8E+04 & 3.3E+13 \\
      \ce{CO2}   & STO-3G  & 15  & 19 & 185 & 3.7E+03 & 141 & 1.6E+03 & 326  & 5.3E+03 & 2.2E+09 \\
      22 elec    & 6-311G  & 39  & 22 & 237 & 5.0E+03 & 176 & 4.2E+03 & 413  & 9.2E+03 & 3.9E+10 \\
      Model      & cc-pvdz & 42  & 22 & 237 & 5.0E+03 & 174 & 4.6E+03 & 411  & 9.6E+03 & 3.6E+10 \\
      system     & cc-pvtz & 90  & 25 & 263 & 5.6E+03 & 204 & 1.0E+04 & 467  & 1.6E+04 & 6.2E+11 \\
                 & cc-pvqz & 165 & 28 & 289 & 6.2E+03 & 224 & 2.0E+04 & 513  & 2.6E+04 & 5.7E+12 \\
                 & cc-pv5z & 273 & 30 & 315 & 6.9E+03 & 240 & 3.8E+04 & 555  & 4.5E+04 & 4.2E+13 \\
      \ce{EC}    & STO-3G  & 34  & 21 & 453 & 9.6E+03 & 168 & 3.5E+03 & 621  & 1.3E+04 & 3.9E+10 \\
      46 elec    & 6-311G  & 90  & 26 & 503 & 1.1E+04 & 204 & 1.0E+04 & 707  & 2.1E+04 & 1.1E+12 \\
      Battery    & cc-pvdz & 104 & 27 & 503 & 1.1E+04 & 209 & 1.2E+04 & 712  & 2.2E+04 & 2.2E+12 \\
      materials  & cc-pvtz & 236 & 31 & 553 & 1.2E+04 & 241 & 3.1E+04 & 794  & 4.3E+04 & 7.4E+13 \\
      \ce{LiPF6} & STO-3G  & 44  & 23 & 687 & 1.5E+04 & 179 & 4.4E+03 & 866  & 1.9E+04 & 1.3E+11 \\
      72 elec    & 6-311G  & 112 & 27 & 763 & 1.6E+04 & 212 & 1.2E+04 & 975  & 2.8E+04 & 3.3E+12 \\
      Battery    & cc-pvdz & 116 & 27 & 763 & 1.6E+04 & 212 & 1.3E+04 & 975  & 2.9E+04 & 3.2E+12 \\
      materials  & cc-pvtz & 244 & 31 & 839 & 1.8E+04 & 237 & 2.7E+04 & 1076 & 4.5E+04 & 7.1E+13 \\
    \end{tabular}
  \end{ruledtabular}
  \label{tab:qubitization_mol_mo_sf_1}
\end{table*}

\begin{table*}
  \centering
  \small
  \caption{
    Costings of the select operator along with the qubitization operator for molecular systems with MO basis sets using $\operatorname{PREPARE}_{\text{SF}}$.
    Here, QROM with clean ancilla assistance is used to minimize the number of Toffolis.
    The parameters of the FeMoCo calculations are obtained from~\cite{berryQubitizationArbitraryBasis2019}.
    The clean ancillae of the $\operatorname{PREPARE}$ step can be reused for the $\operatorname{SELECT}$ step.
    As a result, the qubit cost of the $\operatorname{SELECT}$ steps only contains data qubits $N\ltw M$ as opposed to $(N+4)(\ltw M+2) + 3$.
  }
  \sisetup{
    table-format = 1.1e+2,
    parse-numbers = true
  }
  \begin{ruledtabular}
    \begin{tabular}{llcccScScSS}
      &&&\multirow[b]{2}{1cm}[-1.7mm]{QPE steps}&\multicolumn{2}{c}{SELECT}&\multicolumn{2}{c}{PREPARE}&\multicolumn{3}{c}{TOTAL}\\
      \cmidrule{5-6}
      \cmidrule{7-8}
      \cmidrule{9-11}
      System&Basis Set&Orbs&&Qubits&{Toffoli/Step}&Qubits&{Toffoli/Step}&Qubits&{Toffoli/Step}&{Toffoli Total}\\
      \midrule
      FeMoCo     & RWSWT (54 elec)~\cite{reiherElucidatingReactionMechanisms2017} & 54  & 23 & 378 & 1.1E+04 & 457  & 2.6E+03 & 835  & 1.4E+04 & 5.8E+10 \\
      (from~\cite{berryQubitizationArbitraryBasis2019})
                 & LLDUC (113 elec)~\cite{liElectronicComplexityGroundstate2019}  & 76  & 22 & 904 & 2.5E+04 & 464  & 3.2E+03 & 1368 & 2.8E+04 & 8.1E+10 \\
      \ce{H2O}   & STO-3G  & 7   & 17 & 40  & 1.7E+03 & 163  & 6.3E+02 & 203  & 2.3E+03 & 2.6E+08 \\
      10 elec    & 6-311G  & 19  & 20 & 60  & 2.3E+03 & 218  & 1.3E+03 & 278  & 3.7E+03 & 3.4E+09 \\
      Model      & cc-pvdz & 24  & 20 & 60  & 2.3E+03 & 283  & 1.5E+03 & 343  & 3.9E+03 & 5.1E+09 \\
      system     & cc-pvtz & 58  & 24 & 70  & 2.7E+03 & 486  & 3.1E+03 & 556  & 5.8E+03 & 1.0E+11 \\
                 & cc-pvqz & 115 & 27 & 80  & 3.0E+03 & 885  & 5.8E+03 & 965  & 8.9E+03 & 1.3E+12 \\
                 & cc-pv5z & 201 & 30 & 90  & 3.4E+03 & 987  & 1.0E+04 & 1077 & 1.3E+04 & 1.2E+13 \\
      \ce{CO2}   & STO-3G  & 15  & 19 & 110 & 3.7E+03 & 204  & 1.0E+03 & 314  & 4.7E+03 & 1.9E+09 \\
      22 elec    & 6-311G  & 39  & 22 & 154 & 5.0E+03 & 316  & 2.2E+03 & 470  & 7.2E+03 & 3.1E+10 \\
      Model      & cc-pvdz & 42  & 22 & 154 & 5.0E+03 & 314  & 2.4E+03 & 468  & 7.4E+03 & 2.8E+10 \\
      system     & cc-pvtz & 90  & 25 & 176 & 5.6E+03 & 524  & 4.6E+03 & 700  & 1.0E+04 & 4.0E+11 \\
                 & cc-pvqz & 165 & 28 & 198 & 6.2E+03 & 943  & 8.1E+03 & 1141 & 1.4E+04 & 3.2E+12 \\
                 & cc-pv5z & 273 & 30 & 220 & 6.9E+03 & 1806 & 1.3E+04 & 2026 & 2.0E+04 & 1.9E+13 \\
      \ce{EC}    & STO-3G  & 34  & 21 & 322 & 9.6E+03 & 304  & 1.9E+03 & 626  & 1.1E+04 & 3.4E+10 \\
      46 elec    & 6-311G  & 90  & 26 & 368 & 1.1E+04 & 532  & 4.6E+03 & 900  & 1.5E+04 & 8.2E+11 \\
      Battery    & cc-pvdz & 104 & 27 & 368 & 1.1E+04 & 864  & 5.3E+03 & 1232 & 1.6E+04 & 1.6E+12 \\
      materials  & cc-pvtz & 236 & 31 & 414 & 1.2E+04 & 1775 & 1.2E+04 & 2189 & 2.4E+04 & 4.1E+13 \\
      \ce{LiPF6} & STO-3G  & 44  & 23 & 504 & 1.5E+04 & 346  & 2.3E+03 & 850  & 1.7E+04 & 1.2E+11 \\
      72 elec    & 6-311G  & 112 & 27 & 576 & 1.6E+04 & 883  & 5.5E+03 & 1459 & 2.2E+04 & 2.5E+12 \\
      Battery    & cc-pvdz & 116 & 27 & 576 & 1.6E+04 & 883  & 5.7E+03 & 1459 & 2.2E+04 & 2.5E+12 \\
      materials  & cc-pvtz & 244 & 31 & 648 & 1.8E+04 & 1740 & 1.2E+04 & 2388 & 3.0E+04 & 4.7E+13 \\
    \end{tabular}
  \end{ruledtabular}
  \label{tab:qubitization_mol_mo_sf_2}
\end{table*}

\clearpage

\let\oldaddcontentsline\addcontentsline
\renewcommand{\addcontentsline}[3]{}
\bibliography{references}
\let\addcontentsline\oldaddcontentsline

\end{document}